\documentclass{svjour3}                     
\smartqed  
\usepackage{graphicx}
\usepackage{amssymb,amsmath,latexsym}
\newcommand{\aap}{{Astron. Astrophys.}}
\newcommand{\apj}{{Astrophys. J.}}

\newcommand{\mnras}{{MNRAS}}
\newcommand{\pasj}{{PASJ}}
\newcommand{\aj}{{AJ}}
\newcommand{\apjs}{{ApJS}}
\newcommand{\apjl}{{Astrophys. J. Lett.}}
\newcommand{\procspie}{{Proc. SPIE}}
\usepackage[authoryear]{natbib}

\begin{document}

\title{Young and Intermediate-age Distance Indicators}


\author{Smitha Subramanian \and Massimo Marengo  \and Anupam Bhardwaj \and Yang Huang \and Laura Inno \and  Akiharu Nakagawa \and Jesper Storm }


\institute{S. Subramanian \at
              Kavli Institute for Astronomy and Astrophysics\\
              Peking University, Beijing, China\\
              \email{smitha@pku.edu.cn}
           \and
              M. Marengo \at
              Iowa State University \\
              Department of Physics and Astronomy, Ames, IA, USA\\
              \email{mmarengo@iastate.edu}           
                \and
               A. Bhardwaj \at
               European Southern Observatory\\
               85748, Garching, Germany\\
               \email{abhardwaj@eso.org}
                         \and
                Yang Huang \at
                Department of Astronomy, \\
                Kavli Institute for Astronomy \& Astrophysics,\\
                Peking University, Beijing, China\\
                \email{yanghuang@pku.edu.cn}
                 \and
               L. Inno \at
               Max-Planck-Institut f\"ur Astronomy\\
               69117, Heidelberg, Germany\\
               \email{inno@mpia.de}
            \and
                A. Nakagawa \at
                Kagoshima University, Faculty of Science\\
                Korimoto 1-1-35, Kagoshima 890-0065, Japan\\
                \email{nakagawa@sci.kagoshima-u.ac.jp}       
           \and
              J. Storm \at
              Leibniz-Institut f\"ur Astrophysik Potsdam (AIP) \\
	      An der Sternwarte 16, 14482 Potsdam, Germany\\
              \email{jstorm@aip.de}
}

\date{Received: date / Accepted: date}

\maketitle

\begin{abstract}
Distance measurements beyond geometrical and semi-geometrical methods, rely mainly on standard candles. 
As the name suggests, these objects have known luminosities by virtue of their intrinsic proprieties and play a major role in our understanding of modern cosmology. The main caveats associated with standard candles are their absolute calibration, contamination of the sample from other sources and systematic uncertainties. The absolute calibration mainly depends on their chemical composition and age. To understand the impact of these effects on the distance scale, it is essential to develop methods based on different sample of standard candles. Here we review the fundamental properties of young and intermediate-age distance indicators such as Cepheids, Mira variables and Red Clump stars and the recent developments in their application as distance indicators. 
\end{abstract}


\section{Introduction} 

Distance to celestial objects is one of the key astrophysical parameters.  Measurements of many fundamental properties, such as  luminosities, masses, sizes and ages, depend on distances. Moreover, distance measurements play an important role in our understanding of modern cosmology. We have various methods based on geometrical, semi-geometrical, photometric and kinematic methods, to estimate the distances to celestial objects. Each method has limited range of applicability, such as the radar techniques for solar system objects, parallax methods for stars within a few kpc, 
Cepheids standard candles to a few Mpc and Type Ia supernovae to several 1000 Mpc. So each method which makes each rung in the distance ladder, is used to calibrate the next most distant method. This approach enables us to determine the scale of the Universe. 


The parallax method of determining distance is the most direct and fundamental to all others. As the Earth revolves around the Sun, there is an apparent change in position of nearby stars with respect to the distant stars.  This apparent change in position is a measure of the distance to that object. These distances are purely based on geometrical calculation and no assumptions about the physical or chemical properties of the sources are required. Hence, parallax distances are considered to be highly reliable measurements. However, the parallax measurements of giant stars which have large radius are not very precise. For distant sources, the parallax becomes smaller and this limits the application of parallax method to very nearby objects. 

Distance measurement using eclipsing binaries, which is based on a semi-geometrical method, can reach up to the edge of the Local Group with an accuracy of 5\%.  
We can measure the sizes of the stars from the velocities and the eclipse durations, the luminosities from the radii and surface temperatures, and the distance to the system from the magnitudes. This requires good modelling of the binary system. For early-type stars,  this is a challenge due to the problems with accurate flux calibration and due to the degeneracy between stellar temperature and reddening. 
For late-type stars, these problems are minimal as very accurate flux calibration is available and accurate measurements of linear and angular sizes are possible. 
Thus, we can derive a distance which is primarily limited by the accuracy with which we can measure the light curve and the radial velocities. Such a distance is completely independent of the usual distance ladder. The most accurate distances, to date, to the Magellanic Clouds are estimated using eclipsing binaries (\citealt{piet13}, \citealt{g14}). 

Beyond the distances which can be reached using the geometrical and semi-geometrical methods, objects with known luminosity are  adopted. They are called standard candles and they include Cepheids, RR Lyrae stars, Mira stars and Red Clump (RC) stars. Whereas, Type Ia supernovae are standardisable candle as their luminosities are not known. What we know from nearby Type Ia supernovae is that there is a specific relation between their peak brightness and the time it takes for them to decay \citep{Phillips1993}. Intrinsically bright supernovae shine longer than faint supernovae. From the ratio of peak to width of their light curve, we can determine their absolute magnitude. Here, we restrict our review to distance indicators whose luminosities are known, which are standard candles. The main problems associated with standard candles are their calibration, selection and other systematic uncertainties. To mitigate these issues, several detailed studies are going on. 

In this chapter we review the fundamental properties of young and intermediate-age stellar populations such as, Cepheids (15 -- 300 Myr), Mira variables (1 -- 7 Gyr) and RC stars (2 -- 9 Gyr) and the recent developments in their application as distance indicators. 

\section{Cepheids}
\label{sec:cepheids}

Since Henrietta Leavitt discovered their Period-Luminosity (PL)  relation (\emph{Leavitt law}, \citealt{1912HarCi.173....1L}), Classical Cepheid variables have assumed a fundamental role as primary distance indicators. The Leavitt law was indeed the crucial ingredient allowing Edwin Hubble to determine the first reliable distance to M31 \citep{1929ApJ....69..103H}. In doing so he established the technique that in short order would lead to the discovery of the eponymous \emph{Hubble law} \citep{1929PNAS...15..168H}, and modern cosmology.

Almost a century later, Cepheids PL relations are still the cornerstone of the cosmological distance ladder. Cepheids in the Large Magellanic Cloud (LMC), the Small Magellanic Cloud (SMC) and other Local Group galaxies, provide the link between the local universe, distant standard candles (e.g. Supernovae Type Ia) and the Hubble Flow. The Hubble Space Telescope (HST) Key Project on the Extragalactic Distance  \citep{2001ApJ...553...47F}, in particular, tied all this together by delivering the first precise (10\% accuracy) determination for the value of the Hubble constant.


Despite their importance, Cepheid distances are still challenged by several factors. On the observational side, in addition to the statistical photometric uncertainties related to the measurement of Cepheid's mean magnitudes, the main errors in Cepheid distances are due to interstellar extinction and the systematic errors in the chosen PL relation. These systematic errors are related to the uncertainty in the PL slope and zero point, as well as its intrinsic scatter $\sigma_{IS}$, which in turn is the consequence of PL relations not taking into account the width in temperature of the Cepheid Instability Strip (IS). Due to this intrinsic scatter, PL relations should be really considered as statistical relations that can be correctly applied only to a statistical ensemble of stars, and not to individual Cepheids. As such, when Cepheids are used as galactic tracers to derive the three-dimensional structure of an underlying stellar population, the PL relation $\sigma_{IS}$ becomes the dominant source of error. Instead when we measure the distance of a group of Cepheid stars (such as when using Cepheids to derive the distance of a galaxy used to calibrate the next step in the cosmological distance ladder) interstellar extinction and the empirical calibration of the PL zero point remain the largest uncertainties.

The last decade has seen great progress to overcome these limitations, with observations moving towards infrared wavelengths (where extinction is reduced) and the adoption of HST parallaxes for a number of Cepheid calibrators. Great progress has also been made in bringing under control issues regarding the linearity of the PL relation, and its dependence on metallicity. This has allowed a new determination of the Hubble constant, anchored on Cepheids, with accuracy better than 3\% \citep{2012ApJ...758...24F}. The adoption of Period-Wesenheit (PW) relations has reduced the uncertainties related to the $\sigma_{IS}$ and to extinction at optical and near-infrared wavelengths, while templates have been produced to allow greater photometric accuracy even for sparsely sampled Cepheids light-curves. These developments were crucial for allowing Cepheids to assume the important role as galactic tracers described above.

In the theoretical arena the outstanding issue directly affecting the accuracy of Cepheid distances is the so-called ``Cepheids mass discrepancy'' between evolutionary models and Cepheid masses determined from pulsation theory and in binary systems. If this discrepancy is related to mass loss, it could lead to the formation of dusty circumstellar clouds, that would increase local extinction in the visible, and/or excess thermal emission in the infrared. Detailed calculations of the \emph{projection factor} and \emph{limb darkening}, necessary to precisely measure the distance of nearby individual Cepheid calibrators with variants of the the Baade-Wesselink method, such as the infrared brightness method, had also become available (\citealt{2002ApJ...567.1131M,2003ApJ...589..968M,Storm11a,2012A&A...543A..55N} and references therein).

In this section we discuss how these issues are dealt with modern measurements of Cepheid distances. In section~\ref{ssec:cepheids-nir} we discuss the techniques enabling the measurement of accurate distances for individual Cepheids in the near-IR, with particular emphasis to PW relations, independent determinations of reddening and the use of light curve templates. In section~\ref{ssec:cepheids-mir} we analyse the status of mid-infrared Cepheid distances as part of new efforts to improve the cosmological distance scale. In section~\ref{ssec:cepheids-linearity} we explore the issue of linearity in the PL relation and its role for the determination of the Hubble constant. In section~\ref{ssec:cepheids-metallicity} we discuss the effect of metallicity in the slope of the PL relation of Galactic and Magellanic Clouds Cepheids.


 %
\begin{figure}
  \includegraphics[width=\textwidth]{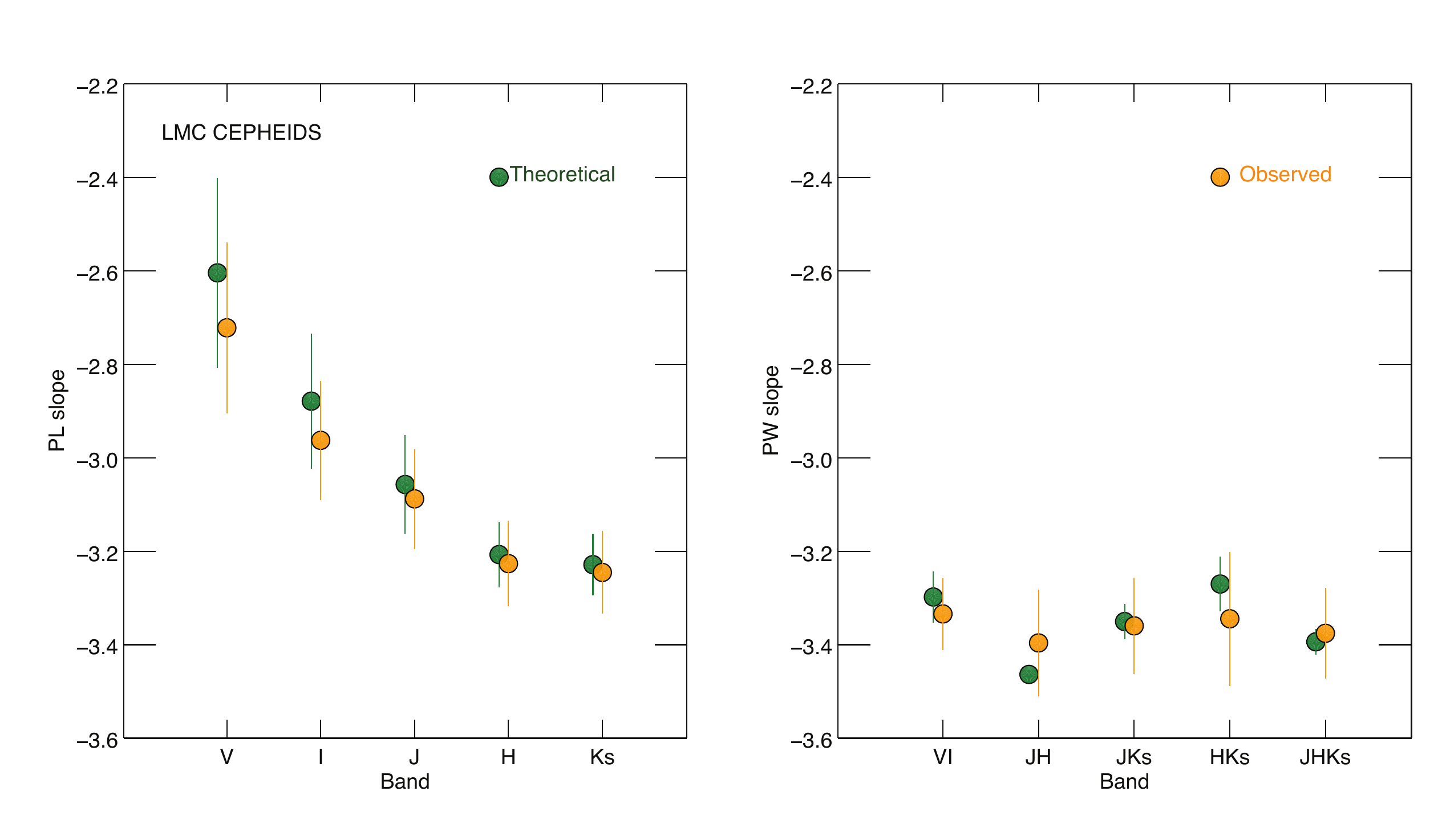}
\caption{Predicted (green) and observed (orange) slopes and dispersions ($\pm 1\sigma$) of 
Period--Luminosity (PL, \emph{left panel})  and Period--Wesenheit (PW, \emph{right panel}) relations of Cepheids in the LMC. 
The models and the data used to produce this plot are described in \cite{inno16}. 
The slope and dispersion of PL relations decrease with increasing wavelength. 
The smallest dispersion is found around the $K_S$ band PL relation,  
which is the less affected by systematics and by uncertainty on the reddening correction (i.e. $A_K\sim A_V/10$). 
However, ($JH$) and ($JHK_S$) PW relations show a smaller predicted dispersion than any PL, but they are reddening free by construction. 
The use of these relations to determine accurate Cepheids' individual distances is still limited by the lower photometric precision of NIR data, which propagates to uncertainty on the colour  term in the Wesenheit definition.}
\label{fig:1}       
\end{figure}
%


\subsection{Accurate individual Cepheid distances from near-infrared observations}
\label{ssec:cepheids-nir}

A comprehensive theoretical framework has been developed over the past 20 years 
to account for all the phenomenological features of pulsating stars
\citep{bono99,bono99b,bono00,marconi05,fiorentino13,marconi15}.
These models allow us to predict $\sigma_{IS}$ and, in turn, to pinpoint
the best tool
to derive Cepheids' individual distances. 
The left panel of Figure~\ref{fig:1} shows the predicted slope (green dots) 
of PL relations in optical ($VI$)  and near-infrared (NIR, $JHK_s$) bands
for $\sim$1,300 synthetic Cepheids in the LMC presented in \cite{inno16}. 
We note that non-linearity in the PL relations is not considered here. As we can see later (section~\ref{ssec:cepheids-linearity}), the implications of non-linearity on the distance scale or the value of Hubble constant is not statistically significant. 
The LMC is indeed a perfect workbench to test the accuracy of PL relations, 
since it is the closest star-forming galaxy almost face-on (inclination $\sim$ 25$^{\circ}$ -- 35$^{\circ}$: \citealt{jac16,inno16,subramanian13}),
with a relatively small extension along the line-of-sight  ($\sim$0.5--10 kpc, \citealt{subramanian09}),
and for which more than 3,000 Cepheids are known,
thanks to the extensive observational campaigns of micro-lensing surveys 
(e.g. MACHO: \citealt{alcock01}, EROS: \citealt{beaulieu95}, OGLE: \citealt{sos2008,sos2010}).
Moreover, the metallicity content of the LMC young stellar population
is well constrained \cite{romaniello08},
and we can adopt the mean metallicity $Z=0.008$ 
to produce a synthetic population of Cepheids.
The green error-bars in Figure~\ref{fig:1} 
indicate the dispersion around each relation.
In the case of the theoretical PL relations, 
the dispersion corresponds exactly to $\sigma_{IS}$. 
The error bars clearly show how
$\sigma_{IS}$ decreases when moving from
optical to the NIR bands. This trend
is a simple consequence of the
dependence of bolometric corrections on temperature,
which produces a narrowing of the IS in the Colour--Magnitude--Diagram (CMD)
at longer wavelengths. 
In the same figure,
the observed slopes and dispersions for the LMC 
Cepheids are also shown \citep{inno16}.
While the slopes are remarkably in agreement
with theoretical predictions, especially in the NIR bands,
the observed dispersions are systematically larger.
Such difference is due to the additional
 sources of scatter that affect the observed mean magnitudes: the photometric error, 
the uncertainty on the reddening correction and the dispersion due to the line-of-sight depth of the LMC. 
The scatter contributing to the latter term is indeed the quantity we want to measure.
Thus, by reducing the scatter due to the PL intrinsic dispersion to the reddening effects and to the photometric error,
we are able to improve the accuracy on the Cepheid's individual distances.

\subsubsection{Period--Wesenheit relations in the NIR}
\label{ssec:cepheids-pw}

Reddening corrections can be performed
by adopting different approaches. 
If the reddening law for a given galaxy or line of sight
is known, then multi-band photometry
of Cepheids allows us to obtain their
distances and colour  excess.
For instance, we can use the $B$ and $V$ band
PL relations to obtain the $E(B-V)$, and then
use the predicted total to selective extinction ratio
$A_V/E(B-V)$, to determine $A_V$.
The above approach is
similar to the use of the so-called PW relations,
where Wesenheit magnitudes \citep{vandenberg75,madore82} are
constructed to be reddening-free.
In fact, if we combine the two observed magnitudes  
m$_{\lambda_1}$ and m$_{\lambda_2}$,
into the Wesenheit magnitude%
\begin{equation}
W(\lambda_2,\lambda_1)=m_{\lambda_1}-\left[\frac{A(\lambda_1)}{E (m_{\lambda_2}-m_{ \lambda_1})}\right]\times (m_{\lambda_2}-m_{ \lambda_1}),
\label{eq: W}
\end{equation}
where  $A(\lambda_i)/E(m_{\lambda_2}-m_{ \lambda_1})$ 
is the total to selective extinction for the given filters,
we find that the observed Wesenheit magnitude is identical to the de-reddened one. 
Wesenheit magnitudes can be defined with any combination of photometric 
bands, but recent theoretical predictions \citep{bono00,marconi05,fiorentino07}
and empirical results  \citep{bono10,inno13,inno16} show that NIR PW 
relations are indeed the best suited tool to derive accurate Cepheids' distances.
In fact, they are linear over the entire period range and,
because they mimic a period-luminosity-colour relation, they have 
a smaller $\sigma_{IS}$ when compared to optical and NIR PL relations,
as demonstrated by comparing the right and left panel of Figure~\ref{fig:1}.
In the right panel the predicted slopes
(green dots) of the optical ($VI$) and NIR PW relations are plotted 
together with their intrinsic dispersions (error bars).
We find that the predicted dispersion around the $VI$
 ($\sigma_{IS}=0.05$ mag) and the $JHK$ ($\sigma_{IS}=0.03$ mag) PW relations
 are 30--60\% smaller than the one around the $K_s$ band PL relation ($\sigma_{IS}=0.07$ mag).
 
\begin{figure}
  \includegraphics[width=\textwidth]{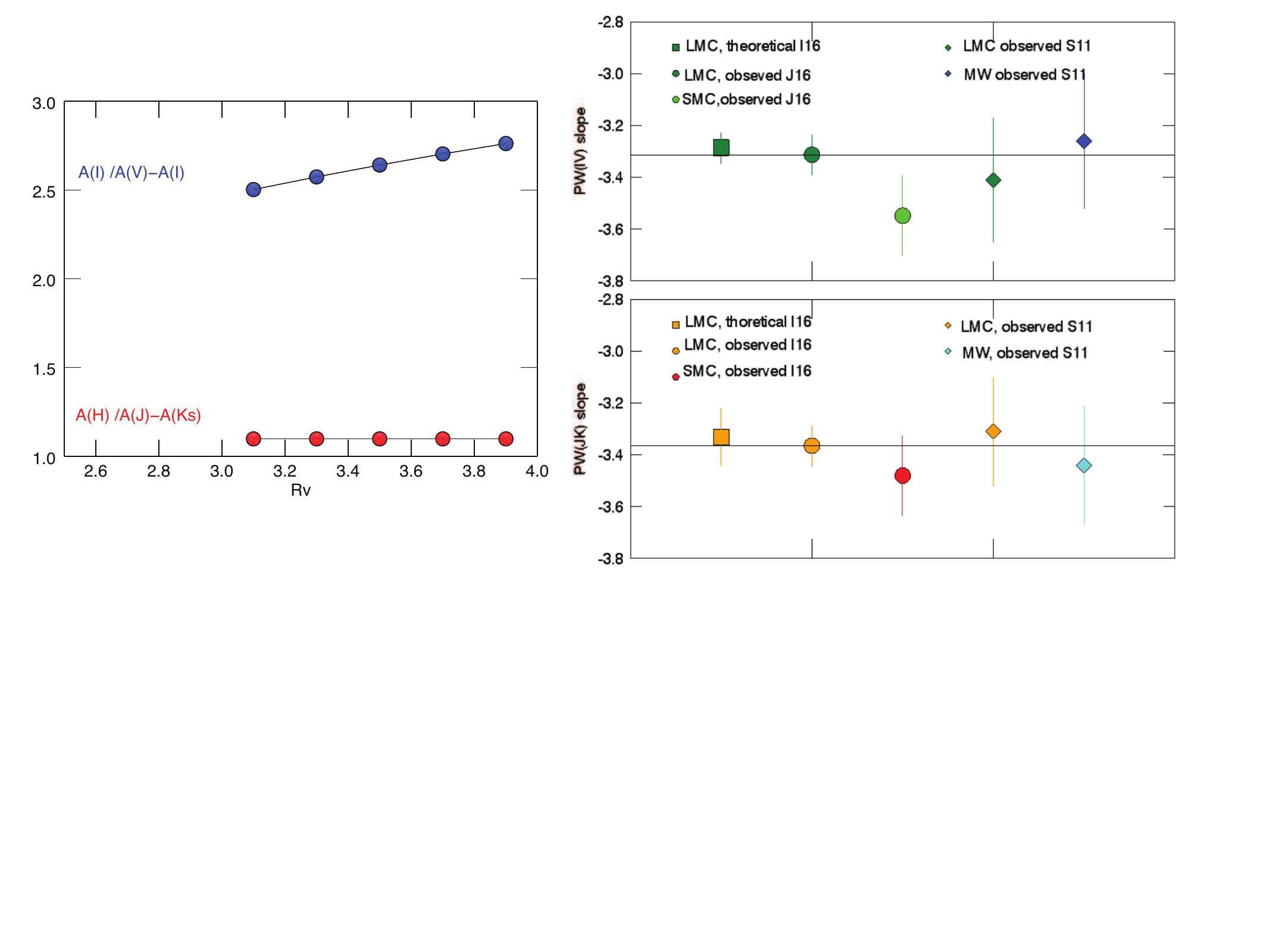}
\caption{\emph{Left} --- The variation of the Wesenheit coefficient  in the optical $I$ and $V$ bands (blue dots) and in the NIR $J$, $H$ and $K_S$ bands
as a function of the assumed total-to-selective absorption ratio ($R_V$). The NIR coefficient is insensitive to the changes in $R_V$.
\emph{Right} --- Comparison between slopes of PW relations in the optical (\emph{top}) and in the NIR (\emph{bottom}) as measured in the LMC, SMC and in the Milky Way.
The observed slopes in the optical are from \cite{jac16} 
for the LMC (green dot) and SMC (lime dot),
and from \cite{storm11} 
for the LMC (green diamond) and the Milky Way (blue diamond). The predicted slope from \cite{inno16} 
is also shown, while the black line
correspond to the observed slope of LMC Cepheids and can be used as a reference.
The NIR slopes are also taken from \cite{inno16} and \cite{storm11}. The comparison shows that the NIR PW relations are in very good agreement inside the dispersion,
thus they seem to be minimally affected by metallicity,  at least in the metallicity range covered by the Milky Way and the Magellanic Clouds.}
\label{fig:2}       
\end{figure}

Moreover, when comparing optical to NIR PW relations, we also 
find that they are minimally affected by uncertainties 
on the adopted reddening law and marginally affected by 
metallicity effects \citep{bono10,freedman10a}. 
For instance, in the left panel of Figure~\ref{fig:2} we show the change of
the Wesenheit coefficients with values of the
total-to-selective absorption ratio ($R_V$), ranging from 3.1 to 4 \citep{mccall04,demarchi16},
in the case of Cardelli's reddening law \citep{cardelli89}. 
Because the ratio $A(\lambda)/A(V)$ is
independent of $R_V$ for $\lambda \geqslant 0.9 \ \mu$m \citep{cardelli89}
and the reddening law can then be approximated by a power law, 
the NIR coefficient results do not change, while
the optical coefficient changes by $\sim$10\%.
In the right panel of Figure~\ref{fig:2} the comparison between slopes of PW relations for the Milky Way \citep{storm11} 
and the Magellanic Cloud Cepheids \citep{jac16,inno16} for the optical (top) and the NIR (bottom) is also shown.
In the case of the NIR PW ($JK_S$) relation, the slopes are common within the error
given by their scatter, thus indicating a small (if any) dependence on the metallicity,
at least in the metallicity range covered by these galaxies. 

In the right panel of Figure~\ref{fig:1} the observed slopes and dispersion of optical and NIR PW relations
for LMC Cepheids (orange dots) are also shown \cite{inno16}. 
Once again, we find that the observed dispersions ($\pm$ 1$\sigma$) are larger than the predicted ones. 
In fact, while the uncertainty on the reddening correction vanishes for the PW relations,
the photometric error becomes larger, since it propagates on the mean colour, 
which appears on the right side of Equation~1. 
Improving the photometric accuracy on NIR measurement is
then a crucial step in order to take full advantage of the use of NIR 
PW relations as tools to derive accurate Cepheids' individual distances.

\subsubsection{Independent reddening estimates}
\label{ssec:cepheids-independent-reddening}

An alternative approach based on the use of multi-band photometry of Cepheids
is the so-called reddening-law fitting method \citep{freedman85,rich14,inno16},
which allows to derive individual Cepheids' distances and  
extinction independently at the same time. 
The above method is based on the following expression 
of the Cepheid true distance modulus as a function of wavelength: 
\begin{eqnarray}
\label{eq4}
\mu_{0}=\mu_{obs}(x)+ \left[ a(x) R_V +b(x) \right] \times  E(B-V)
\end{eqnarray}
where $x\equiv\lambda^{-1}$, $a(x)$ and $b(x)$ are the coefficients of the adopted 
reddening law. Thus, if we obtain $\mu_{obs}$ in different filters
by using the PL relations, the fit will provide simultaneously 
$\mu_{0}$ and the extinction $E(B-V)$. However,
these quantities would still be affected by systematics 
due to the degeneracy in intrinsic colour and extinction.

Reddening maps can also be used to predict the amount of extinction in a given band
for a specific line-of-sight.  
In the case of an external system, such as the Magellanic Clouds, two-dimensional
reddening maps can be adopted to correct the observed magnitude for the individual reddening. 
This method was applied in \citep{haschke12,macri15,ripepi16,bhardwaj16}, who used the
reddening maps from \cite{haschke11}.
In the Galaxy, individual reddening along the line-of-sight can be
obtained from up-to-date three-dimensional reddening maps
\citep{marshall06,green15,schlafly16}. This approach is very promising,
especially when accurate three-dimensional map
of the dust distribution in the Milky Way will become available thanks to Gaia \citep{rezaei16}. 
Individual reddening of Cepheids can also be
determined from their spectra, by using empirically calibrated
relations between the equivalent width of Diffuse Interstellar Bands (DIB)
and $E(B-V)$ \citep{friedman11}. 
For instance, this method has been applied to derive de-reddened magnitudes
of Cepheids in the inner disk by \citep{martin15,andrievsky16}. 
Note that all above methods still rely on the assumption of the reddening
law and thus total-to-selective absorption ratio, thus they are still prone to
systematic errors.

\subsubsection{NIR templates as tools to optimize NIR variability surveys}
\label{ssec:cepheids-templates}

The main drawback in using NIR PL or PW relations to determine
Cepheid distances is that NIR light curves
are available for a limited number of stars.
For instance, in the Galaxy only 
$\lesssim$200 \citep{laney92,monson11} out of $\gtrsim$600 \citep{dambis15}
known Cepheids have accurate NIR photometry. 
The reasons is that variability surveys have been traditionally 
performed in optical bands, where the pulsation amplitude is 
typically larger, (e. g. $\sim$1.2 mag in $B$ vs. $\sim$0.2 mag in $K_S$),
thus allowing a more robust classification of the variables.
On the other hand, the lower amplitudes in the NIR bands enable us 
to determine accurate mean magnitudes  
from a smaller number of phase points ($\sim$10--12 vs. $\sim$30).
This is a sensitive issue, especially for surveys 
aiming at discovering new Cepheids
in the most extincted region of the Galaxy, 
where NIR photometry is necessary to cope with such high extinction 
(e.g. VISTA Variables in the 
Via Lactea (VVV), \citealt{minniti10}, InfraRed Survey Facility (IRSF), \citealt{matsunaga11,matsunaga13,matsunaga16}). 
However,  NIR ground--based observations are still more time-consuming than 
optical observations, because of sky subtraction. 
A cheap solution to this problem in terms of observing time, is the use  of light-curve templates
 to determine accurate NIR mean magnitudes from a few or even only one measurement.
Such templates are available in the literature for both RR Lyrae 
\citep{jones96,sesar10} and classical Cepheids \citep{sos2005,pejcha12,inno15,ripepi16}
and rely on empirical calibrations.

\begin{figure}
  \includegraphics[width=\textwidth]{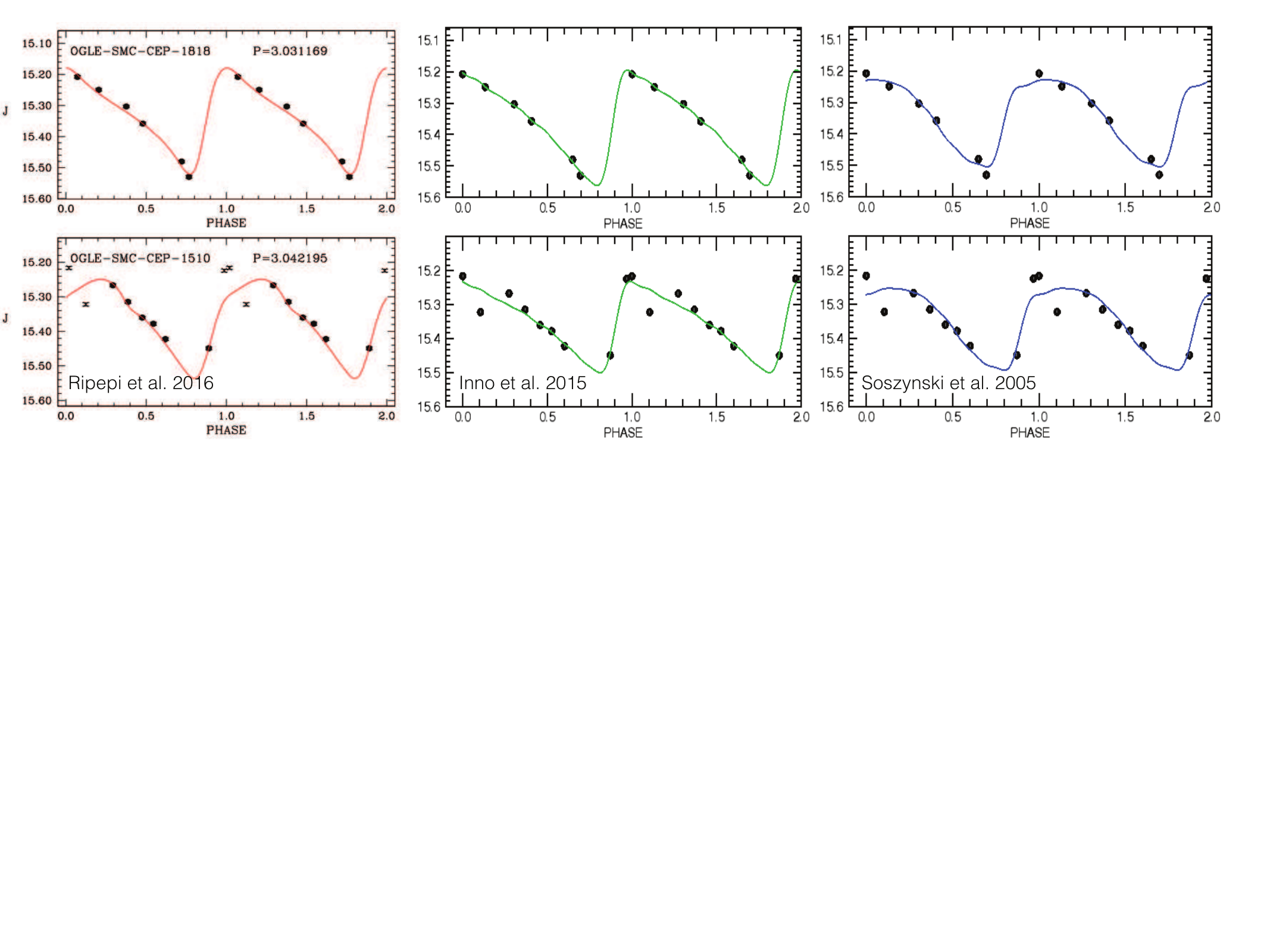}
\caption{A comparison between $J$ band templates available in the literature applied to the VMC data of two SMC Cepheids with periods of about 3 days. 
The left panels are directly taken from Figure~11 in \cite{ripepi16}, while the middle and right panels have been obtained using the approach
by \cite{inno16} and by \cite{macri15} respectively. In the case of the Cepheid OGLE-SMC-CEP-1818, shown in the top row,
the templates by \cite{ripepi16} (\emph{left panel}) and by \cite{inno16} (\emph{middle panel}) are more accurate than the one from \cite{sos2005},
which predicts a shallow bump on the rising branch, that is not consistent with the data.  The differences in the fit produces a difference in
the resulting mean magnitude of 0.03 mag, or $\gtrsim$ 1.5\% in distance.
The bottom row shows similar results in the case of the Cepheid OGLE-SMC-CEP-1510. Note that the
method implemented by \cite{ripepi16} rely on an outlier cut, which removes three observation (crosses), while this interaction
is not necessary in the case of the template-fitting method used by \cite{inno16}, which provides a mean-magnitude smaller of $\lesssim$0.01 mag,
 or 0.5\% in distance.
The use of the new accurate templates available in the literature allows us to derive accurate mean magnitude, and hence distances,
even for sparsely observed Cepheids.}
\label{fig:3}       
\end{figure}

\paragraph{NIR templates to fit sparsely sampled light-curves.}
Recent and ongoing surveys, such as the VVV, the VISTA Survey of Magellanic Clouds System (VMC, \citealt{cioni11}),
the NIR Synoptic Survey \citep{macri15} and the IRSF surveys,
use NIR bands to access farther or more obscured stars.
In order to optimise their scientific outcome respect to the 
surveyed area and the cadence, 
these surveys collect sparsely sampled light-curves in one
or more NIR bands. 
However, if the period of a given Cepheid is known with sufficient accuracy, 
it is possible to perform a template-fitting to the multi-epoch observations 
in order to accurately determine their mean magnitudes.
For instance, VMC provides NIR magnitudes 
measured at on average $\sim$5.7, $\sim$6.3 and $\sim$16.7 epochs 
in the $Y$, $J$ and $K_{\rm{S}}$ band respectively, for the SMC 
Cepheids in the OGLE-III catalog. \cite{ripepi16} used the best sampled light-curves in the VMC dataset
to build new templates, interpolated the data of the sparsely sampled curves with splines and then applied a template-fitting procedure in
three steps. They obtained first the optical-to-NIR scaled amplitudes, 
then the  phase-lags between
the OGLE optical and the NIR light curves and finally
the mean magnitudes. 
\cite{inno16} applied directly the templates from \cite{inno15} to 
similar VMC data for LMC Cepheids and solved independently for amplitudes, 
mean magnitudes and phase-lags at the same time. 
A comparison between of the two mentioned template-fitting techniques applied to
the VMC light curves for Cepheids in the SMC is shown in the left and middle panels of
Figure~\ref{fig:3}.
The templates allow to predict the shape of the light curve and
then determine accurate mean-magnitudes from six epochs. Differences
on the mean magnitude obtained by adopting the templates by \cite{ripepi16} (red line)
and the ones by \cite{inno16} (green line) are smaller than 0.01 mag, or 0.5\% in distance to the SMC.
The template-fitting approach is also to be preferred when
the photometric error on individual observations is 
so large ($\gtrsim$0.1 mag) to introduce
spurious features in the light-curve shape.  
This is for instance the case of light curves from the Near-Infrared Synoptic Survey.
\cite{macri15} uses the templates by \cite{sos2005} and a template-fitting
procedure similar to the one described above to determine
LMC Cepheids $J$, $H$ and $K_S$ mean magnitudes.
For completeness, we also applied the same approach to the VMC data in Figure~\ref{fig:3} (right panel).
The fit obtained by using the templates by  \cite{sos2005} (blue line) seems to be
 less accurate in reproducing the light-curve shape of the SMC Cepheids, with
 a difference on the mean magnitude of $\sim$0.03 mag, or 1.5\% in distance to the SMC.

\paragraph{NIR templates to correct for random-phase effects.}
If the light curve in the optical band is available,
the NIR templates can also be used to obtain mean-magnitude
from NIR single-epoch observations \citep{sos2005,inno15}.
The accuracy on these mean-magnitudes  is limited by  
$i)$ the photometric error on the NIR single-epoch observation;
$ii)$ the precision to which we can predict the phase of the NIR observation and the NIR amplitude; and
$iii)$ the accuracy of the template itself.
Indeed, the templates by \cite{inno15} provide mean magnitude with an error lower than 2\%
for single-epoch measurements with 1\% photometric precision or better.
Thus, the uncertainty on Cepheid's individual distance determination
is already dominated by the $\sigma_{IS}$ ($\gtrsim$ 3\%).
This means that these templates allow us to 
already reach the precision limit of the method, 
even with single-epoch NIR observations, if the photometric precision is sufficient.

\subsubsection{Summary}
\label{ssec:cepheids-nir-summary}

NIR PW relations allow us to derive Cepheid distances 
with the accuracy to meet the precision of ongoing cosmological experiments.
In fact, they appear to be linear over the entire period range and their slopes appear to be common between the Magellanic Clouds, 
indicating that they are universal in the considered metallicity range (see section~\ref{ssec:cepheids-linearity} and \ref{ssec:cepheids-metallicity} for an in-depth discussion of the linearity of the NIR PL relations and their dependence on metallicity). 
Moreover, PW relations are minimally affected by uncertainties in the adopted reddening law. 
For instance, by using NIR PW relations, \cite{inno13} obtained 
the most precise estimate of distances to the Magellanic Clouds 
based on Cepheids, with random errors at 1\% level 
and systematics at 5\% level. 
The systematic errors are dominated by the uncertainties in the PW zero point. 
This means that the new independent calibrations from Gaia and HST
will play a crucial role to reach a better accuracy. 

Moreover, new NIR light-curve templates can provide accurate mean magnitudes 
from a few or even one observation, allowing a significant saving of telescope time 
and an optimal usage of data already available (e.g. from 2MASS, VVV, VMC, etc.). 
The use of templates to predict the amplitude and the shape of the variation of physical quantities of Cepheids
along the pulsation cycle, such as luminosity and  radial velocity, is a powerful tool.
In fact, this approach (single-epoch photometry and spectroscopy + light and radial velocity curve templates for Cepheids) 
is very efficient to investigate the spatial distribution and
the kinematics of the resolved stellar population in nearby galaxies,
especially in the approaching E-ELT era, 
since we will be able to apply it to more distant spiral galaxies in the Local Group and beyond ($>$4Mpc).


\begin{figure*}
\includegraphics[width=0.99\textwidth,keepaspectratio]{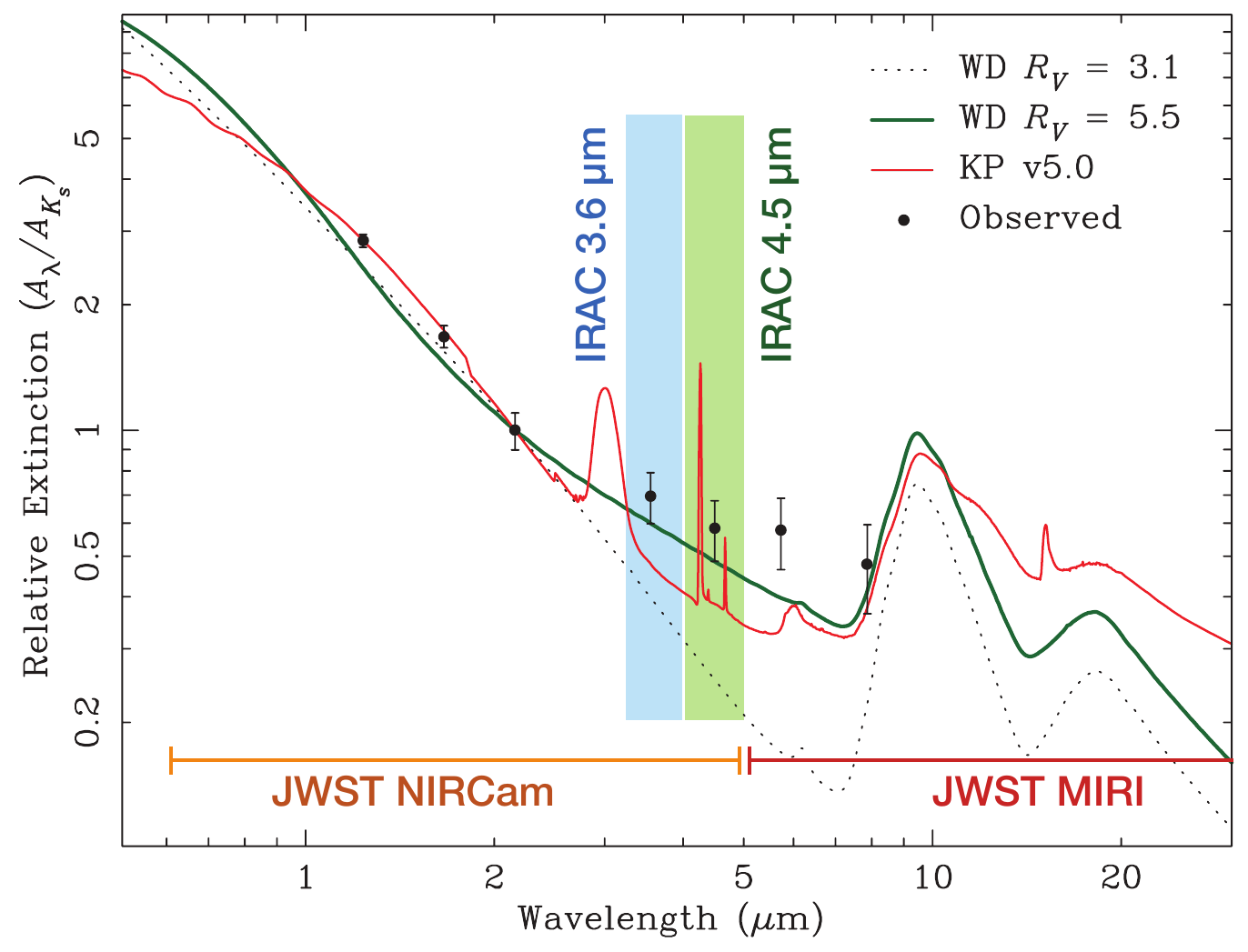}
\caption{Infrared average extinction law in the Perseus molecular cloud (data), compared with a variety of dust models. Warm Spitzer/IRAC and JWST bandpasses are indicated on the plot. The minimum extinction is found in the spectral region covered by warm IRAC and the red edge of JWST NIRCam. Adapted from \cite{2009ApJ...690..496C}.}
\label{fig:chapman}
\end{figure*}

\subsection{Mid-infrared Cepheid distances and the cosmological distance scale}
\label{ssec:cepheids-mir}

Observing Cepheids in the mid-infrared (MIR, $L$, $M$, $N$ and $Q$ photometric bands) leads to significant advantages with respect to optical wavelengths. Recent determinations of the interstellar and extragalactic extinction curve (see e.g. \citealt{2009ApJ...690..496C,1985ApJ...288..618R,2005ApJ...619..931I,2007ApJ...663.1069F,2007ApJ...664..357R,2009ApJ...696.1407N}, Figure~\ref{fig:chapman}) show that in the thermal infrared, the dust extinction is reduced by more than an order of magnitude  compared to the $V$ band ($A_\lambda/A_V \simeq 0.065$ and $\simeq 0.052$ in the $L$ and $M$ spectral bands respectively \citealt{2005ApJ...619..931I}). Furthermore, for stars with G and K spectral type the shape of the spectrum closely follows the temperature-independent Rayleigh-Jeans distribution. As a consequence, the brightness variations related to the pulsations are largely insensitive to the effective temperature variation, being dominated by the smaller changes in surface area consequence of the stars' radial pulsations. This leads to small amplitudes for wavelengths longer than the $K$ band ($\sim 0.3$-0.4~mags for the average 10~days period), which reduces the number of epochs necessary to measure accurate average magnitudes. Finally, infrared Cepheid spectra are mostly free from line blanketing, in principle reducing the dependence of the Leavitt law from metallicity, even though CO absorption can still play a significant role (see discussion in section~\ref{ssec:co} below). For these reasons the last decade has seen multiple efforts aimed to develop robust Cepheid distance determinations using their PL relation at MIR wavelengths.

\subsubsection{Empirical PL relations and the CHP program}
\label{ssec:empiricalPL}

Mid-infrared astronomy received a boost in sensitivity and accuracy by many orders of magnitude with the launch of the Spitzer space telescope \cite{2004ApJS..154....1W}, and in particular its InfraRed Array Camera (IRAC, \citealt{2004ApJS..154...10F}). IRAC is currently operating in the extended ``warm mission'' following the exhaustion of its cryogenic liquid Helium (LHe), still acquiring images and providing better than 1\% photometry in its 3.6 and 4.5~$\mu$m bands. IRAC enabled obtaining accurate magnitudes of Cepheid stars as far as the Magellanic Clouds, in addition to the Milky Way Galaxy. These capabilities have been exploited by several teams that derived an empirical calibration of MIR Cepheids PL relations \citep{2008ApJ...679...76N,2010ApJ...709..120M,2011ApJ...743...76S,2012ApJ...759..146M,2015ApJ...813...57N}.

\begin{figure*}
\includegraphics[width=0.99\textwidth,keepaspectratio]{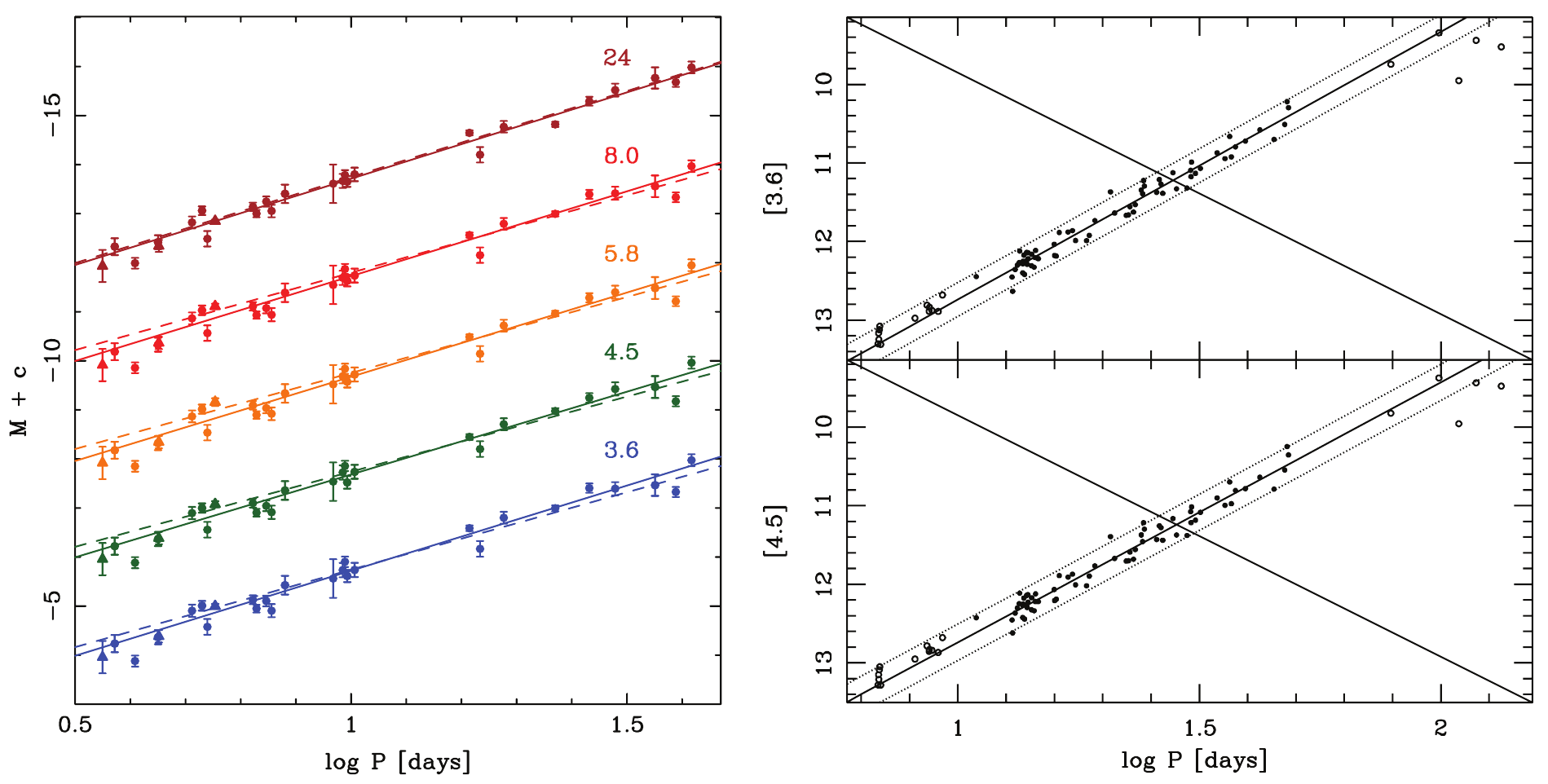}
\caption{Empirical Leavitt law of Classical Cepheids in the Galaxy (left panel, adapted from \citealt{2010ApJ...709..120M}) and in the LMC (right panel, adapted from \citealt{2011ApJ...743...76S}). These Galactic Classical Cepheids relations have been obtained with the Spitzer space telescope IRAC (3.6 to 8.0~$\mu$m) and MIPS (24~$\mu$m) instruments during the cryogenic mission, and are based on two epochs measurements. Solid lines show the best fit PL relation for all stars, while the dashed line was derived using only stars with astrometric distances. The period of first overtone Cepheids was ``fundamentalised'' using the relation from \cite{1997MNRAS.286L...1F}. The LMC relations have instead been obtained during the Spitzer warm mission, and are based on carefully carefully reconstructed light curves (24 epochs each) for each Cepheid. A similar strategy was used by \cite{2012ApJ...759..146M} to derive accurate zero points for the Galactic 3.6 and 4.5~$\mu$m Classical Cepheids PL relation.}
\label{fig:pl-ir}
\end{figure*}

Figure~\ref{fig:pl-ir} shows the 3.6 to 24~$\mu$m Leavitt law (\citealt{2010ApJ...709..120M}) measured for Galactic Classical Cepheids with IRAC and MIPS \citep{2004ApJS..154...25R} during the Spitzer cryogenic mission. Each star was observed in two random epochs six months apart. The PL relation zero point was then determined using parallaxes measured with the HST Fine Guidance Sensors \citep{2007AJ....133.1810B} and Hipparcos (for Polaris, \citealt{2007MNRAS.379..723V}), or distances derived with the InfraRed Surface Brightness (IRSB) technique by \cite{2007A&A...476...73F} (see detailed description of the method in section~\ref{sec:IRSB}). The calibration of the Classical Cepheids PL relation was then refined at 3.6 and 4.5~$\mu$m during the Spitzer/IRAC warm mission with the Carnegie Hubble Program (CHP, \citealt{2011AJ....142..192F}). As part of this program \cite{2012ApJ...759..146M} measured well sampled (24 epochs) light-curves of 37 Galactic Classical Cepheids with accurate astrometric or IRSB distances, calibrating the PL zero points with a $\sim 3$\% accuracy. \citet{2011ApJ...743...76S,2016ApJ...816...49S} then used the Galactic zero points to measure the distance and geometry of the LMC and SMC with a $\sim 4$\% accuracy. This provided a new, direct way of anchoring the extragalactic distance scale to Classical Cepheid distances, allowing a new measurement of the Hubble constant and other cosmological parameters (\citealt{2012ApJ...758...24F}).

\begin{figure*}
\includegraphics[width=0.99\textwidth,keepaspectratio]{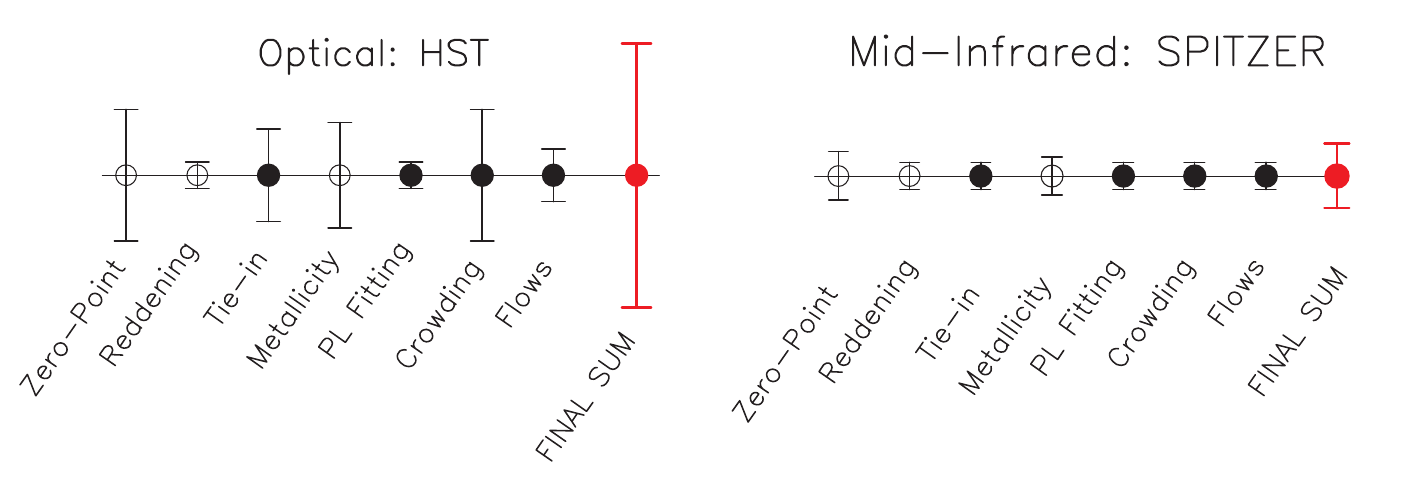}
\caption{Comparison of individual error sources in the Hubble constant determination between the HST Key project (left) and the CHP (right). Open symbols are the uncertainties most directly affected by the change in wavelength from optical to infrared bands. The final sum shows how the overall uncertainty decreased from $\sim 10$\% (H$_0$ = 72 $\pm$ 7 km s$^{-1}$ Mpc$^{-1}$) in the HST Key Project to $\sim 2.8$\% (H$_0$ = 74.3 $\pm$ 2.1 km s$^{-1}$ Mpc$^{-1}$ ) in the CHP. Adapted from \cite{2012ApJ...758...24F}.}
\label{fig:chp}
\end{figure*}

Figure~\ref{fig:chp} breaks down the error budget in the CHP (right) and HST Key Project determination of $H_0$. The open symbols indicate the sources of uncertainty more directly related to the change in wavelength from the optical to the MIR range. The largest improvement is in the PL zero point calibration. This is the consequence of being able to use Galactic calibrators with astrometric distances to anchor the Leavitt law, rather than relying to independent estimates of the LMC distance. As mentioned above, this is only possible in the infrared where the extinction of individual calibrator stars is one order of magnitude less than in the visible. The effect of reddening itself on the overall $H_0$ determination is however not dramatically reduced, since it contributes in the form of a statistical uncertainty, and was already limited to less than $\sim 1$\% in the HST Key Project by observing a large number of distance indicators. The uncertainty due to metallicity effects is instead halved from $\sim 3.5$\% to $\sim 1.4$\%, again a consequence of being able to directly calibrate the Cepheids PL relation at high metallicity in the Galaxy, rather then relying to the intermediate step of calibrating the zero point using the lower metallicity LMC (see detailed discussion in \citealt{2012ApJ...758...24F}).

The overall effect of moving from visible to infrared bands results in both the reduction of the uncertainty in the $H_0$ determination (from $\sim 10$\% in the HST Key Project to $\sim 2.8$\% in the CHP) and an increase in the value of the constant (from $H_0 = 72 \pm 3 \pm 7$ to $74.3 \pm 2.1$~km~s$^{-1}$~Mpc$^{-1}$ in the HST Key Project and CHP, respectively). The increase, while still within the uncertainties, is significant because it goes in the opposite direction of recent determinations of $H_0$ from the analysis of Planck Cosmic Microwave Background (CMB) radiation and Baryonic Acoustic Oscillations (BAO), which are instead favouring lower values (e.g. $67.8 \pm 0.9$~km~s$^{-1}$~Mpc$^{-1}$ and $68.8 \pm 2.2$~km~s$^{-1}$~Mpc$^{-1}$ in \citealt{2016A&A...594A..13P} and \citealt{2015MNRAS.448.3463C} respectively). A careful assessment of the systematics present in the ``direct'' determination of the distance scale (anchored on stellar indicators such as Cepheids and SN Type Ia) vs. the CMB/BAO analysis, reveals that this \emph{tension} in the value of $H_0$ is approaching $3 \sigma$ \citep{2016ApJ...826...56R}. Solving this issue has important consequences for cosmology, since the CMB analysis is dependent on the specifics of the $\Lambda$CDM cosmological model used to fit the data. Any discrepancy with a direct determination of $H_0$ and the other cosmological parameters would be an indication of new physics, such as the number or relativistic species and/or the mass of neutrinos (see e.g. \citealt{2014PhRvD..90h3503D,2014PhRvL.112e1302W}). For this reason it is worth analysing in detail what are the possible systematics that may still affect the role of Cepheids as infrared standard candles, as well as to develop complementary ``direct'' routes to the distance scale, anchored to Population II distance indicators such as RR~Lyrae and the Tip of red giant branch (RGB) in Globular Clusters, or other standard candles such as RC stars and Miras.

\subsubsection{Effects of variable CO absorption on Cepheids infrared distances}
\label{ssec:co}

The method followed by the CHP to calibrate the Leavitt law for Classical Cepheids relies on the assumption that the PL slope (determined by fitting LMC Cepheids) and zero point (measured from Galactic Cepheids with known distances) do not depend in a significant way from metallicity. An in-depth discussion of metallicity effects on Cepheids PL relations, with special emphasis on the NIR, is presented in section~\ref{ssec:cepheids-metallicity}. In the MIR, as mentioned before, the influence of metallicity is generally believed to be lower, due to the far less importance of line blanketing at wavelengths where the stellar spectrum approaches the Rayleigh-Jeans limit. Indeed, tests performed as part of the CHP showed no significant correlation in the residuals between the adopted PL relations and the 3.6 and 4.5~$\mu$m magnitudes of individual stars over the metallicity range $-0.6 < \textrm{[Fe/H]} < +0.2$ (see \citealt{2012ApJ...758...24F}) of the observed stars.

\begin{figure*}
\includegraphics[width=0.99\textwidth,keepaspectratio]{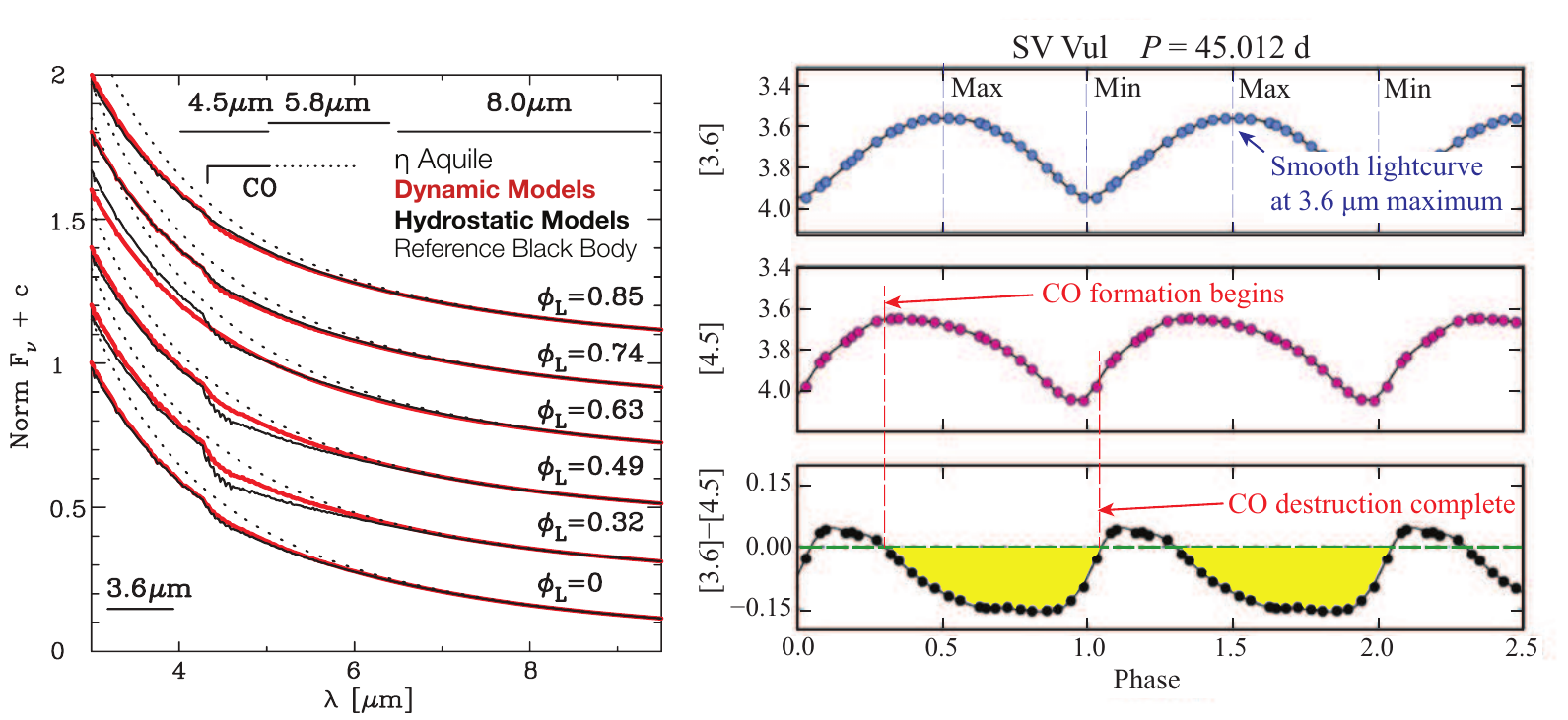}
\caption{\emph{Left} --- Series of time-dependent models of the Galactic Classical Cepheids $\eta$~Aql showing the changes in the broad CO absorption feature straddling the IRAC 4.5~$\mu$m band (adapted from \citealt{2010ApJ...709..120M}). \emph{Right} --- Colour curve of the long period ($P = 45.012$~days) Galactic Classical Cepheid star SV Vul, adapted from \cite{2016MNRAS.459.1170S}. Note the flattening of the 4.5~$\mu$m curve at maximum light, with a pronounced $[3.6]-[4.6]$ blue colour as the star progresses towards minimum radius.}
\label{fig:co}
\end{figure*}

Figure~\ref{fig:co} however shows that the spectral brightness in the $M$ band (straddling across the IRAC 4.5 and 5.8~$\mu$m bands) is severely affected by variable CO absorption. Time-dependent hydrodynamic models of pulsating Cepheids (left panel, calculated for $\eta$~Aql by \citealt{2002ApJ...567.1131M}) reveal that CO absorption is greatly increased for pulsation phases following maximum radius, where the atmosphere is expanding and the effective temperature is at its minimum. As the star approaches minimum radius (corresponding to minimum infrared luminosity) the CO molecule is chemically destroyed, removing a significant fraction of its absorption feature in the star's spectrum. The effect of this variable CO abundance on the infrared light curve and colour is studied in great detail by \cite{2016MNRAS.459.1170S}, and shown in the right panel of Figure~\ref{fig:co}. Since at 3.6~$\mu$m the stellar spectrum is largely unaffected by CO absorption, the light curve is smooth during the entire pulsation cycle. At 4.5~$\mu$m, however, the formation of the CO molecule sets a ceiling in how bright the star could become at maximum light, flattening the light curve, until CO is destroyed near minimum radius. This behaviour leads to a well defined negative-slope period-colour-metallicity relation for Classical Cepheids with period between $\sim 6$ and 60~days ($0.8 < \log P < 1.8$), characterised by \cite{2016MNRAS.459.1170S} for stars in the Galaxy, LMC and SMC. For Cepheids with period shorter than 6~days the relation flattens, due to the higher effective temperature ($T_{eff} \simeq 6309~K$) that prevents the formation of CO spectral headbands for the entire pulsation cycle. The period-colour relation for ultra-long period Cepheids ($P$ greater than $\approx 60$~days) is instead dramatically inverted (large positive slope). The reasons why this happen is not well understood. Several lines of evidence points towards the effects of lower rotation, suppressing rotational mixing that in turn would decrease the abundance of CNO elements in the photosphere  \citep{2014A&A...564A.100A}, as well as the lack of first dredge-up for Cepheids in this period range \citep{2000ApJ...543..955B}. Both phenomena would prevent the formation of significant amount of CO.

Based on this analysis, \cite{2016MNRAS.459.1170S} concluded that the  IRAC band at 3.6~$\mu$m is the safest choice for measuring precision Cepheids distances in the MIR, since at this wavelength the effects of metallicity (in terms of absorption features from CO and other molecules) are minimised. The 4.5~$\mu$m band dependence on a variable CO absorption, on the other hand, suggests the enticing possibility of adopting the $[3.6]-[4.5]$ period-colour-metallicity relation as a metallicity indicator. The expected dispersion of such indicator appears to be on the order of $\approx 0.20$~dex \cite{2016MNRAS.459.1170S}, comparable with the current precision in [Fe/H] measurements for individual Cepheids in the Galaxy, LMC and SMC.

\subsubsection{Mid-IR excess around Cepheids}
\label{ssec:excess}

Cepheids are intermediate mass stars (typically 4-6~$M_\odot$), and as such they enter the core He-burning phase at a relatively young phase, before they have the chance to wander far from the regions where they form. This means that a large fraction of them is found at small Galactic latitude (of the 10 Cepheid calibrators with HST parallax, all except $\beta$~Dor are found within $\approx \pm 10$~deg from the Galactic plane). As a consequence, Galactic Cepheids, as well as Cepheids detected in other galaxies seen edge-on, are projected over regions with high extinction (in the visible) or contaminated by ``cirrus'' of diffuse infrared emission. MIR is the ideal wavelength range to limit the impact of both effects, since at these wavelengths extinction is minimised (see discussion above) while the emission from galactic filaments starts to pick-up only at wavelengths longer than $\sim 25$~$\mu$m. The situation is however different in case of circumstellar emission, produced by circumstellar dust heated at temperatures warm enough to emit copious amount of thermal MIR radiation.

Circumstellar dusty envelopes can occur for two different reasons: \emph{i}) the dust is part of a cloud that formed independently from the star, and it just so happen that the star drifted into it, or \emph{ii}) the dust condensed in an outflow originating from the star. The existence of outflows from Cepheids is controversial, but can be related to the ``Cepheids mass discrepancy'' mentioned before. As was first noted more than 40 years ago by \cite{1972ApJ...171..593F}, current evolutionary models predict Cepheid masses 10-15\% larger than masses derived from pulsation theory (see e.g. \citealt{2005ApJ...629.1021C,2006ApJ...642..834K}) or measured from Cepheids in binary systems \citep{2008AJ....136.1137E,2010Natur.468..542P,2013AJ....146...93E}. This discrepancy could be at least in part explained if the mass deficit is the result of mass loss before or during the Cepheid phase (see e.g. \citealt{2012ApJ...760L..18N} and references therein). Mass loss during earlier evolutionary phases could result in circumstellar clouds still lingering as the star enters the instability strip, while active outflows would interact with the surrounding interstellar medium (ISM), piling-up ISM dust and gas in a ``bow shock'' aligned with the direction of motion of the star. It is not clear what could be triggering Cepheids mass loss: perhaps a pulsation-driven wind, or the influence of a close-in binary companion \citep{2015ApJ...804..144A}. This range of phenomena is however observed in several Cepheids and should be taken into consideration when assessing the accuracy of the Cepheids PL relation (and its calibration with Galactic Cepheids) at MIR wavelengths.

\begin{figure*}
\includegraphics[width=0.99\textwidth,keepaspectratio]{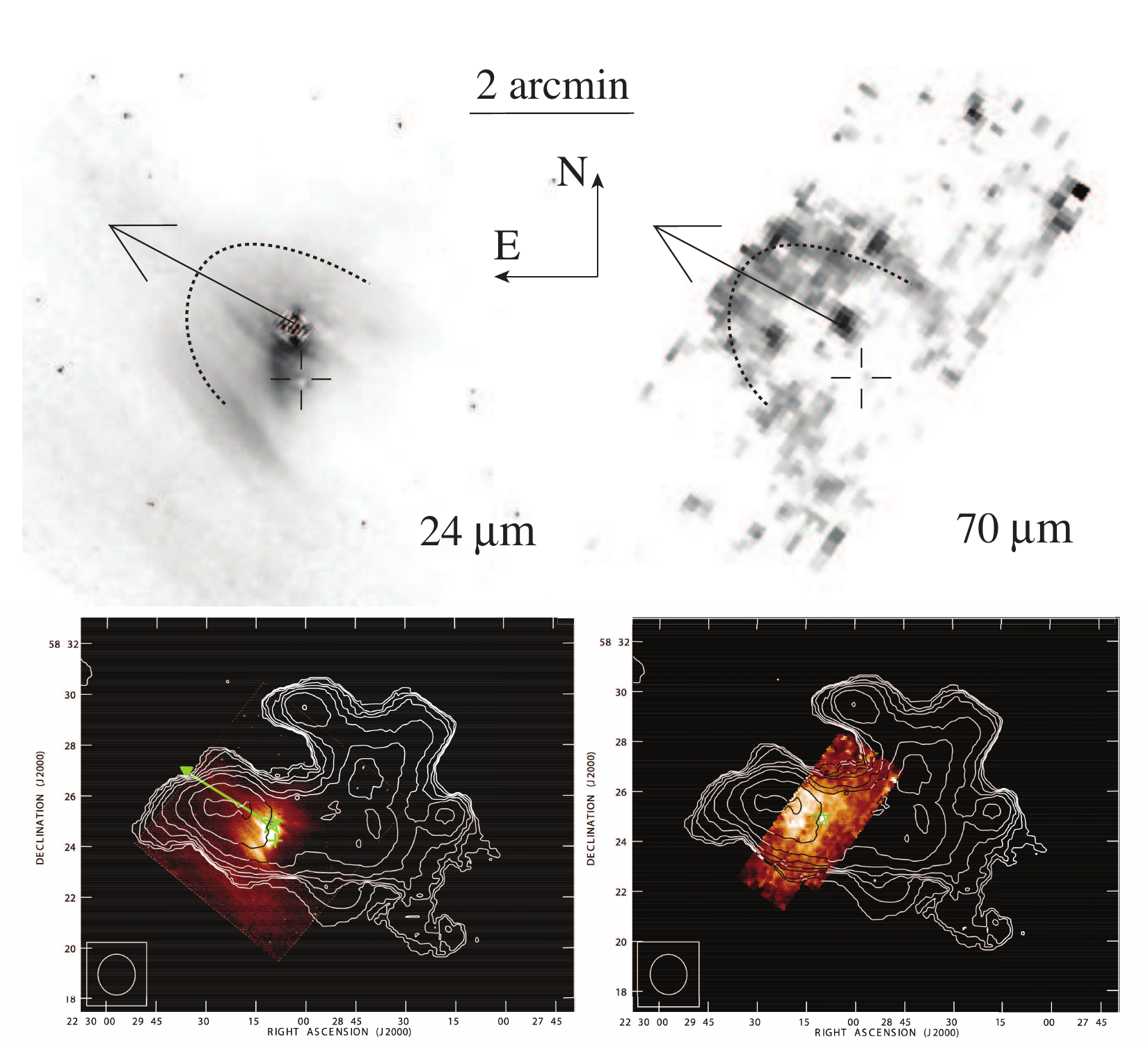}
\caption{$\delta$~Cep circumstellar nebula observed by Spitzer at 24 and 60~$\mu$m (top row, adapted from \citealt{2010ApJ...725.2392M}), and at the VLA in the H~\textsc{i} 21-cm line (bottom row, adapted from \citealt{2012ApJ...744...53M}). The shape of the far-IR emission resembles a bow shock aligned with the direction of the space motion of the star (indicated by an arrow). The same morphology is found in the much larger structure detected in the radio.}
\label{fig:deltacep}
\end{figure*}

The extreme case of a Cepheid embedded in a thick circumstellar envelope is RS~Pup, a long period Classical Cepheid ($P = 41.39$~days) located at the centre of a well known reflection nebula $2^\prime$ across \citep{1961PASP...73...72W}. The nebula is also bright at infrared wavelengths as short as 5.8~$\mu$m \citep{1986AJ.....91.1209M,2009A&A...498..425K,2011AJ....141...42B}. The bulk of this nebula is believed to be comprised of cold, dusty ISM material, shaped by a stellar wind that originated from the star, possibly in the earlier evolutionary phase when RS~Pup was a rapidly rotating B dwarf \citep{2009A&A...498..425K}. Warm emission detected at 10~$\mu$m, however, hints to the possibility of ongoing mass loss \citep{2009A&A...498..425K,2011A&A...527A..51G}.

The most convincing case for ongoing mass loss is however offered by the Cepheid class namesake itself, $\delta$~Cep. This star has been found to posses a bright circumstellar nebula, extending as much as $\sim 1.3^\prime$ ($\sim 2 \times 10^4 \ AU$ at the distance of the star; top panels in Figure~\ref{fig:deltacep}, \citealt{2010ApJ...725.2392M}). Observed at 70~$\mu$m, the shape of the nebula resembles a bow shock aligned with the space velocity of the star. Subsequent radio observations with the VLA found that this morphology repeats at much larger spatial scales ($\sim 13^\prime$ or 1~pc) in the 21-cm H~\textsc{i} line \citep{2012ApJ...744...53M}, consistent with a stellar wind with outflow velocity of $\sim 35 \pm 1.2$~km/s. Combined with the infrared observations, this detection suggest the presence of ongoing mass loss with a rate up to $10^{-7}$--$10^{-6} \ M_\odot$/yr. The estimated gas to dust mass ratio is very high ($\sim 2300$  \citealt{2010ApJ...725.2392M,2012ApJ...744...53M}), suggesting that the dust-poor wind is originated by some dynamic process different than the dust-driven winds found in Miras. Assuming that a significant portion of this circumstellar gas is originating from the star (as opposed to have been collected from the local ISM), such mass loss rates can help explain the Cepheids mass discrepancy described earlier.

Excess emission, if not taken properly into account, could bias Cepheid distances, or the calibration of the empirical Leavitt law, by artificially increasing the infrared brightness of the stars. However, it appears to be a relatively rare phenomenon, with less than $\sim 25$\% of Cepheids surveyed with Spitzer showing any evidence of infrared emission \cite{2011AJ....141...42B} (this result is supported by a similar rate of detection of H~\textsc{i} nebulae in a follow-up VLA survey, \citealt{2016arXiv160906611M}). Even in extreme cases like RS~Pup, the infrared excess emission is not detected at wavelengths shorter than 5.8~$\mu$m: this suggests that any bias due to circumstellar emission is not affecting observations at 3.6 and 4.5~$\mu$m, which remain safe for determining distances with PL relations with the techniques explained in \ref{ssec:empiricalPL}.

\subsubsection{The future of MIR Cepheid distances}
\label{ssec:cepheids-mir-future}

The advent of MIR space astronomy has finally opened the possibility of measuring accurate Cepheid distances at wavelengths less affected by interstellar extinction, metallicity effects and contamination from interstellar and circumstellar thermal emission. The $L$ band appears to be the ideal wavelength range for this measurements, lacking the presence of variable CO absorption that is instead affecting in the $M$ band. The CHP project \citep{2011AJ....142..192F} has demonstrated that the 3.6~$\mu$m Cepheid PL relation can be effectively used to measure the Hubble constant and other cosmological parameters with a precision comparable to the accuracy (better than $\sim 3$\%) in Planck's CMB and BAO analysis \citep{2012ApJ...758...24F}. A ``tension'' between Cepheid distances and the Planck analysis, however, remain, and needs to be investigated for its potential to reveal new physics in the accepted $\Lambda$CDM model. Accurate ($\mu$as) parallaxes for hundreds of Galactic calibrators, expected by 2022 with the final Gaia data release,  will provide the next significant jump in reducing the systematic errors in the calibration of the Cepheid Leavitt law. Further improvement will come with the launch, in 2018, of the James Webb Space telescope (JWST), whose NIRCam instrument will allow $L$ band observation of Cepheids in host galaxies of Supernov\ae{} Type Ia beyond the Local Group. Combined together, these improvement will enable 1\% measurements of $H_0$ as well as competitive measurements of other cosmological parameters such as the equation of state of dark energy $w_0$, key to validate current cosmological models \citep{2012SPIE.8442E..28G}.



\subsection{Linearity of the Cepheids PL Relation}
\label{ssec:cepheids-linearity}

The linearity of the Cepheid PL relation has been under debate in the past decade with empirical evidences of fundamental-mode Cepheid PL relation displaying a break at 10 days in optical bands \citep{tammann03,smkc05,ccn05,varela13}. It was suggested that the LMC Cepheid PL relation is best described by two separate slopes for short ($P < 10$~days) and long ($P > 10$~days) period Cepheids, therefore, a ``Non-linear relation''. The argument for a fiducial break period at 10 days was attributed to the resonance $P_2/P_0 = 0.5$, in the normal mode spectrum \citep{slee81,ccn05}. The non-linearity analysis of PL relations was extended to multiple wavelengths $BVI_{c}JHK_{s}$ \citep{cchoong08}, advocating that PL relations at longer wavelengths are supposedly linear, being less sensitive to metallicity and extinction. Recently, a statistical framework to study the non-linearity in LMC Cepheid PL relations at multiple wavelengths was developed \citep{bhardwaj2016} and it was found that optical band fundamental-mode Cepheid PL relations are indeed non-linear at 10 days while NIR PL relations provide evidence of a statistically significant non-linearity around 18 days.


In this section, we will discuss the application of a statistical framework on the multiwavelength data for Cepheids in the LMC. We use optical $V$ and $I$ band mean magnitudes for Cepheids in the LMC from the Optical Gravitational Lensing Experiment (OGLE - III) survey \citep{sos2008}. We include 1849 fundamental (FU) and 1238 first-overtone (FO) mode Cepheids in this analysis. The near-infrared ($JHK_S$) mean magnitudes for 775 FU and 474 FO mode Cepheids are adopted from LMC near-infrared synoptic survey \citep{macri15}. The extinction corrections were applied using reddening maps of \cite{haschke11} and \cite{cardelli89} extinction law.

\subsubsection{The Statistical Framework}
\label{ssec:2}

We discuss the test-statistics in brief. The detailed mathematical formalism is provided in \cite{bhardwaj2016}.

\paragraph{F-test :}  The F-test works under the null hypothesis that a single regression line over the entire period range is a better model to fit the data. The alternative hypothesis consists of a full model with two different slopes for short and long period Cepheids:

\begin{eqnarray}
\label{eq:breakpl}
m_0 &=& a_{S} + b_{S}\log(P) ; ~~\mbox{where $P$ $<$ $P_b$},\nonumber \\
  &=& a_{L} + b_{L}\log(P) ; ~~\mbox{where $P$ $\geq$ $P_b$},
\end{eqnarray} 

\noindent where $m_0$ is the extinction-corrected mean magnitude and $P_b$ is the fiducial break period. The difference in the residual sum of squares of the two fits, given the number of degrees of freedom, is estimated as the observed F-statistics. This F-value is compared with theoretical F-distribution with 95\% significance level. An observed F-value greater than the theoretical F-value leads to the rejection of null hypothesis.

\paragraph{Random-walk :} The robust non-parametric random-walk test generates the distribution of the residuals from the data itself. At first, the data is sorted according to the period and a single regression line is fitted over the entire period range. The amplitude of the partial sum of residuals is taken as observed R-statistics. If the partial sums are a random-walk, R will be small, suggesting a linear relation. In order to generate a theoretical R-distribution, the residuals are permuted $10^4$ times to estimate R-values. The fraction of permuted R-values that exceed the observed R-value, gives the probability of acceptance of a linear relation.

\paragraph{Testimator:} For testimator analysis, the sample is sorted and divided into several subsets of period bins. Slope of the PL relation in the first subset is adopted as initial estimate for the next subset. The null hypothesis is that the two slopes under consideration are equal. The standard $t$-test is applied to test the consistency of the slopes. If the slopes are equal, a smoothed slope over the two subsets is adopted as initial estimate for next subset. The process is repeated for all subsets unless the null-hypothesis is rejected.

\paragraph{Segmented lines and the Davies test:} An alternative approach to the problem of identifying the existence of a break point is presented in \cite{muggeo2003}. The method first performs a linear piecewise regression considering the
existence of the break. Thereafter, the Davies test is used to evaluate if the two segments are
different enough to account for two separate linear behaviours. The method is implemented in
the R package segmented \citep{muggeo2008}.

\subsubsection{Analysis and results}
\label{ssec:2}

\begin{figure*}
\includegraphics[width=0.99\textwidth,keepaspectratio]{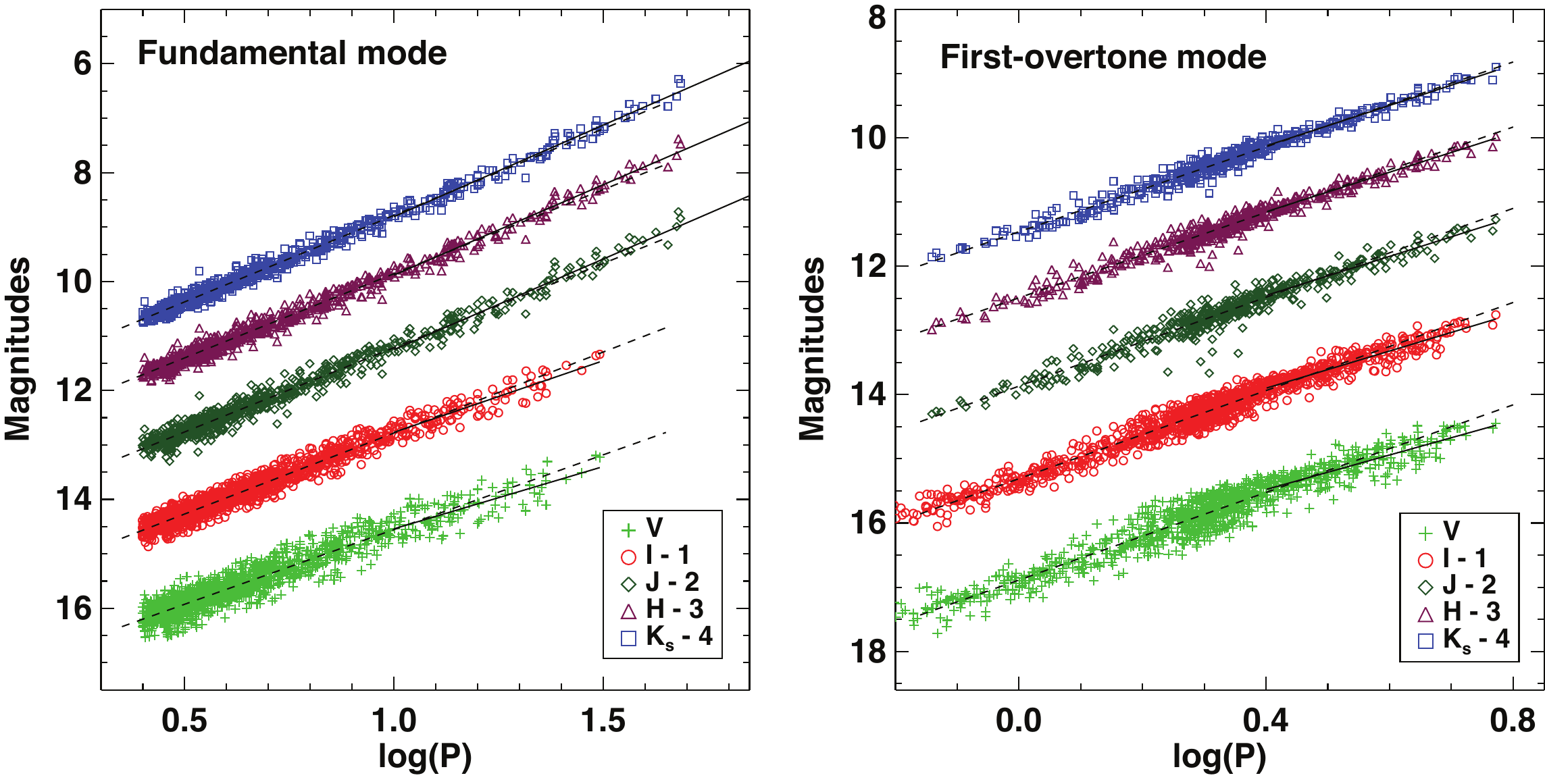}
\caption{ Optical and near-infrared PL relations for LMC FU and FO mode Cepheids. The dashed/solid 
lines represent the best fit regression for Cepheids with periods below and above the break period (10 days for FU and 2.5 days for FO mode Cepheids).}
\label{fig:nl_pl}
\end{figure*}

\begin{table}
\caption{The results of statistical tests on LMC Cepheid PL relations at multiple wavelengths.}
\smallskip
\label{tab:nl_pl}
\resizebox{\textwidth}{!}{%
\begin{tabular}{cccccccccc}
\hline\noalign{\smallskip}
Band & $\log(P_{\mathrm{b}})$ & $b_S$ & $\sigma_S$ & $b_L$ & $\sigma_L$ & $p(F)$ & $p(R)$& TM & $p(D)$\\
\noalign{\smallskip}\hline\noalign{\smallskip}
\multicolumn{10}{c}{FU}\\
\noalign{\smallskip}\hline\noalign{\smallskip}
$V$    &1.004& -2.743$\pm$0.032     &0.172&    -2.285$\pm$0.185     &0.232&     0.002&     0.168& Reject &0.006\\
$I$    &1.014& -2.973$\pm$0.022     &0.117&    -2.628$\pm$0.127     &0.161&     0.001&     0.128& Reject &0.003\\
$J$    &1.262& -3.118$\pm$0.031     &0.109&    -3.336$\pm$0.071     &0.144&     0.004&     0.037& Reject &0.003\\
$H$    &1.259& -3.106$\pm$0.026     &0.090&    -3.320$\pm$0.060     &0.120&     0.001&     0.044& Reject &0.000\\
$K_{s}$&1.275& -3.192$\pm$0.022     &0.079&    -3.361$\pm$0.055     &0.111&     0.003&     0.076& Reject &0.001\\
\noalign{\smallskip}\hline\noalign{\smallskip}
\multicolumn{10}{c}{FO}\\
\noalign{\smallskip}\hline\noalign{\smallskip}
$V$    &0.446&    -3.410$\pm$0.045&     0.187&	-2.688$\pm$0.108     &0.155&     0.000&     0.000& Reject &0.000\\
$I$    &0.449&    -3.430$\pm$0.032&     0.132&	-2.894$\pm$0.078     &0.112&     0.000&     0.000& Reject &0.000\\
$J$    &0.427&    -3.455$\pm$0.064&     0.117&	-3.107$\pm$0.103     &0.102&     0.043&     0.122& Accept &0.059\\
$H$    &0.422&    -3.322$\pm$0.049&     0.097&	-3.117$\pm$0.075     &0.075&     0.136&     0.284& Accept &0.156\\
$K_{s}$&0.019&    -3.303$\pm$0.042&     0.087&	-3.179$\pm$0.062     &0.062&     0.376&     0.277& Accept &0.735\\
\noalign{\smallskip}\hline\noalign{\smallskip}
\end{tabular}}
{\footnotesize {\bf Notes:} $P_{\mathrm b}$, $b$ and $\sigma$ refer to the break period, slope and dispersion, respectively.
The subscripts $S$ and $L$ refer to the short period range and long period range, respectively. $p(F)$, $p(R)$ and $p(D)$ represent the probability of acceptance of null hypothesis i.e. linear relation. ``TM'' displays the results of testimator analysis: ``Accept/Reject'' represents a linear/non-linear relation.}
\end{table}

The optical and near-infrared band PL relations for FU and FO mode Cepheids are displayed in Figure~\ref{fig:nl_pl} and the results
of statistical analysis are presented in Table~\ref{tab:nl_pl}. If the probability of the acceptance of null hypothesis is less than 0.05, the PL relation under consideration is non-linear. For FU mode Cepheids, optical band PL relations are found to be non-linear at 10 days, according to the $F$-test, testimator and the Davis test. The random walk does not provide evidence to support a non-linearity at $95\%$ significance level. If more than two statistics suggest a break, we consider the PL relation to be non-linear. At NIR wavelengths, all four test statistics suggest a statistically significant deviation in the slope for short and long period Cepheids, except for random walk in $K_S$ band. Interestingly, the break period is around $\log(P)\sim1.25\textrm{-}1.30$ in the near-infrared PL relations. It is important to note that a distinct variation in Fourier amplitude parameters around 20 days was also observed for redder bands as compared to optical bands \cite{bhardwaj2015}. This suggest that the changes in light curve structure can also be associated with these non-linear features in PL relations. Although the multi-wavelength LMC PL relations exhibit a non-linearity, the SMC PL relations are found to be linear at 10 days \citep{ngeow15}.

We also test the FO Cepheid data in the LMC for possible statistically significant non-linearities. The optical band PL relations present evidence of a non-linearity based on all four test statistics. The Davis test suggests a break at $\log(P) = 0.45~(P\sim2.8)$~days 
in PL relations. The evidence of non-linearity decreases for longer wavelengths and $HK_S$ bands are found to be linear with all four test-statistics. The robustness of these results was also validated under various assumptions, such as independent and identically distributed random observations, normality of residuals and homogeneity of variances. 

The LMC FU mode Cepheid sample was used together with Cepheids in supernovae host galaxies to estimate the impact of these non-linearities on the distance scale \citep{bhardwaj2016}. The additional Cepheid and supernovae data was adopted from the SH0ES {\it (Supernovae and $H_0$ for the Equation of State of dark energy)} project \citep{riess11}. The Cepheids in the LMC were calibrated using the late-type eclipsing binary distance \citep{piet13} and a global matrix formalism was adopted similar to SH0ES. The global slope of Cepheid PL relations is found to be consistent with linear version of LMC PL relations. However, the two slope model of LMC PL relations provides a stronger constrain on the global slope and the metallicity coefficients. This model was also adopted in recent SH0ES paper with additional supernovae hosts, leading to a 2.4\% local determination of Hubble constant.

\subsubsection{Conclusions}

The statistical analysis of non-linearities in the LMC Cepheid PL relations provides evidence of a break in fundamental-mode Cepheid PL relations at $VIJHK_S$ wavelengths. The first-overtone mode Cepheids display a break at 2.8 days, only at optical wavelengths. The observed non-linearities suggest a correlation with sharp changes in the light curve structure of Cepheids. This needs a theoretical investigation to look for additional resonance feature in the light curves for Cepheids. The global slope for Cepheids in SH0ES galaxies using the 
linear and non-linear versions of the LMC PL relations are found to be very similar. Therefore, these non-linearities will not have any statistically significant impact on the distance scale or the value of Hubble constant. This is due to the fact that the dispersion of
the {\it HST}-based PL relations for Cepheids in supernovae host galaxies ($\sim0.3\textrm{-}0.4$~mag) is more than three times the dispersion of the calibrator LMC PL relation. Therefore, such small but statistically significant changes in slope of PL relation may only have a major impact on distance parameters with more precise PL relations, for example with the {\it JWST}.


\subsection{Metallicity Effects}
\label{ssec:cepheids-metallicity}

The metallicity of a star usually affects its luminosity to a certain
degree and there is no reason to believe that Cepheids should differ
from other stars in this regard. The question is if the effect is
significant and, if  significant, whether it can be properly corrected for.
\cite{riess11} pointed out that to determine the Hubble constant to
better than 2\% one of the major obstacles is the effect of metallicity
on the Cepheid luminosities.

In preparation for the HST Key project on the
extragalactic distance scale \citep{2001ApJ...553...47F} the metallicity
effect was determined empirically by comparing high metallicity Cepheids
in the inner parts of M101 with low metallicity Cepheids in the
outer part of the galaxy \citep{Kennicutt98}.  They found an effect of $0.1\pm0.1$~mag/dex in the $V$ band, metal-poor stars being brighter. Many different empirical
efforts trying different avenues followed in the next years but even the
sign of the effect was disputed (e.g. \citealt{romaniello08}).

The approach used by  \cite{Kennicutt98} which relies on extragalactic Cepheids has
the obvious advantage that the stars are at the same distance and span
a large range of metallicities but it suffers from the problem of
disentangling reddening and blending effects and
the necessity to infer the metallicities from the local gas in the
galaxy. 
The Gaia satellite will surely provide very accurate distance to
galactic Cepheids but the sample will be limited to Cepheids with
metallicities not too different from solar, so only a limited
metallicity range can be studied and this will eventually limit the 
accuracy with which the slope of the period-luminosity relation can 
be determined.

Here we present the status of an alternative approach 
based on the infrared surface brightness method, which is a 
Baade-Wesselink type method. In this approach the stellar luminosity is
determined directly for each star individually and a PL  
relation can be constructed on this basis. The added advantage of this
method is that it can be applied to Cepheids not only in the Milky Way
but also in the Magellanic Clouds (e.g. \citealt{Storm04}). In this way
we can compare stars spanning a significant range of metallicities in a
direct way: in fact we can delineate period-luminosity relations for
Cepheids of three different metallicities representing the young
populations in the SMC, LMC and solar neighbourhood.

\subsubsection{The near-IR surface brightness method}
\label{sec:IRSB}

\begin{figure}
  \includegraphics[width=0.99\textwidth,keepaspectratio]{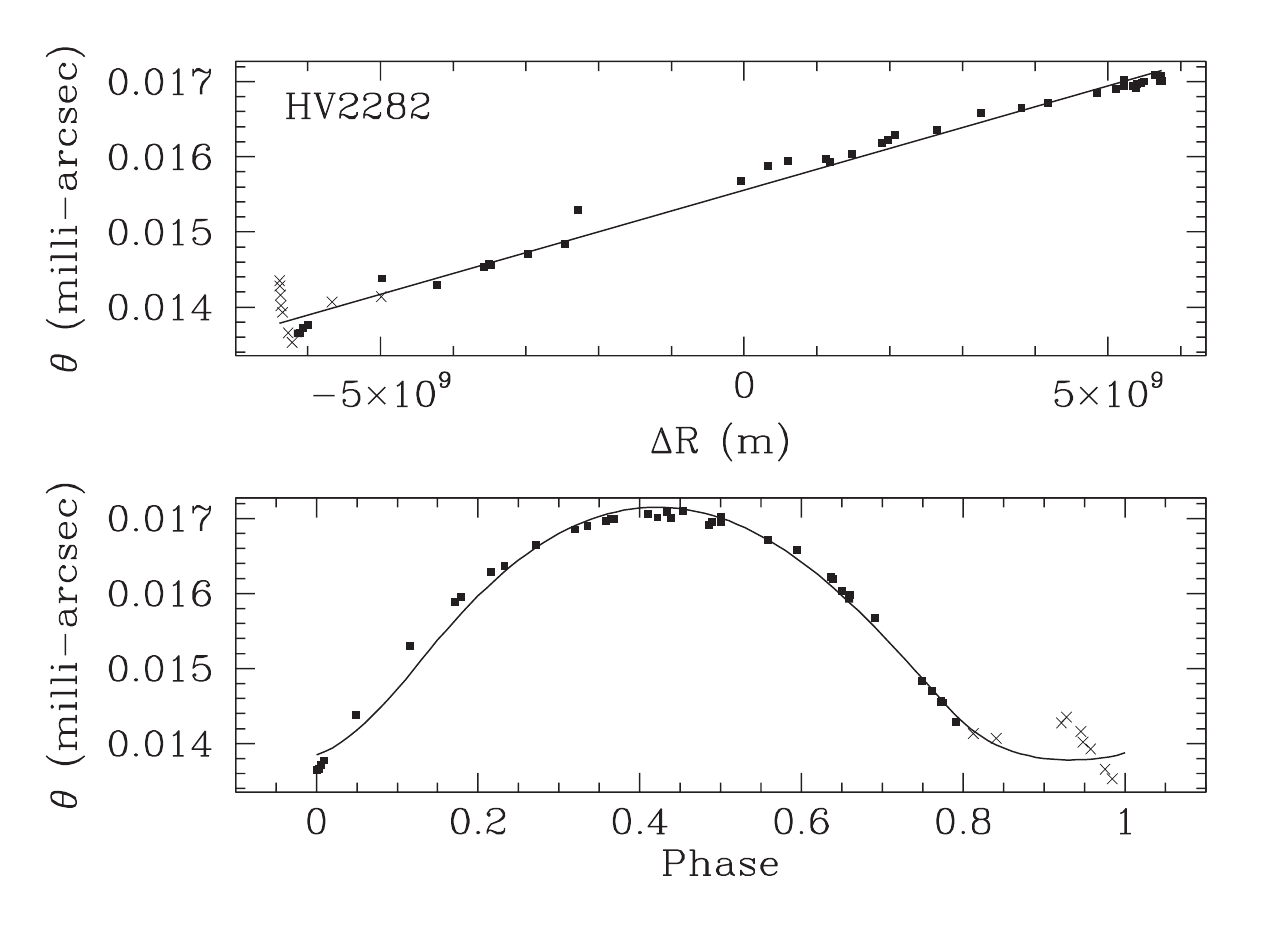}
\caption{\label{fig.HV2282fit}Example of the fit of the angular diameters derived from
photometry and spectroscopy for the LMC star HV2282 \citep{storm11}.}
\end{figure}

The IRSB method is a geometrical
method as it relates the stellar angular diameter, $\Theta$, at  a given
phase, $\phi$, with the stellar radius, $R$, and the distance, $d$ through 

\begin{equation}
\label{eq.geom}
\Theta (\phi) = 2R(\phi)/d = 2[R_0 - \Delta R(\phi)]/d
\end{equation}

If we can observe the angular diameter, $\Theta(\phi)$, and the radius 
variation, $\Delta R(\phi)$ at different phases we can solve the
equation for the two variables, radius $R_0$, and distance $d$.

The angular diameter is related to the stellar surface brightness
$F_V$, through the relation $F_V=4.2207-0.1V_0-0.5 \log (\Theta)$.
\cite{BarnesEvans76} 
found that  a linear relation between colour and
stellar surface brightness could be established. 
\cite{FG97} expanded this to the NIR $(V-K)$ colour, based on
interferometric measurements of static stars. This relation was refined
by  \cite{diB98} and \cite{Kerv04}
first deriving the relation:

\begin{equation}
F_V = -0.1336(V-K)_0 + 3.9530
\end{equation}
based on interferometric measurements of Cepheids.

Next we can measure the absolute radius variation $\Delta R$, by
integrating the pulsation velocity curve. This differs from the
directly observed radial velocity curve as it takes out the systemic
velocity $v_{\mbox{\scriptsize sys}}$, and correct for the 
projection effect as the
radial velocity measures a weighted mean across the surface of the star
and not just the velocity exactly on the point nearest to the observer.
The projection effect is corrected by multiplying by the so-called
$p$-factor. For a more detailed discussion of the calibration of the
$p$-factor, see \cite{Storm11a}.
This can then be  written:

\begin{equation}
\Delta R (\phi) = \int -p(v_{\mbox{\scriptsize rad}}(\phi) - v_{\mbox{\scriptsize sys}})d\phi
\end{equation}

These equations shows that all that is needed is good light curves in
$V$ and $K$ and a good radial velocity curve. These are all obtainable
for stars in the Milky Way as well as in the Magellanic Clouds and
potentially in M31.

In Figure~\ref{fig.HV2282-cep1797-data} an example of a fit to
Equation~\ref{eq.geom} for the LMC star HV2282 is shown. We note that
the phase interval between 0.8 and 1.0 is usually disregarded as
the stars often shows certain spectral lines in emission suggesting
shocks passing through the atmosphere. This can lead to the
surface-brightness relation being invalid and it can affect the radial
velocity measurement at some level.

\subsubsection{The data set}

\begin{figure}
  \includegraphics[width=0.50\textwidth,keepaspectratio]{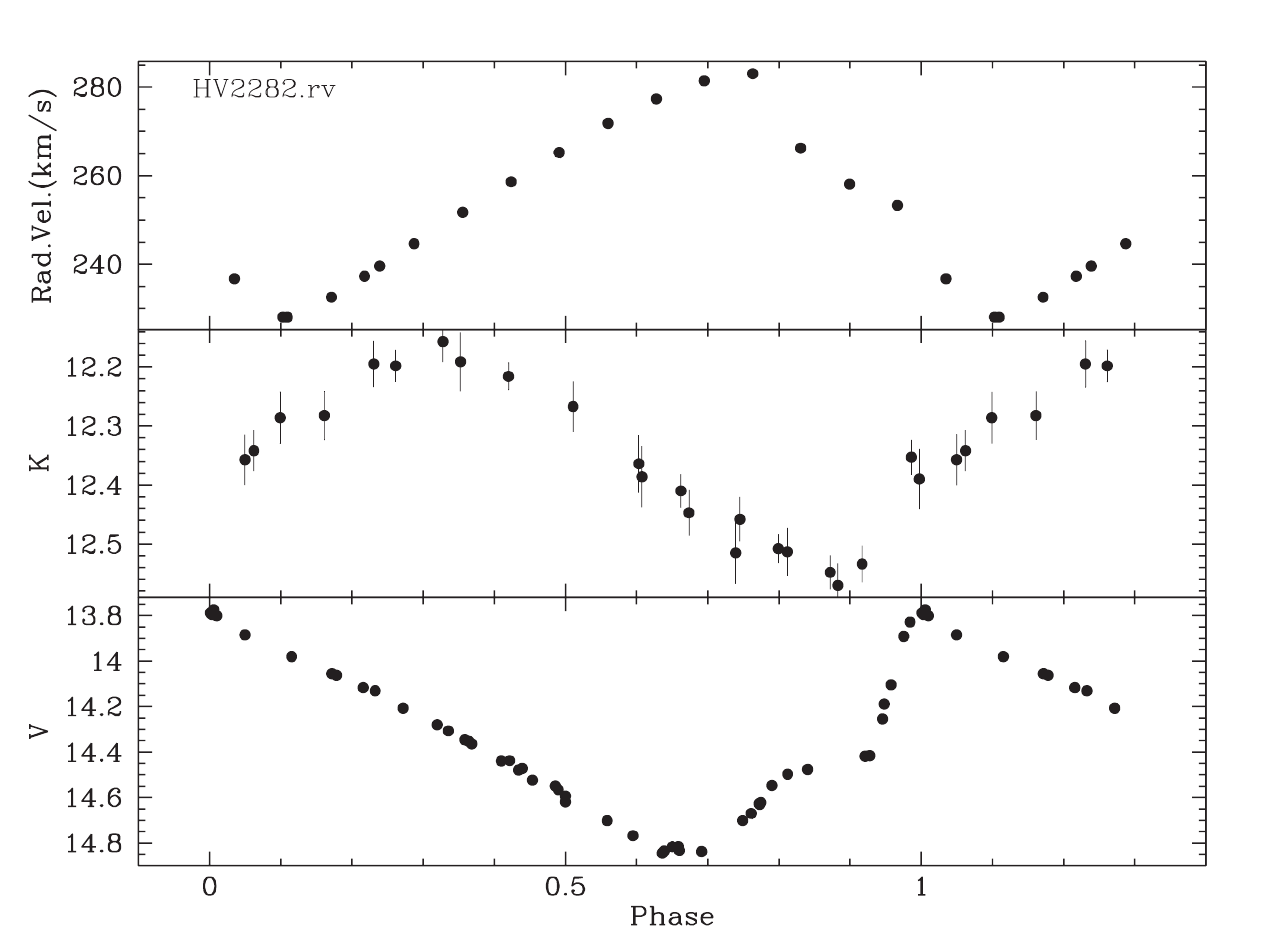}
  \includegraphics[width=0.50\textwidth,keepaspectratio]{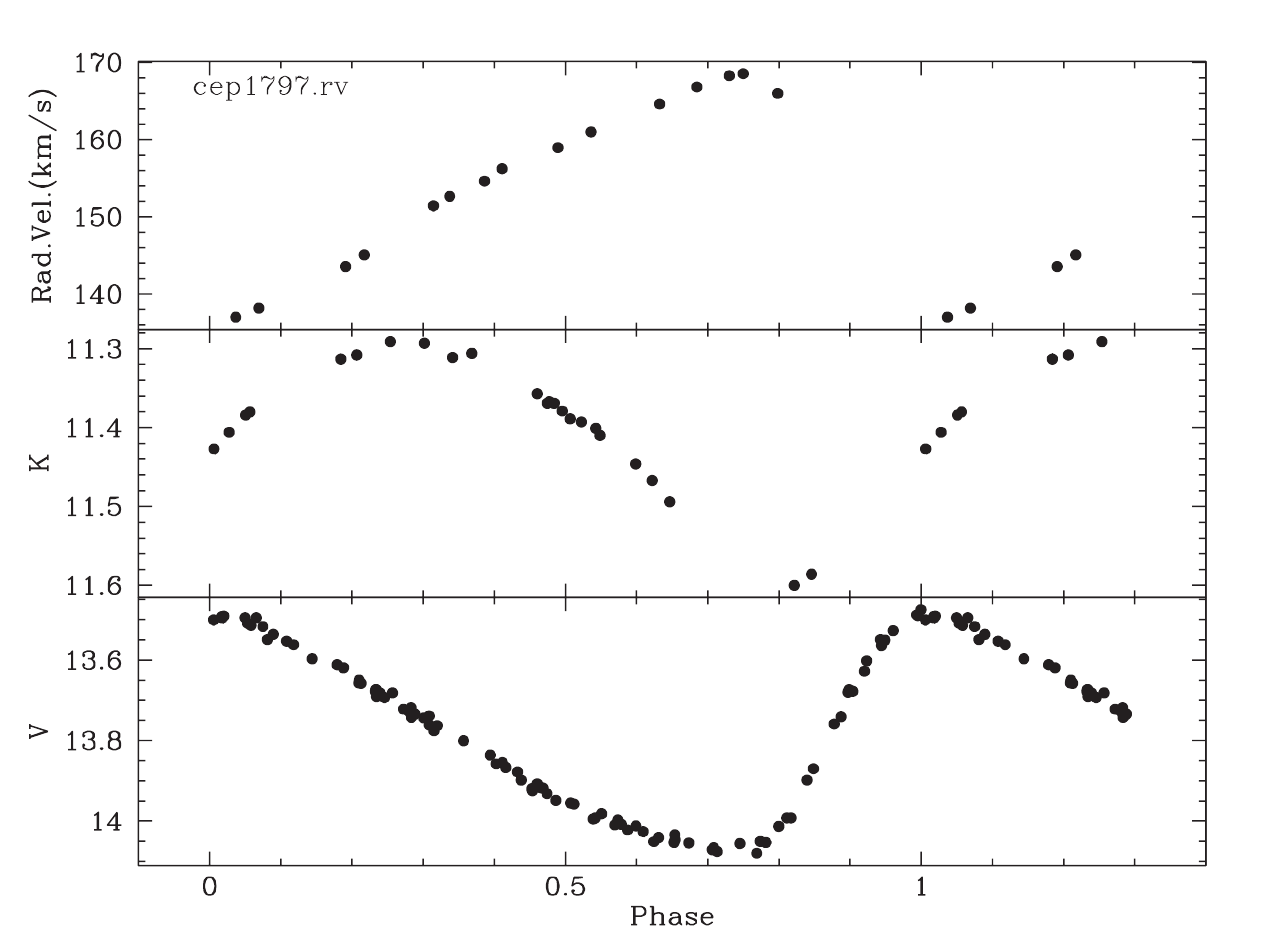}
\caption{\label{fig.HV2282-cep1797-data} \emph{Left} --- Example of data set for the LMC Cepheid HV2282. The $V$ band data from OGLE-III \cite{sos2008}, the $K$-band data is from \cite{Persson04},  and the radial velocity data is from the HARPS spectrograph at ESO \cite{storm11}. \emph{Right} --- Example of a data set for the SMC Cepheid SMC-1797. The $K$ band data is a preview from an on-going program at the ESO-NTT and the radial velocity data is a preview of an ongoing program with HARPS at the ESO 3.6m telescope.}
\end{figure}
%
%

 \cite{Storm11a} presented an analysis based on 76 
Milky Way stars,
36 LMC stars and 5 SMC stars. In Figure~\ref{fig.HV2282-cep1797-data} (left panel) a typical data
set for one of these LMC stars is shown. The project has been expanded 
to include an additional 25 SMC stars to bring down the error bar on 
the lowest metallicity data point. In Figure~\ref{fig.HV2282-cep1797-data} (right panel) an example
of this new data set is shown. The data acquisition is still in progress
so the phase coverage will eventually be better, but the high quality of
the data is evident. The optical data is taken from the OGLE-IV
\cite{Udalski15} survey while the $K$ band data is being acquired
at the ESO-NTT and the radial velocities with HARPS at the ESO 3.6m
telescope.

In addition to the work on the SMC Cepheids we have also been acquiring
new accurate radial velocity curves for more than 50 Milky Way Cepheids
for which good $V$ and $K$ band light curves exist. This should also
further constrain the most metal rich data point.

\begin{figure}
  \includegraphics[width=0.50\textwidth,keepaspectratio]{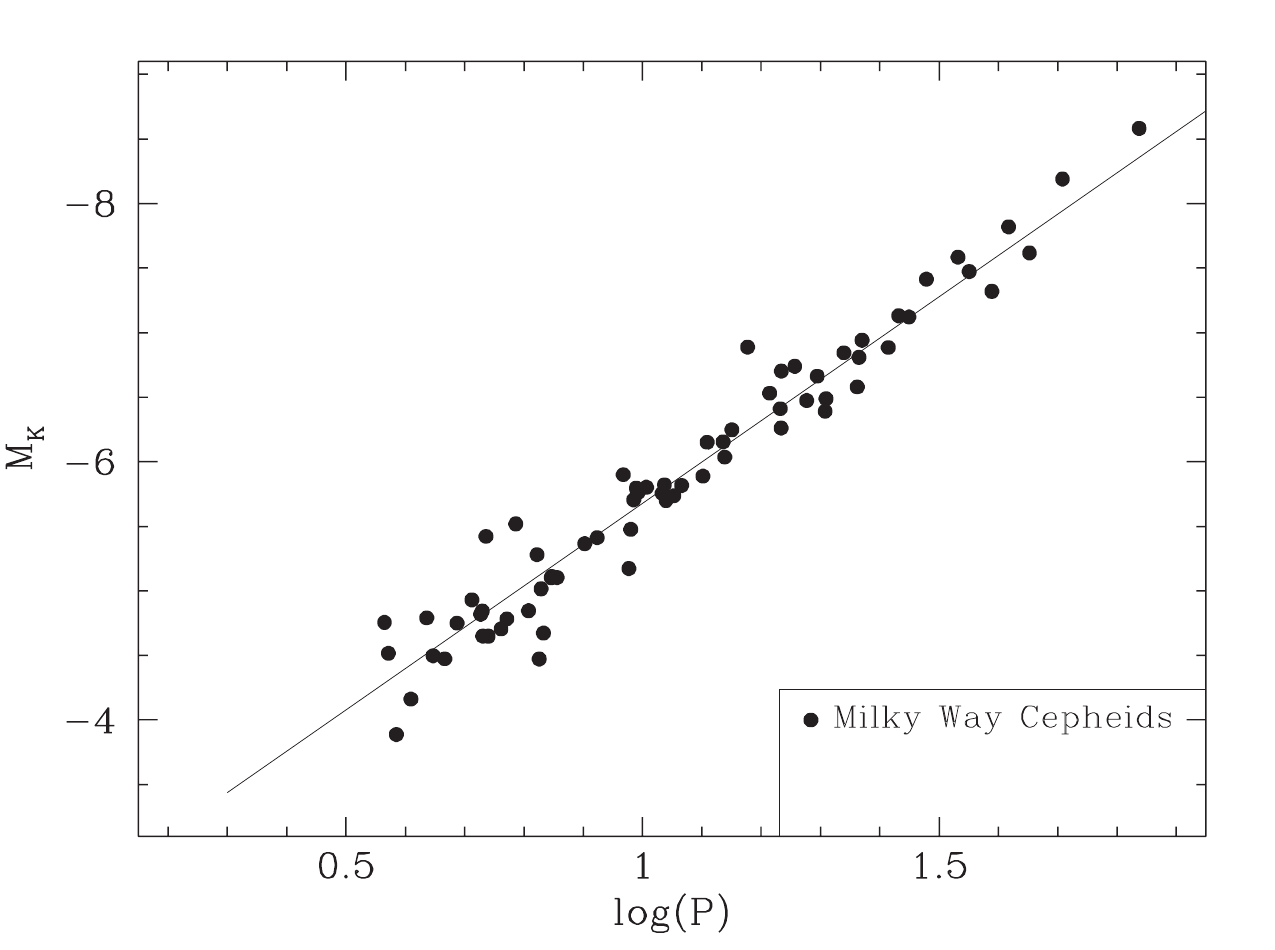}
  \includegraphics[width=0.50\textwidth,keepaspectratio]{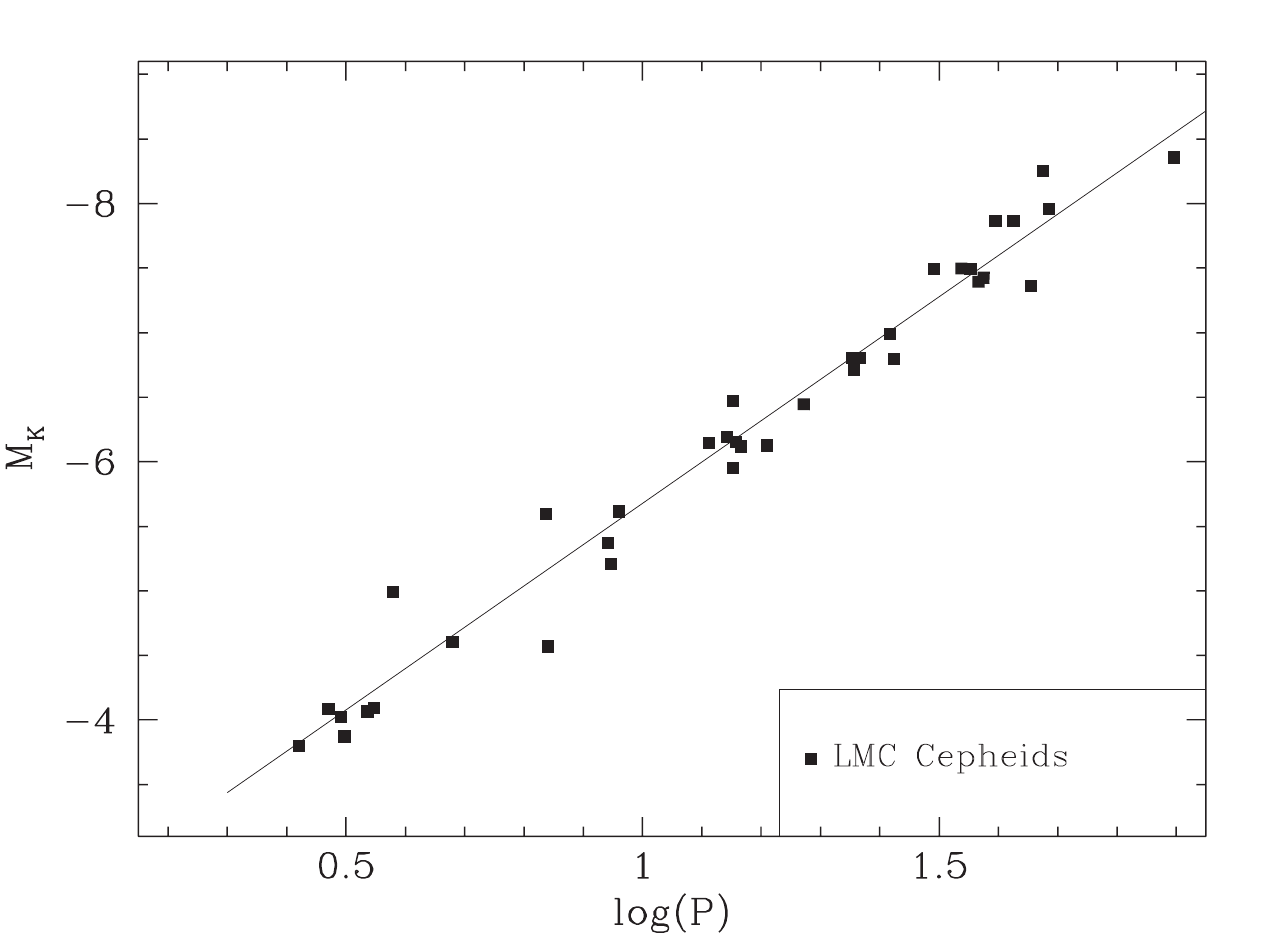}
  \includegraphics[width=0.50\textwidth,keepaspectratio]{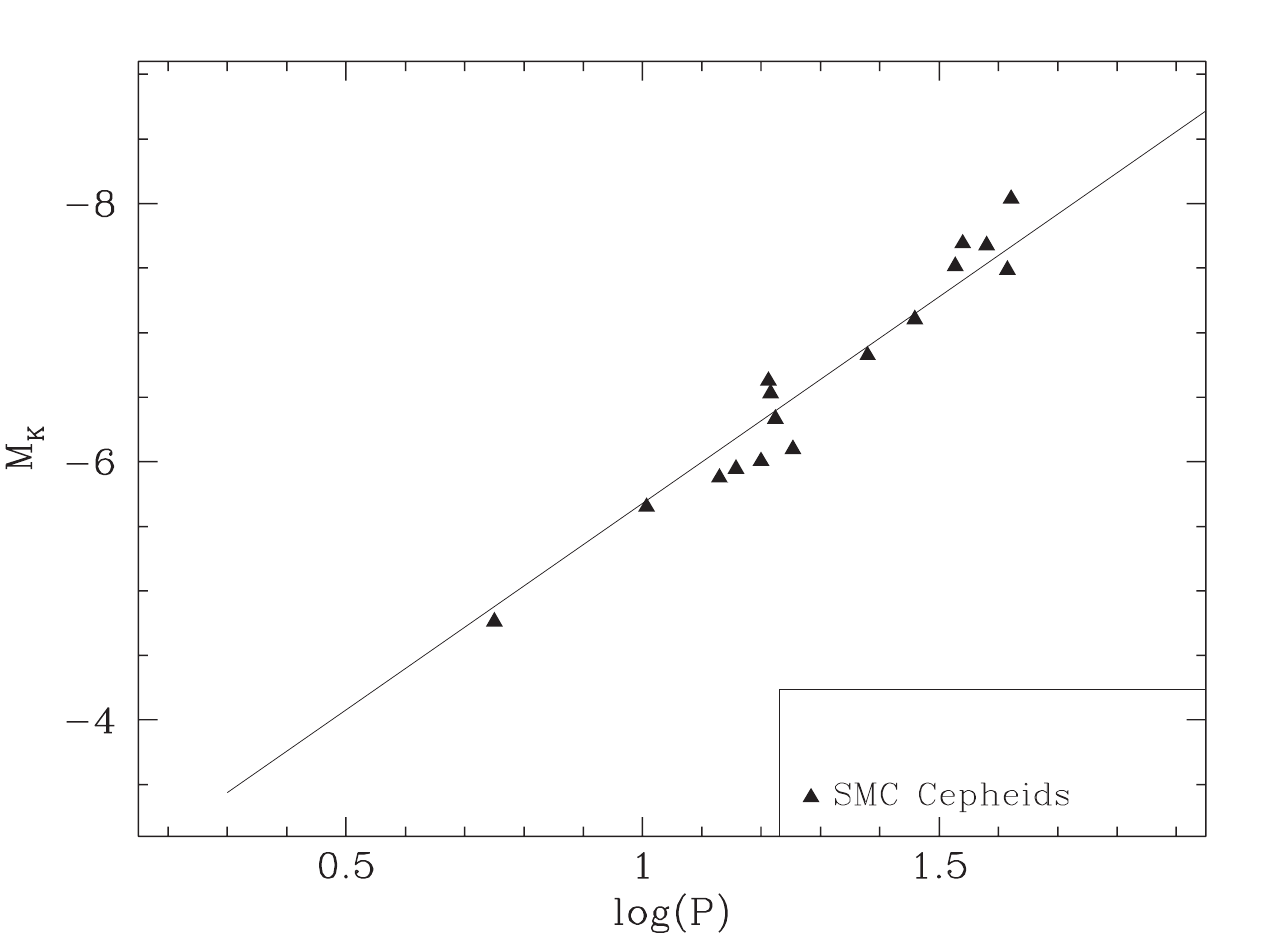}
  \includegraphics[width=0.50\textwidth,keepaspectratio]{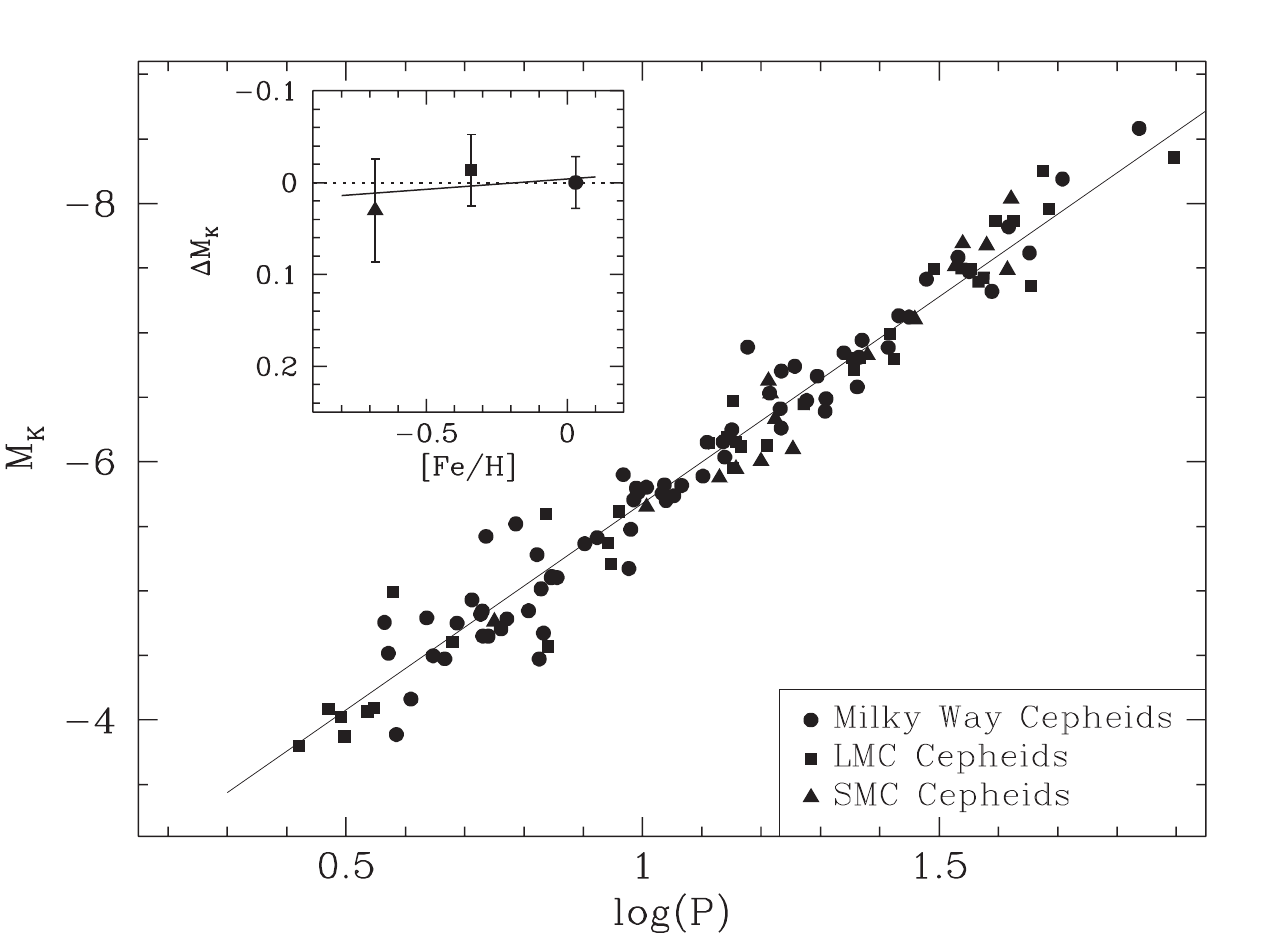}
\caption{\label{fig.PLall} \emph{Top} --- The PL relation in the $K$ band for a sample of Cepheids in the Milky Way (\emph{left}) and in the LMC (\emph{right}), from \cite{Storm11a,storm11}. \emph{Bottom left} --- Preliminary PL relation in the 
$K$ band for part of the new sample of SMC Cepheid. \emph{Bottom right} --- Combined PL relation with all the stars presented in the previous three panels. The inset shows the offset of the three different samples with respect to the overall PL relation at a
period of 10 days.}
\end{figure}

\subsubsection{The Period-Luminosity relations at different metallicities}
\label{ssec:PLZ}

In Figure~\ref{fig.PLall} (top panels) we have plotted the PL relations
in the $K$ band for the samples of Milky Way and LMC stars from \cite{Storm11a,storm11}. 
In both plots the line over-plotted is the best fit to their
complete sample of stars. It is evident that the two relations are very
similar. In \cite{storm11} only 5 SMC stars with a period of around 15~days
were available so the low metallicity relation was poorly constrained.
Work is now ongoing to remedy this limitation by extending the sample by
an additional 25 stars (PI: Gieren). In the bottom left panel of Figure~\ref{fig.PLall}
we plot the preliminary PL relation for the sub-sample of stars where
the data is sufficient to perform an initial analysis. Again the
over-plotted line represents the fit to the global sample in 
\cite{storm11}. We note that at least the preliminary agreement is very good. 

We can then determine the offset for each sample at a
period of 10~days with respect to the reference relation in each of the
plots.  We have adopted metallicities of [Fe/H] =+0.03 for
the Milky Way, $-0.34$ for the LMC, and $-0.68$ for the SMC, following \cite{Luck98}.
In the last panel of Figure~\ref{fig.PLall} we have over-plotted all the stars with the
reference relation. In the inset the individual offsets for the three
samples is plotted against the adopted metallicity. It is evident that
the effect of metallicity in the $K$ band appears to be small, consistent with
zero, and that the error
estimate on the slope will be significantly less than 0.1~mag/dex. The
final estimate and associated error bar will have to wait the completion 
of the data acquisition and final analysis, but the preliminary
results are looking very promising.


\section{Miras}
Mira variables are M-type stars pulsating with periods of 100 to 1000 days, showing high mass loss rates ($10^{-5}-10^{-7} M_{\odot}$yr$^{-1}$). 
They are thought to be experiencing a pulsating phase while on the asymptotic giant branch (AGB) evolutionary phase. 
Mira variables are also well known for their remarkable periodicity and its large amplitude in V-band magnitude. 
Mira variables in the LMC exhibit a PL relation between their pulsation period and K-band apparent magnitude ($K$).
This PL relation allow these stars to be used as distance estimators, and accurate distance of these sources helps us to understand the nature of their variability. 

Although a narrow PL relation for Miras in the LMC was found by \cite{fea89}, a similar relation for the Galactic Miras has not been precisely obtained because of large errors in absolute magnitudes. 
Such large errors arise from ambiguities of absolute magnitudes suffering directly from errors in distances estimations. \cite{van97} reported the relation of Galactic Miras based on the parallax measured with Hipparcos satellite \citep{per97}. 
Although the Hipparcos mission has been an historical milestone in astrometry, distances of the Galactic Mira variables are too far for an accurate determination of their parallaxes. Therefore, a calibration of the PL relation based on Galactic Mira variables is not well established even today. 

The very long baseline interferometer (VLBI) consists of a number of antennas separated by several thousand Kilometers. Angular resolution of the order of milli-arc seconds is obtained by imaging synthesis  of VLBI at radio wave bands of cm to mm wavelength. Because of its high resolution, the VLBI has a strong advantage for parallax measurements compared to other current astronomical instruments. Using absolute magnitudes derived from accurate distances measured with astrometric VLBI observations, we can investigate precise PL relation in the Galaxy. 
In this sub-section, we present our recent on-going efforts to calibrate the PL relation of Galactic Mira variables using the VLBI Exploration of Radio Astronomy (VERA) array. 

\begin{figure}
\begin{center}
\includegraphics[width=73mm, angle=0]{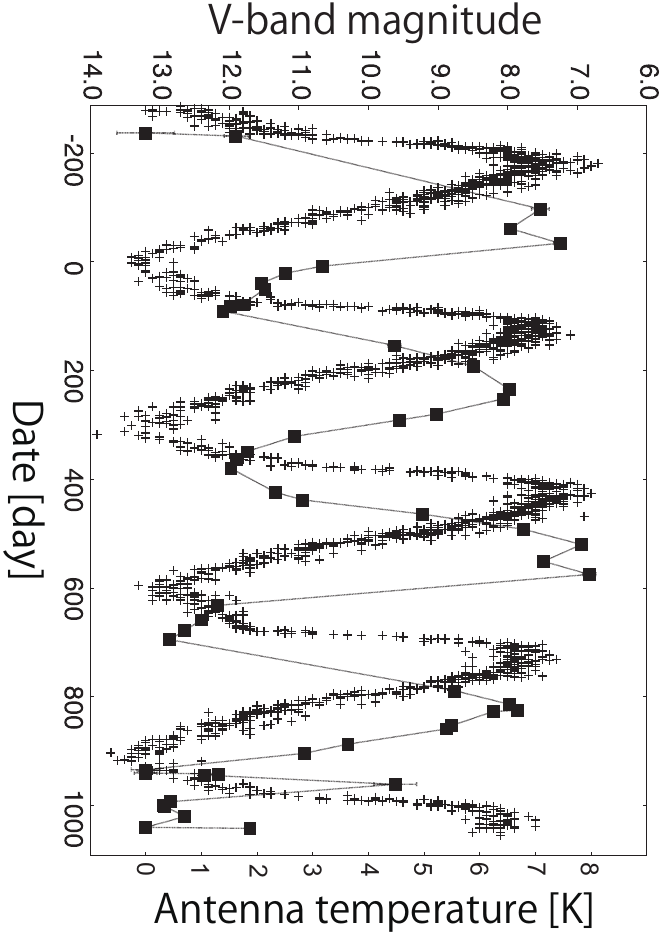}
\caption{Optical and radio variations of a Mira variable 'R UMa'. Filled squares and crosses indicate H$_{2}$O maser and V-band variations of the source. Magnitudes from American Association of Variable Star Observers (AAVSO) are used for the V-band light curve. }
\label{fig_lightcrve_ruma}
\end{center}
\end{figure}

\begin{figure}
\begin{center}
\includegraphics[width=73mm, angle=0]{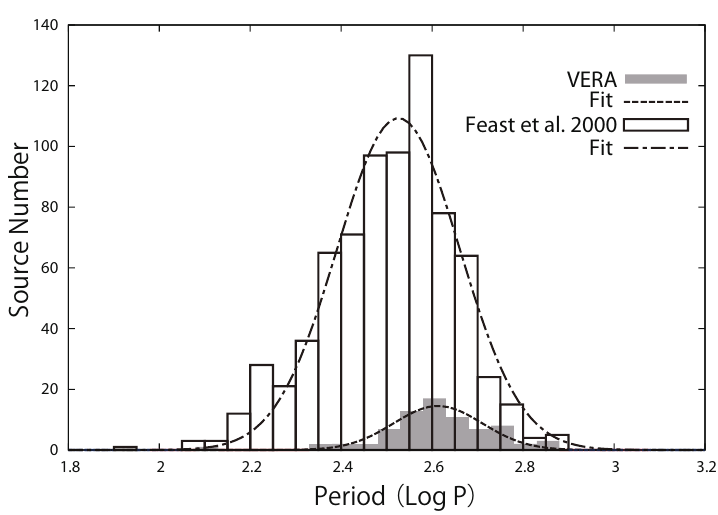}
\caption{Period distribution of the Galactic Mira variables. Open blocks indicates $\sim$800 Mira variables studied in \cite{fea00}. 
The period distribution of Mira variables accompanied with H$_2$O maser emission seems to have slight shift to longer period. }
\label{fig_hist}
\end{center}
\end{figure}

\subsection{Source selection and single-dish monitoring}
Mira variables show maser emission around them. Maser stands for Microwave Amplification by Stimulated Emission of Radiation and hence is a non-thermal radiation representing very compact structure and high brightness temperature. Many different masers occur in astrophysical environments. SiO, H$_2$O, OH, and CH$_3$OH masers are detected around evolved stars or star forming regions. They are frequently used as probes to study kinematics of matters around the stellar objects. The milli arc second level high resolution achieved by the VLBI technique is well suited for observation of the maser emissions. 
H$_2$O maser emission around Mira variables is so bright and compact that they can be a good target for VLBI astrometry.
But the intensity of the maser emission usually shows large variability. 
Like the periodic pulsation at optical band, intensity of the maser emission sometimes represents quasi-periodic variability with the same pulsation periods defined in the visible. 
Figure~\ref{fig_lightcrve_ruma} shows a result from monitoring the Mira variable  'R UMa'. 
This is a case that represents a good period consistency in optical ($V$ band) and maser emission (22 GHz).

In order to derive the parallax from maser emission, we have to find sources with bright maser emission (all through their cycle) enough to be detected with the VLBI method. 
We are monitoring a sample of candidate targets with the IRIKI 20 m telescope in VERA array with one month interval. 
Early results of the monitoring program are reported by \cite{shi08}. 
The results from the single-dish monitoring are used to select the sample for our VLBI program. 

The distribution of the pulsation periods of our targets is presented in Figure~\ref{fig_hist}. 
Open blocks indicates $\sim$800 nearby Mira variables studied by \cite{fea00}. 
Filled blocks indicate the sources of our VLBI study which are accompanied by H$_2$O maser emission. 
We fitted these two distributions with Gaussian models and obtained peak periods of 338 days ($\log P=2.53$) and 407 days ($\log P=2.61$) for optically defined Miras and maser emitting Miras, respectively.  Although there is a difference between two peaks, a selection bias can not be ruled out at present. More careful consideration about both populations is needed before we conclude them as two  separate populations.

The diffuse and weak components of the maser emission is resolved by the small synthesised beam of the VLBI. 
Therefore, it is important to find target sources that have bright and compact components that can be detected with the VLBI method. 
As a threshold of a total power intensity, $\sim$10 Jy is adopted for detection in our VLBI monitoring program. 
We are observing $\sim$250 maser sources to find targets whose maser emission stably shows  intensity with  $\geq$ $\sim$10 Jy. 
From the point of view of pulsation periods, sampling a wide periods range is required to obtain better solutions of the PL relation. 

\subsection{Astrometric VLBI observations with VERA}
\begin{figure}
\begin{center}
\includegraphics[width=73mm, angle=0]{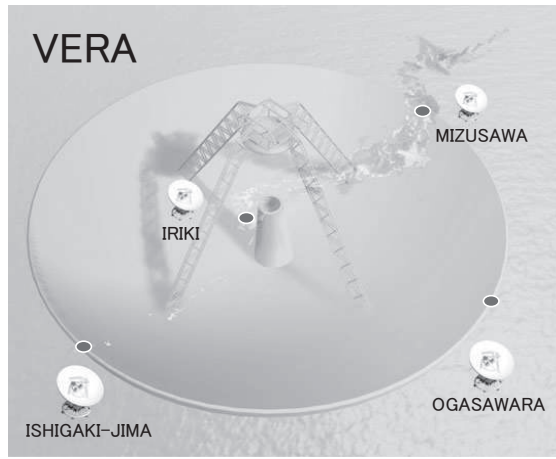}
\caption{Schematic figure of the VLBI Exploration of Radio Astrometry (VERA) antennas. 
Four antennas with 20 m aperture spanning Japan are dedicated to astrometric observation at frequencies of 22 and 43 GHz. 
}
\label{fig_veraarray}
\end{center}
\end{figure}

We started a series of VLBI observations with VERA after a careful selection of the targets. 
The VERA array consists of four antennas with 20 m aperture located in Japan \citep{kob03}. 
The maximum baseline length is $\sim$2300\,km (Figure~\ref{fig_veraarray}). 

We observe target maser source and reference continuum source simultaneously using the dual beam system installed in the VERA antenna \citep{kaw00}. At first, we solve phase solutions using a reference source, which usually is a bright QSO. Data reduction of the target source can be realised by applying the phase solution from the QSO to the target. Two dimensional Fourier transformation of the data provide a map of the target maser source, from which we measure the positions of the maser spots. Since core emission of QSO can be treated as a position reference fixed to the sky plane in milli arc second scale astrometry, the motion of the maser spots detected on the phase reference map can be considered as a combination of the linear proper motion and parallactic oscillation. As the effect of the solar motion is same in the target and reference source, this motion is cancelled out in the phase referencing technique. By numerical fitting of the data using a function consisting of an oscillating term and a linear term, we derive the parallax. We note that the linear motion term is a combination of internal motion of the maser spots and systemic motion of the star. Both these motions are treated as linear motion in the data analysis. Typically, we need 1.5 to 2.0 years to measure a parallax for one source. As part of this program we have observed 33 sources, with 6 more under monitoring. 
Table 2 shows the target sources of our study. 
Coordinates, variable types, pulsation periods, and the logarithms of the periods are also presented.


\begin{table}[htbp]
\caption{Target LPVs of our study.}
\label{table_verasrc}
\begin{center}
\begin{tabular}{llllll} 
\hline
ID.& Source & RA       & DEC                & Period & Maser \\ 
   &        &[\,\,\,h\,\,\,\,m\,\,\,\,s\,\,\,] & [\,\,\,$^{\circ}$\,\,\,\,'\,\,\,\,"\,\,\,] & [day] &       \\ \hline\hline 
1 & SY~Scl & 00 07 36.25 &  $-$25 29 39.9       & 413 & H$_2$O \\
2 & WX~Psc & 01 06 25.984 &  $+$12 35 53.05     & 660 & H$_2$O \\
3 & RU~Ari & 02 44 45.50 &  $+$12 19 03.0       & 354 & H$_2$O \\
4 & IK~Tau & 03 53 28.87 &  $+$11 24 21.7       & 470 & SiO \\
5 & V637~Per & 03 54 02.28 &  $+$36 32 17.6     & $\cdots$ & H$_2$O \\
6 & R~Tau & 04 28 18.0004 &  $+$10 09 44.770    & 321 & H$_2$O \\
7 & BX~Eri & 04 40 32.754 &  $-$14 12 02.39     & 165 & H$_2$O \\
8 & T~Lep & 05 04 50.84569 &  $-$21 54 16.5172  & 368 & H$_2$O \\
9 & BW~Cam & 05 19 52.56 &  $+$63 15 55.8       & $\cdots$ & H$_2$O \\
10 & RW~Lep & 05 38 52.737 &  $-$14 02 26.84    & 145 & H$_2$O \\
11 & BX~Cam & 05 46 44.29 &  $+$69 58 24.2      & 454 & H$_2$O \\
12 & U~Ori & 05 55 49.16994 &  $+$20 10 30.6872 & 368 & H$_2$O \\
13 & AP~Lyn & 06 34 33.92 &  $+$60 56 26.2      & 450 & H$_2$O \\
14 & U~Lyn & 06 40 46.487 &  $+$59 52 01.64     & 434 & H$_2$O \\
15 & NSV17351 & 07 07 49.38 &  $-$10 44 05.9    & $\cdots$ & H$_2$O \\
16 & Z~Pup & 07 32 38.05674 &  $-$20 39 29.0936 & 509 & H$_2$O \\
17 & OZ~Gem & 07 33 57.75 &  $+$30 30 37.8       & 598 & H$_2$O \\
18 & QX~Pup & 07 42 16.947 &  $-$14 42 50.20     & 551 & H$_2$O \\
19 & V353~Pup & 07 46 34.151 &  $-$32 18 16.26   & $\cdots$ & H$_2$O \\
20 & HU~Pup & 07 55 40.160 &  $-$28 38 54.84     & 238 & H$_2$O \\
21 & R~Cnc & 08 16 33.82789 &  $+$11 43 34.4557 & 362 & H$_2$O \\
22 & X~Hya & 09 35 30.26615 &  $-$14 41 28.6002 & 301 & H$_2$O \\
23 & R~LMi & 09 45 34.28304 &  $+$34 30 42.7839 & 372 & H$_2$O \\
24 & R~Leo & 09 47 33.48791 &  $+$11 25 43.6650 & 310 & SiO \\
25 & R~UMa & 10 44 38.47292 &  $+$68 46 32.7016 & 302 & H$_2$O \\
26 & VX~UMa & 10 55 39.88 &  $+$71 52 09.8       & 215 & H$_2$O \\
27 & HS~UMa & 11 35 30.70408 &  $+$34 52 04.1775 & 517 & H$_2$O \\
28 & S~Crt & 11 52 45.09863 &  $-$07 35 48.0810 & 155 & H$_2$O \\
29 & T~UMa & 12 36 23.46459 &  $+$59 29 12.9746 & 257 & H$_2$O \\
30 & U~CVn & 12 47 19.61 &  $+$38 22 30.5       & 346 & H$_2$O \\
31 & W~Hya & 13 49 01.99810 &  $-$28 22 03.4881 & 361 & H$_2$O \\
32 & RU~Hya & 14 11 34.39861 &  $-$28 53 07.4089 & 332 & H$_2$O \\
33 & RX~Boo & 14 24 11.62662 &  $+$25 42 13.4091 & 278 & H$_2$O \\
34 & RS~Vir & 14 27 16.38997 &  $+$04 40 41.1432 & 354 & H$_2$O \\
35 & FV~Boo & 15 08 25.766 &  $+$09 36 18.19     & 306 & H$_2$O \\
36 & Y~Lib & 15 11 41.30861 &  $-$06 00 41.3727 & 276 & H$_2$O \\
37 & S~CrB & 15 21 23.95608 &  $+$31 22 02.5730 & 360 & H$_2$O \\
38 & WX~Ser & 15 27 47.043 &  $+$19 33 51.71     & 425 & SiO \\
39 & SW~Lib & 15 55 33.42 &  $-$12 51 05.5       & 292 & H$_2$O \\
40 & FS~Lib & 16 00 23.76 &  $-$12 20 57.6       & 415 & H$_2$O \\
41 & IRC$+$10374 & 18 43 36.47 &  $+$13 57 22.8  & 514 & H$_2$O \\
42 & IRC$-$20540 & 19 08 54.62 &  $-$22 14 19.4  & 510 & H$_2$O \\
43 & SY~Aql & 20 07 05.40 &  $+$12 57 06.3       & 356 & H$_2$O \\
44 & SV~Peg & 22 05 42.08385 &  $+$35 20 54.5280 & 145 & H$_2$O \\
45 & R~Peg & 23 06 39.16689 &  $+$10 32 36.0892 & 378 & H$_2$O \\
46 & R~Aqr & 23 43 49.46201 &  $-$15 17 04.1385 & 387 & SiO \\
47 & R~Cas & 23 58 24.87336 &  $+$51 23 19.7011 & 430 & SiO \\
\hline
\end{tabular} 
\end{center}  
\end{table}   

\subsection{Results of our observation with VERA}
\label{sec_results}
The main goal of our VLBI observations with VERA is measuring the parallaxes (and hence, the distances) to our sources. 
Once the distance is determined, we study the properties of the target sources and their PL relation.
Here, we present some results on stellar properties, such as effective temperature, size of the photosphere, and kinematics of the maser spots. A preliminary result of the PL relation study is also presented. 

\subsubsection{Semi-regular variable S~Crt}
\begin{figure}
\begin{center}
\includegraphics[width=72mm, angle=0]{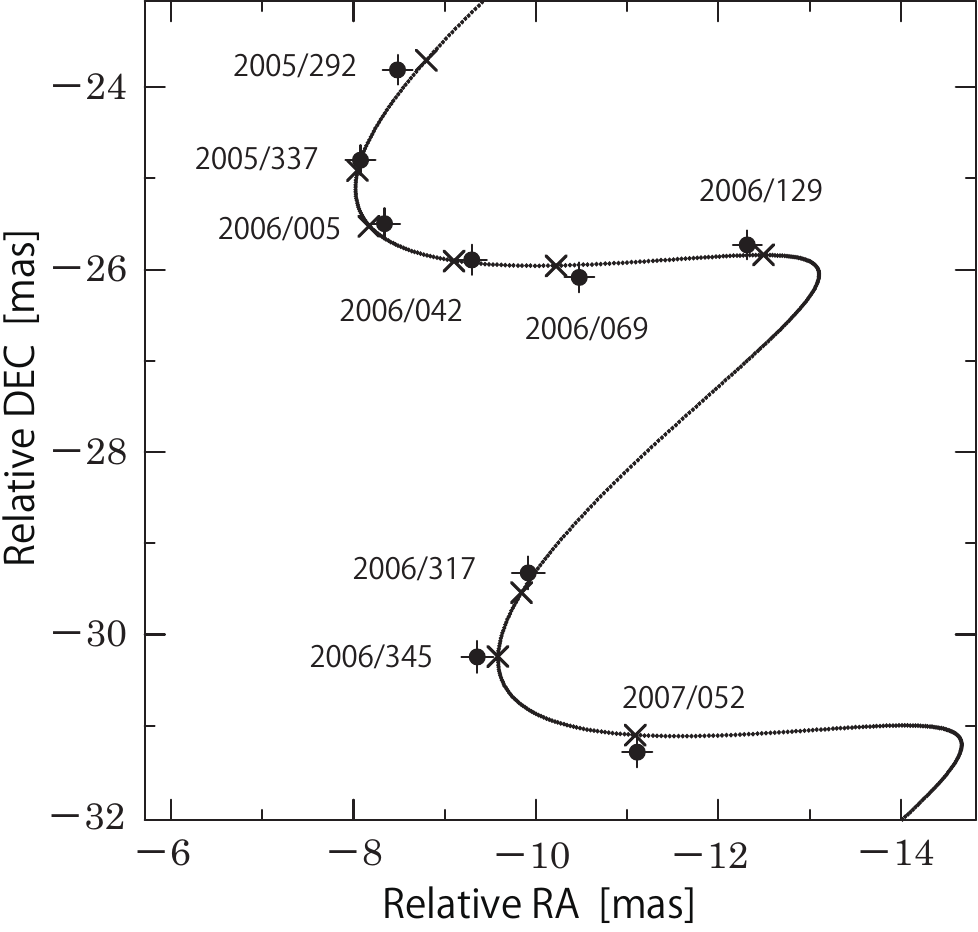}
\caption{Serpentine trajectory of a maser spot in S Crt. Filled circles indicate observed positions, cross symbols indicate best-fit model positions. Combination of a linear proper motion and a parallactic motion gives the serpentine motion.}
\label{fig_scrt_sky}
\end{center}
\end{figure}

\begin{figure}
\begin{center}
\includegraphics[width=72mm, angle=0]{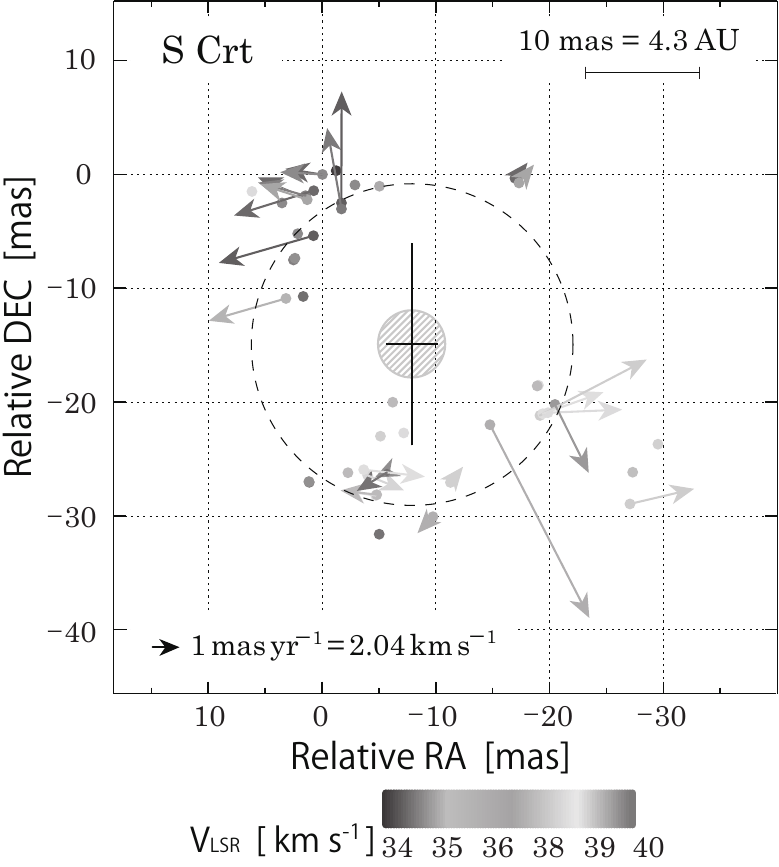}
\caption{Distribution and motions of the maser spots detected around a Mira variable S Crt. 
Inner small circle on the map centre indicates the estimated photospheric size of the star. 
Arrows indicate internal motion of the maser spots, and radial velocity is indicated by gray scale.
The arrow at the bottom-left corner indicates a proper motion of 1\,mas\,yr$^{-1}$, corresponding to 2.04\,km\,s$^{-1}$ at the distance of 430\,pc. The dashed line represents a circular fit to the maser distribution, and has a radius of 14.9\,mas\,(see \citealt{nak08} for more details).  }
\label{fig_scrt_map}
\end{center}
\end{figure}

Figures~\ref{fig_scrt_sky} and \ref{fig_scrt_map} show the results of our VERA observation of the semi-regular variable S~Crt. 
We see a clear serpentine motion of a maser spot on the sky plane (Figure~\ref{fig_scrt_sky}). 
Since the motion of the maser spot is a combination of the linear proper motion and the parallactic ellipse, a resultant motion is observed as the curvature. 
A parallax of 2.33$\pm$0.13\,mas was obtained, and it was converted to the distance of 430$^{+25}_{-23}$\,pc. 
From these series of our VLBI observations, we also estimated the motion of dozens of maser spots in a fixed system around the central star (Figure~\ref{fig_scrt_map}). 
This figure shows the distribution of maser spots on an area of $\sim$ 40 $\times$ 40 mas$^2$ (ˆ$\sim$ 17 $\times$ 17 au$^2$). 
The distribution of the maser spots was found to be relatively compact as much as $\sim$ 10 au in diameter. 
A cross symbol on the map centre indicates the position of the central star, estimated from the distribution and motions of the maser spots. 
Outwards motions and inhomogeneous distribution of maser spots are described in \cite{nak08}. 
\cite{ari99} determined an effective temperature of the photosphere ($T_{\mathrm{BB}}$) of S~Crt to be 3097$\pm$100\,K by fitting two blackbodies to its infrared spectrum. Another temperature of 496 K was attributed to the circumstellar gas and dust shell.
A shaded circle at the center indicate the photosphere size as estimated from the effective temperature of the photosphere ($T_{\mathrm{BB}}$) and observed infrared magnitudes. 
The estimated photosphere radius ($R_{*}$) is $1.81\pm0.14\times10^{13}$\,cm\,(260$\pm$20\,$ R_{\odot}$). 
If we assume that the optical magnitude variation is attributed only to the photospheric radius variation and also assume a constant temperature, then an acceptable radius range of 213 -- 309 $R_{\odot}$ can be estimated. 
In previous studies by \cite{han95} and \cite{van97}, the photospheric radii of dozen of Mira variables are reported to be larger than 300\,$R_{\odot}$. 
This suggest that the photospheric radius of S~Crt may be close to the lower limits to those of Mira variables reported in the literature. 
This result shows once more, how accurate distance measurements are important for better understanding of the difference between Mira variables and semi-regular variables. 

\subsubsection{Mira variable SY~Scl}
The next example of our study is related to the kinematics of a Mira variable SY~Scl. 
The parallax of this source is measured to be 0.75$\pm$0.03\,mas and its corresponding distance is 1.33$\pm$0.05\,k\,pc \cite{nyu11}. 
We also derived the source systemic motion ($\mu^{\star}_x, \mu^{\star}_y$) $=$ (5.57$\pm$0.04, $-$7.32$\pm$0.12) [mas\,yr$^{-1}$]. 

\begin{figure*}
\begin{center}
\includegraphics[width=120mm, angle=0]{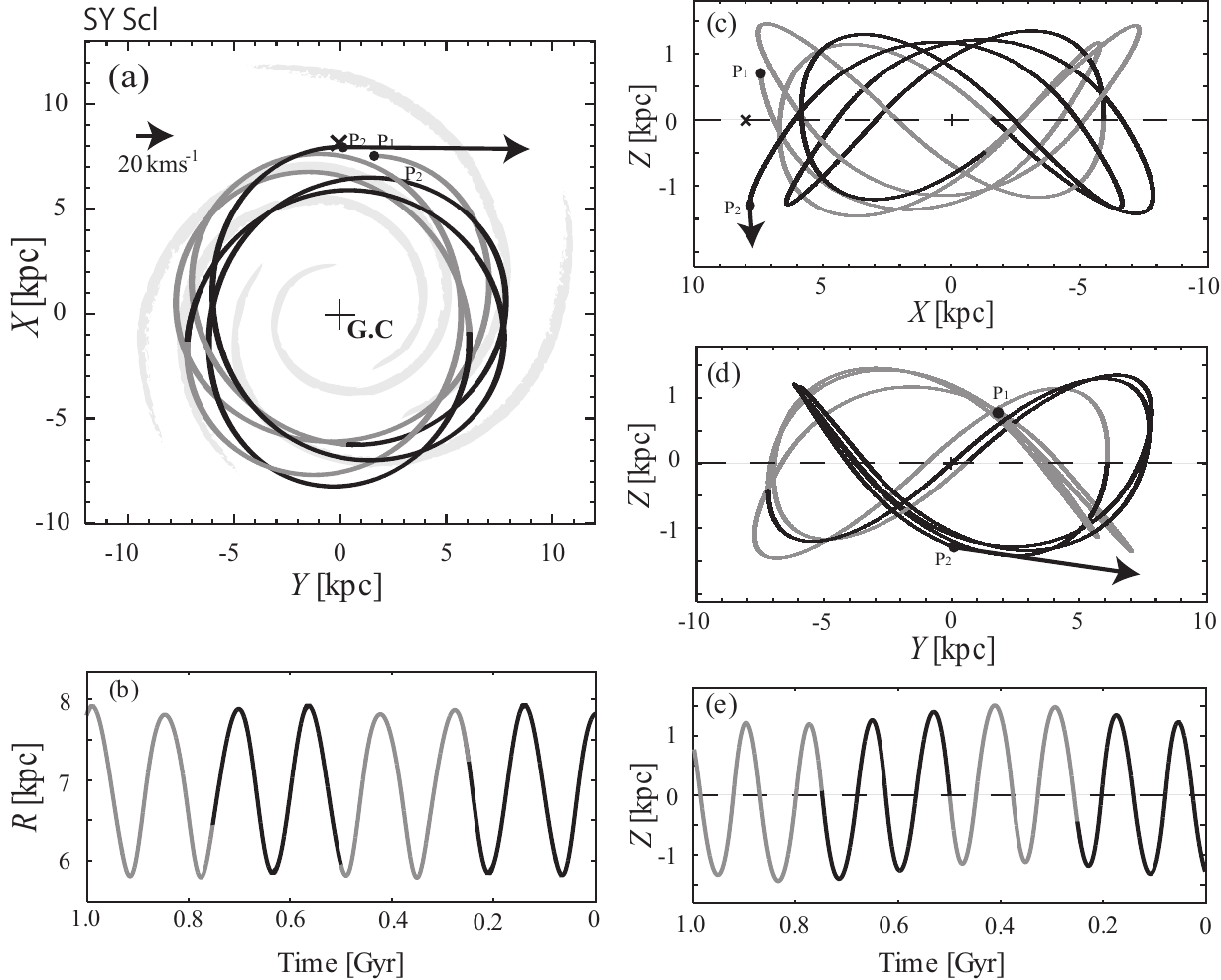}
\caption{
Three dimensional orbit of the Mira variable SY~Scl during the last 1\,Gyr on the schematic view of the Galaxy. 
Black and gray lines indicate the orbit of SY~Scl over 1\,Gyr in which black/gray line changes every 250\,Myr. 
(a) orbit in the face-on view of the Galaxy. 
P$_1$ and P$_2$ indicate the position of SY~Scl 1\,Gyr ago and present, respectively.  
The Sun's location is presented with a cross at ($X_{\odot}, Y_{\odot}, Z_{\odot}$) $=$ (8.0, 0, 0) kpc. 
Spiral arms reported by  \cite{nak06} are shown as light-gray lines in the background of the figure. 
(b) Time variation of radial distance of SY~Scl from the Galactic center.
(c) Orbit in the edge-on view ($X - Z$ plane).
(d) Same as (c) but in the $Y - Z$ plane. 
(e) Time variation of hight from the Galactic plane.  
Dotted lines at $Z =$0 in (a), (b), and (c) represent the Galactic plane. 
}
\label{fig_syscl_orbit}
\end{center}
\end{figure*}

Using these results and radial velocity  of 22 km\,s$^{-1}$ for this source, we obtained a 3-dimensional motion from a numerical simulation. 
This spatial systemic motion is converted to that with respect to the Local Standard of Rest (LSR), by adopting a solar motion of ($U_{\cdot}, V_{\cdot}, W_{\cdot}$) $=$ (10.0$\pm$0.4, 5.2$\pm$0.6, 7.2$\pm$0.4) [km\,s$^{-1}$] \citep{deh98}. 
We found the star has a motion with respect to the LSR ($U$s, $V$s, $W$s) $=$ ($-$5.2$\pm$0.3, $-$53.8$\pm$2.1, $-$33.3$\pm$0.3) [km\,s$^{-1}$], where $U$ points toward the Galactic centre, $V$ in the direction of the Galactic rotation, and $W$ toward the northern Galactic pole.
We can see that SY~Scl is moving approximately southward from the Galactic plane with a peculiar velocity of 63$\pm$2 km\,s$^{-1}$.  
Figure~\ref{fig_syscl_orbit} illustrates the orbits of SY~Scl in the ($X, Y, Z$) coordinate system in the last 1\,Gyr. 
One can see that the orbit shows large distortions and oscillations of the order of kpc scale. 
The radial oscillation has a period of 142\,Myr and shows an amplitude of 1.6\,kpc (Figure~\ref{fig_syscl_orbit}\,b). 
The orbit projected in the $Z$-direction has an amplitude of 1.5\,kpc and the period is 120\,Myr (Figure~\ref{fig_syscl_orbit}\,e). 
This spatial motion suggests that SY~Scl is apparently orbiting the Galaxy as a member of the Galactic thick disk rather than the thin disk. 

\begin{figure}
\begin{center}
\includegraphics[width=73mm, angle=0]{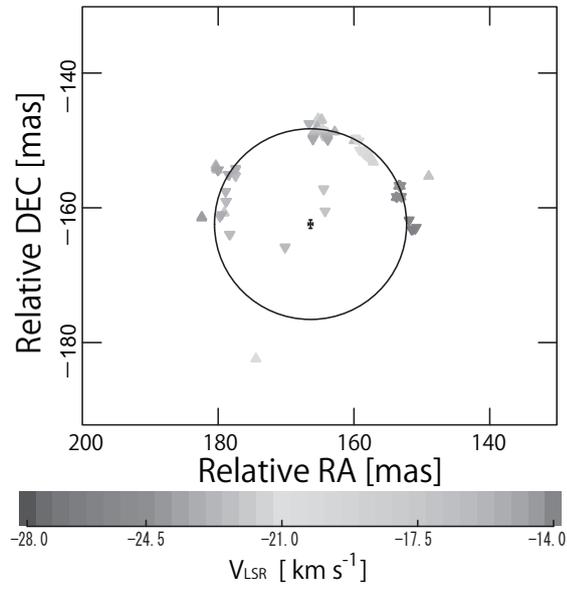}
\caption{
Distribution of SiO masers around R Aqr~\citep{kam10}.
Relative positions of maser spots are given with respect to the J2000 Hipparcos coordinates. 
Triangles pointing upwards represent the distribution of the v $=$ 1, J $=$ 1$-$0 transition and those pointing downwards show the spots of the v $=$ 2, J $=$ 1$-$0 transition.
The circle fitted to the maser spots is shown and the central cross marks its centre and the uncertainty of the central position (at 2$\sigma$-level). 
}
\label{fig_raqr_map}
\end{center}
\end{figure}

\subsubsection{Semi-regular variable RX~Boo}
Long period variables sometime show multiple pulsation periods. 
The semi-regular variable RX~Boo is an example of this case. 
We determined a parallax of 7.31$\pm$0.50\,mas and its corresponding distance of 136$^{+10}_{-9}$\,pc. 
An apparent $K-$band magnitude of $-$1.85 is converted to an absolute magnitude of $-7.35^{+0.15}_{-0.14}$ mag based on the parallax measurement.  
This source is reported to have different pulsation periods pulsation periods, which are summarised in table 4 of \cite{kam12}. 
We found that they can be classified into two typical values, $\sim$160 days and over $\sim$ 320 days. 
From the position of the source on $\log P - M_K$ plane, one can study the stellar structure and pulsation properties. 
 \cite{kam12} compared the $M_K$ and dual periods with the P-L relation in the Magellanic Clouds \citep{ita04-1}. 
Then it was found that RX~Boo belongs to the sequence C (pulsating in fundamental mode) at longer period, and belongs to the sequence C' (pulsating in first-over tone mode) at shorter period.  
This fact indicates that both periods of RX~Boo appear to be related to the properties of RX Boo stellar structure, with each period corresponding to a different pulsation mode. 
\cite{kam12} concluded that the simultaneous enhancement of two modes of pulsation may be evidence for the transient nature of RX~Boo between the two modes. 

\subsubsection{43\,GHz observation of a Mira variable R~Aqr}
Figure~\ref{fig_raqr_map} represents a result of VLBI observation of SiO maser in the Mira variable R~Aqr observed on 24 December 2005 \citep{kam10}.  
Although the observing duration of R~Aqr was longer than one year, it was difficult to find long-lived maser spots which can be used to solve an annual parallax.
\cite{kam10} fitted the spatial distribution of the maser spots with a ring and derived the center position at each observation. 
Then, by tracing the ring centres, a parallax of 4.7$\pm$0.8\,mas was determined. 
The method worked well and a distance of  214$^{+45}_{-32}$\,pc was derived. 
Several years later, \cite{min14} re-analysed the same data set and successfully detected one long-lived maser spot with a lifetime of 1.1 year.
Then, the parallax of R~Aqr was revised to the more accurate value of 4.59$\pm$0.24\,mas with its corresponding distance of 218$^{+12}_{-11}$\,pc. 
The two measurements show consistency within their errors. 


\begin{figure}[htpb]
\begin{center}
\includegraphics[width=80mm, angle=0]{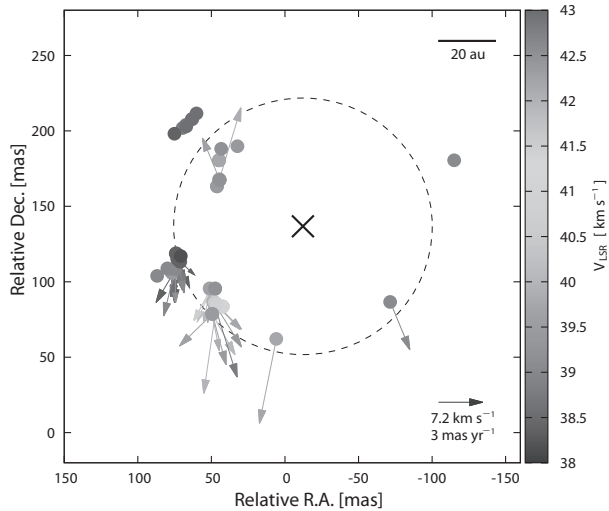}
\caption{ 
H$_2$O maser distribution and internal motion in R~UMa \citep{nak16}. 
A 300\,mas square area, which corresponds to 152 au square, is presented. 
Maser spots are indicated with filled circles. 
H$_2$O maser shell with a radius of 85 mas is presented with a dotted circle. 
This motion is derived by subtracting the VLBI astrometric result from the Hipparcos data. 
See \cite{nak16} for more details. 
}
\label{fig_ruma_map}
\end{center}
\end{figure}

\begin{figure}
\begin{center}
\includegraphics[width=73mm, angle=0]{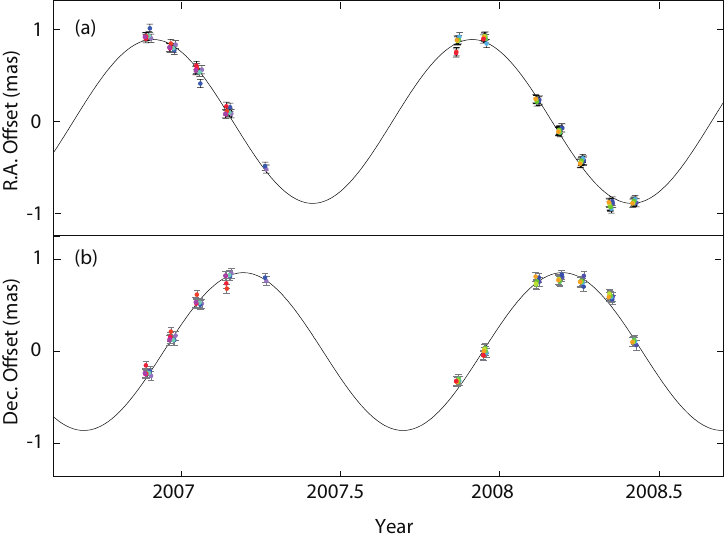}
\caption{Parallactic oscillation along the axes of RA (a) and DEC (b) of the maser spots in a Mira variable R~UMa \citep{nak16}. 
Filled circles and solid lines represent observed data and best fit models, respectively. 
}
\label{fig_ruma_parallax}
\end{center}
\end{figure}

\subsection{Combined analysis of VERA and Hipparcos} 
With our VLBI method, it is very difficult to reveal maser spot motions at a rest frame fixed to the central star because we cannot detect any emission directly from the star. Therefore, we introduce some reasonable assumption in order to reveal the kinematics of the maser spots.  
For example, the isotropy of the circumstellar kinematics was usually assumed. 
In this subsection, we show the results obtained from a new method which combines astrometric measurements from the Hipparcos satellite~\citep{per97} and astrometric VLBI.

In VLBI observations, the detected proper motions of each maser spot ({\boldmath $\mu$}$^{\mathrm{VLBI}}$) consist of three kinds of motions, such as the Galactic rotation, a systemic motion of the star, and their circumstellar motions. 
As for the Hipparcos data, proper motions ({\boldmath $\mu$}$^{\mathrm{HIP}}$) also include the same kinematics as {\boldmath $\mu$}$^{\mathrm{VLBI}}$ except for circumstellar motions of the maser spots. 
So, if we subtract {\boldmath $\mu$}$^{\mathrm{VLBI}}$ from {\boldmath $\mu$}$^{\mathrm{HIP}}$, we can obtain the circumstellar motions of the maser spots on the rest frame fixed to the central star. 

Figure~\ref{fig_ruma_map} shows the distribution and circumstellar motions of maser spots around the Mira variable R~UMa derived with the procedure described above. 
In the revised Hipparcos catalog~\cite{van07}, the proper motion of R~UMa is reported to be {\boldmath $\mu$}$^{\mathrm{HIP}}$ $=(-40.51\pm0.79, -22.66\pm0.78)$ mas\,yr$^{-1}$. 
By subtracting this value from the proper motions of each maser spot obtained from VERA observations, we obtained circumstellar motions for 38 out of all 72 maser spots in R~UMa. 
The motions are indicated with arrows in the figure. 
Angular motion of 1 mas\,yr$^{-1}$ corresponds to 2.41 km\,s$^{-1}$ at the source distance of 508 pc. 
Parallax of this source was measured to be 1.97$\pm$0.05\,mas with its corresponds to the distance of 508$\pm$13\,pc \cite{nak16}.
Figure~\ref{fig_ruma_parallax} represents oscillating terms of the parallactic motion in RA (a) and DEC (b) of the 15 maser spots.

\begin{table}
\caption{VLBI parallaxes of the Galactic long period variables}
\label{table_parallax}       
\begin{tabular}{lllllllll}
\hline\noalign{\smallskip}
Source & Type & Parallax & $P$ & $\mathrm{Log}P$ & $K$ & $M_K$ &Maser & Reference$\ast$  \\
       &      & [mas]          &[day]&           & [mag] & [mag] &      & (Parallax, $K$) \\ 
\noalign{\smallskip} \hline 
\noalign{\smallskip}
RW~Lep & SRa   & 1.62$\pm$0.16   & 150 & 2.176 & 0.639 &    $-8.31\pm0.22$ &H$_2$O& \cite{kam14}, a\\ 
S~Crt  & SRb  & 2.33$\pm$0.13    & 155 & 2.190 &    0.786&   $-7.38\pm0.12$ &H$_2$O& \cite{nak08}, a\\ 
RX~Boo & SRb  & 7.31$\pm$0.5    & 162 & 2.210 & $-$1.96 &  $-7.64\pm0.15$ &H$_2$O& \cite{kam12}, b\\ 
R~UMa  & Mira & 1.97$\pm$0.05  & 302 & 2.480 &  1.19  &  $-7.34\pm0.06$ &H$_2$O&   \cite{nak16} \\ 
FV~Boo& Mira  & 0..97$\pm$0.06 & 340 & 2.531 & 3.836  &  $-6.23\pm0.13$ & H$_2$O&   \cite{kam16-2}, a\\
W~Hya  & SRa  &10.18$\pm$2.36   & 361 & 2.558 & $-$3.16 &  $-8.12\pm0.51$ &OH    & \cite{vle03}, c\\
S~CrB  & Mira & 2.39$\pm$0.17   & 360 & 2.556 &    0.21   &   $-7.90\pm0.15$ &OH    & \cite{vle07}, c\\
T~Lep  & Mira & 3.06$\pm$0.04    & 368 & 2.566 & 0.12&        $-7.45\pm0.03$ &H$_2$O& \cite{nak14}, c\\ 
R~Aqr  & Mira & 4.7$\pm$0.8         & 390 & 2.591 & $-$1.01&   $-7.65\pm0.37$ &SiO   & \cite{kam10}, c\\
R~Aqr  & Mira & 4.59$\pm$0.24      & 390 & 2.591 & $-$1.01&   $-7.70\pm0.11$ &SiO   & \cite{min14}, c\\
RR~Aql & Mira & 1.58$\pm$0.40    & 396 & 2.598 &    0.46 &    $-8.55\pm0.56$ &OH    & \cite{vle07}, c\\
U~Her  & Mira & 3.76$\pm$0.27    & 406 & 2.609 & $-$0.27 &   $-7.39\pm0.16$ &OH    & \cite{vle07}, c\\
SY~Scl & Mira & 0.75$\pm$0.03     & 411 & 2.614 &    2.55&     $-8.07\pm0.09$ &H$_2$O& \cite{nyu11}, b\\ 
R~Cas  & Mira & 5.67$\pm$1.95    & 430 & 2.633 & $-$1.80 &   $-8.03\pm0.78$ &OH    & \cite{vle03}, c\\
U~Lyn  & Mira & 1.27$\pm$0.06     & 434 & 2.637 & 1.533 &     $-7.95\pm0.10$ &H$_2$O& \cite{kam16-1}, a\\ 
UX~Cyg & Mira & 0.54$\pm$0.06   & 565 & 2.752 &    1.40   &   $-9.94\pm0.24$ &H$_2$O& \cite{kur05}, a\\
S~Per & SRc & 0.413$\pm$0.017   & 822 & 2.915 & 1.33 &       $-10.59\pm0.09$ &H$_2$O& \cite{asa10}, b\\
PZ~Cas & SRc &  0.356$\pm$0.026  & 925 & 2.966 & 1.00 &      $-11.24\pm0.16$ &H$_2$O&\cite{ kus13}, b\\
VY~CMa & SRc & 0.88$\pm$0.08   & 956 & 2.980 & $-$0.72 &  $-11.00\pm0.20$ &H$_2$O& \cite{cho08}, b\\
NML~Cyg & --- & 0.62$\pm$0.047  & 1280 & 3.107 & 0.791 &   $-10.25\pm0.16$&H$_2$O & \cite{zha12}, a\\
\noalign{\smallskip}\hline
\end{tabular}
References of the apparent magnitudes ($K$) are as follows : 
(a) The IRSA 2MASS All-Sky Point Source Catalog \citep{cut03}, 
(b) Catalogue of Stellar Photometry in Johnson's 11-colour  system \citep{duc02}, 
(c) Photometry by \cite{whi00}. 
\end{table}

\begin{figure}
\begin{center}
\includegraphics[width=73mm, angle=0]{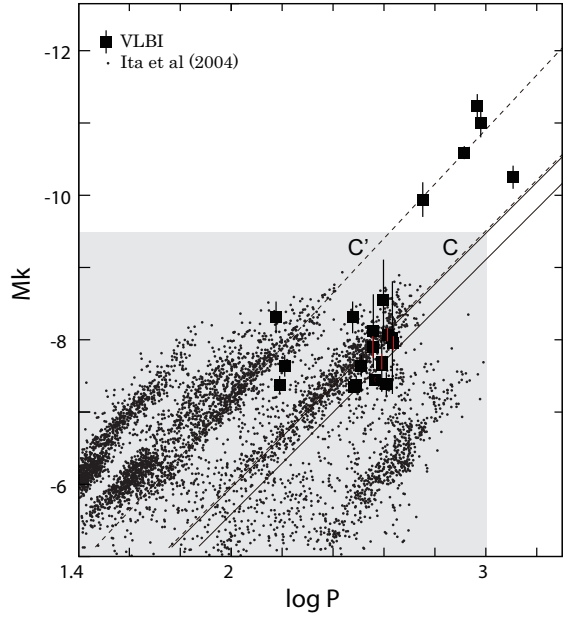}
\caption{
Absolute magnitudes ($M_K$) $-$ $\log P$ diagram of the Galactic long period variables. 
Filled squares indicate sources in table~\ref{table_parallax}, whose distances are derived from VLBI astrometry. 
Solid lines (upper: unweighted, lower: weighted) show the fits of $M_K - \log P$ relations for Galactic Mira variables. 
Small dots in the shaded area represent the LPVs in the LMC \citep{ita04-1}. Dashed lines corresponding to the labels, C and C' represent the fundamental mode and first over-tone sequences given in \cite{ita04-1}. 
}
\label{fig_plr}
\end{center}
\end{figure}

\subsection{Period luminosity relation of the Galactic Mira variables}
In table~\ref{table_parallax}, we summarise the parallax measurements of the Galactic LPVs derived from astrometric VLBI observations in the order of increasing pulsation period. 
It include not only Mira variables, but also semi-regular (SRa, SRb) and red supergiants (SRc).
Maser molecules used in the parallax measurements are also shown. 
References for parallaxes and apparent magnitudes, $K$ are given in the table~\ref{table_parallax} footnote. 
Absolute magnitudes $M_K$ are obtained from the distances and $K$ values. 
Only distance errors were considered for error estimation of $M_K$. 

Here, we define a P-L relation in the form of $M_K = -3.52 \, \mathrm{log}P + \delta$, where we assume a fixed slope of $-3.52$ as determined by \cite{ita04-1}. 
Using Miras and four semi-regular variables (RW~Lep, S~Crt, RX~Boo, and W~Hya) in table~\ref{table_parallax}, we solved for the constant $\delta$. 
Unweighted and weighted least squares fitting to the data gives $\delta$ of 1.09$\pm$0.14 and 1.45$\pm$0.07, respectively. 
In this fitting, the periods of semi-regular variables are being 'fundamentalised' by being multiplied by two \citep{kam12}. 
Since $M_K$ errors of a few sources are quite small compared to other source, it gives a $\delta$ discrepancy of 0.36 between two fittings. 

In Figure~\ref{fig_plr}, filled squares represent sources in table~\ref{table_parallax} on $\log P - M_K$ plane. 
The $\log P - M_K$ relation obtained from unweighted (upper) and weighted (lower) fitting is presented with two solid lines. 
Two dashed lines indicate relations derived by \cite{ita04-1} for sequence C' (first-overtone) and $C$ (fundamental tone), respectively. 
The LPVs in LMC in \cite{ita04-1} are also presented with small dots in a shaded area. 
We used a distance modulus of 18.49 (\citealt{van07}, \citealt{piet13} and \citealt{degrijs14}) to estimate $M_K$ values of LMC stars. 
Since R~UMa falls on the sequence C, it is likely that the star pulsates in a fundamental mode. 
We find a consistency of the relations for Mira variables between our Galaxy and the LMC within the accuracy of $\delta$. 

In summary, we have presented the VLBI measured astrometric results of LPVs, mainly focusing on Mira variables in our Galaxy. 
Better calibration of period-luminosity relation of the Galactic Mira variables is important to estimate their distances free from an assumption that the relation of the LMC and our Galaxy is same. 
We are continuing our VLBI observations for this purpose.

\section{Red Clump Stars}
\label{sec:red-clump}
\begin{figure}
\includegraphics[height=0.99\textwidth,width=1.0\textwidth]{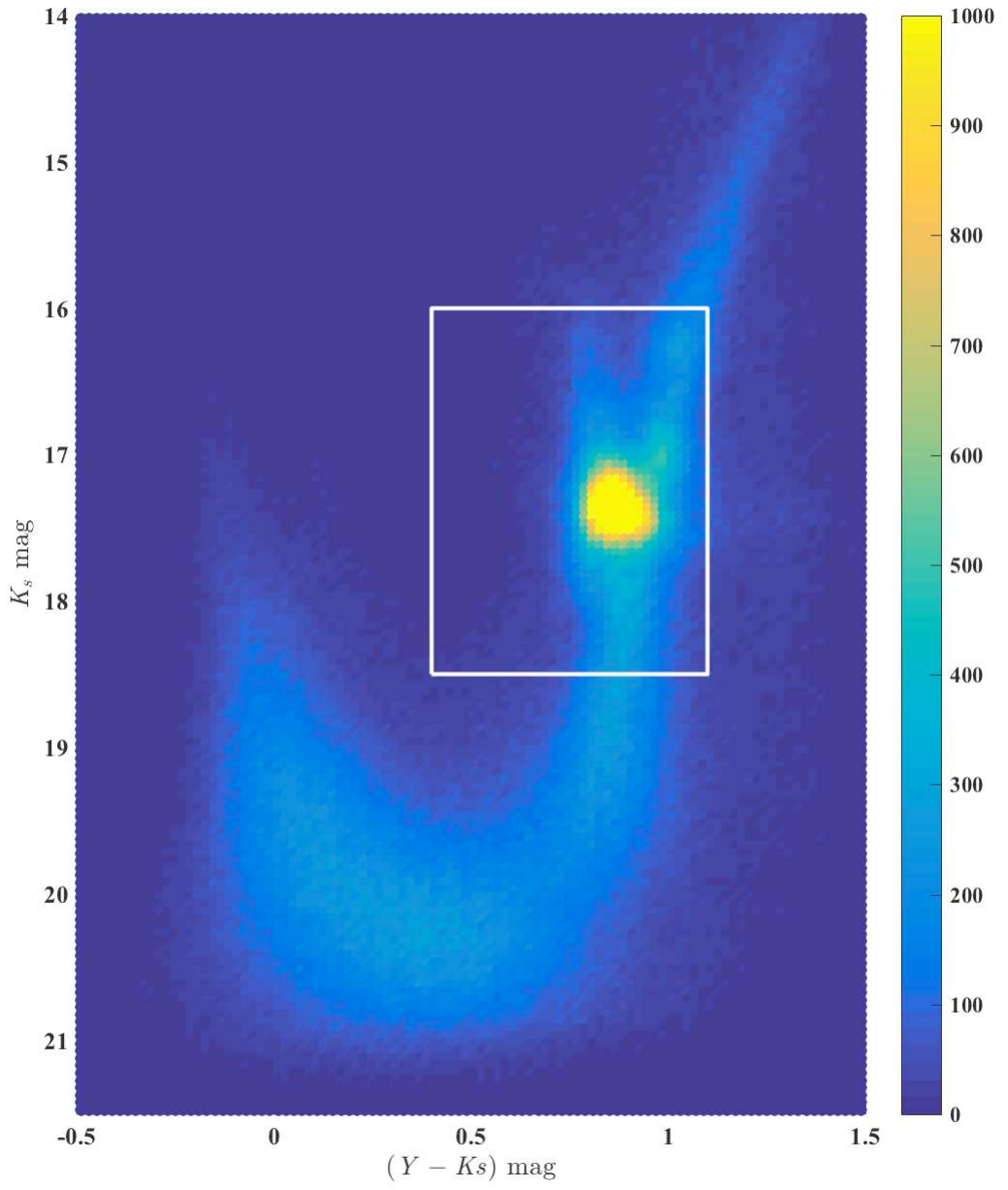}
\includegraphics[height=0.99\textwidth,width=1.0\textwidth]{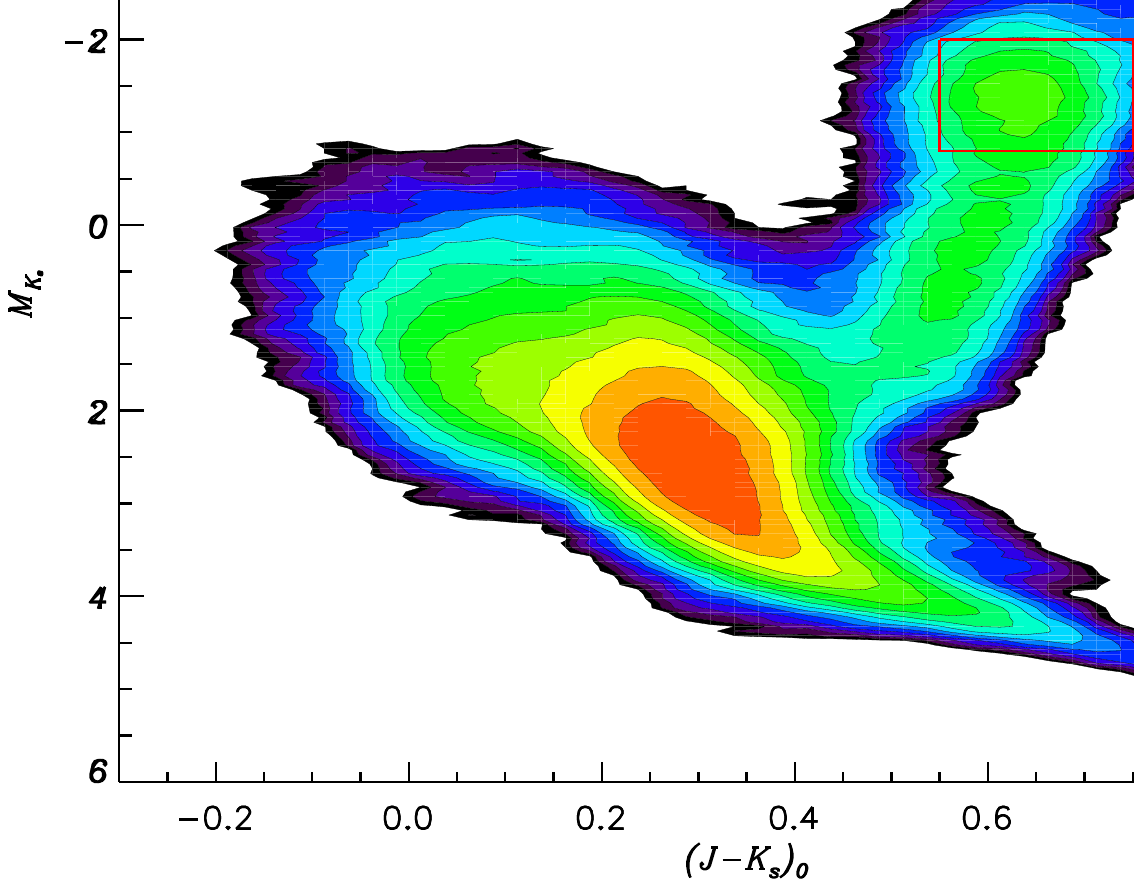}
\caption{Top panel: Hess diagram representing the stellar density in one of the sub-regions of the central VMC tile, SMC4\_3. The colour code represents the increasing stellar density. The white box represents the initial selection box of the RC stars. Bottom panel: Hess diagram corresponding to the Galactic stars that are common in LAMOST spectroscopic survey and Gaia DR1 TGAS data (TGAS parallax err $<$ 30\% and magnitudes from 2 MASS). The red box represents the location of spectroscopically selected RC stars.}
\label{rc_cmd}

\end{figure}

RC stars are intermediate-age ($\sim$ 2 -- 9 Gyr) core helium burning stars, which are easily identifiable in the CMD, close to the RGB, of star clusters and nearby galaxies. The location of RC in the $\it{(Y-K_\textnormal{s})}$ vs $\it{K_\textnormal{s}}$ CMD of a region in the SMC is shown in the upper panel of Fig~\ref{rc_cmd}. The absolute magnitude ($M_{Ks}$) vs $(J-K_s)$ CMD of Galactic stars which are common in Large Sky Area Multi-Object Fibre Spectroscopic Telescope (LAMOST) survey data and Gaia DR1 Tycho-Gaia Astrometric Solution (TGAS) data, with the location of RC (selection of RC stars is described in Sect.\ref{mw_rc}), is shown in the bottom panel of Fig~\ref{rc_cmd}. RC stars have a mass range of 1 -- 3 solar masses and are the younger and metal-rich counter parts of horizontal branch stars. 

Stellar evolutionary models indicate that RC stars should have constant luminosity. The near constancy of the clump absolute magnitude is the result of He ignition in an electron-degenerate core for low mass stars. He burning cannot start until the stellar core mass attains a critical value of about 0.45 M$\odot$. Hence, all these stars have similar core masses at the beginning of He burning, and hence similar luminosities. 

\cite{Cannon1970} compared the observed CMDs of intermediate-age clusters with that derived from stellar evolutionary models and identified RC as a distance indicator. Benefited from the {\it Hipparcos} (ESA\,1997), the absolute magnitudes of RC stars are well calibrated with accurate parallaxes (typically with errors smaller than 10\%) of several hundred local RC stars. Since then, RC stars (which are numerous and bright) have been widely used as standard candles, to estimate distances to the Galactic Centre \citep{PS98} and the Local Group of galaxies (e.g. LMC by \citealt{Laney12}; SMC by \citealt{cole98}; M31 by \citealt{StGr98}). The study of RC stars in the Galactic Bulge revealed double RC feature, which indicated populations at different distances, implying the X-shaped structure of the Bulge \citep{WG13}.  Apart from being used as absolute distance indicators, RC stars are also used to map the extinction (eg: \citealt{haschke11,ben13}) and to study the three dimensional structure (eg: \citealt{os2002}; \citealt{K09}; \citealt{ss10,ss12,subramanian13}) of Magellanic Clouds. 

The general method adopted in the estimation of distance using RC stars is by modelling the RC luminosity function (LF) with a Gaussian + quadratic polynomial term. The  mean and dispersion of the Gaussian function represent the mean RC magnitude and its dispersion respectively. The quadratic polynomial term is to model the RGB stars. Instead of quadratic polynomial, various studies also use power law or exponential function. The mean RC magnitude is compared with the absolute magnitude of solar neighbourhood sample to estimate the distance to star clusters and nearby galaxies. 

Despite the advantages of being numerous, easily identifiable in the CMD and a simple approach to measure the distances using single epoch photometric data in two bands, distance estimation using RC suffers from various caveats and limitations. 

The main caveat of the RC method is the effect on the mean absolute magnitude of RC stars due to the differences in ages and metallicities of the sample with respect to the local Hipparcos sample. This effect is termed as population effect and it is also dependent on the wavelength in which RC stars are observed. One of the earlier striking results from the RC method (with similar modelling of the $\it{I}$ band LF) was the determination of 15\% shorter distance to the Large Magellanic Cloud (LMC) (\citealt{szh98,u98}) than the values assumed to date. This created concerns on the Cepheid distance scale and determination of Hubble constant.  The theoretically predicted population corrections, for the Hipparcos $-$ LMC system, in the $I$ band were $\sim$  0.2 -- 0.3 mag (\citealt{Girardi1998,cole98,gs01}) and hence reconciling with the LMC distance of about 50 kpc.  This suggests the importance of population correction in the estimation of absolute distance using RC stars. 

The other caveat is related to the accuracy of the estimation of the mean RC magnitude from the LF. The LF is contaminated by RGB stars and it is also skewed towards brighter magnitudes due to the presence of large number of young ($\sim$ 400 Myr) RC stars (which appear as vertical extension above the classical RC in the CMD). The presence of 1 Gyr old secondary clump (\citealt{Girardi1998}, compared to classical RC they occupy a fainter and bluer location in the CMD) also makes it difficult to accurately measure the mean magnitude of the RC. The secondary RC stars are those in which the He burning is initiated in a non-degenerate core. Refer the detailed review by \cite{Girardi2016} on the properties of RC stars and its use as distance indicator. 

Again, the advantage of RC stars as distance indicator using photometric observations is mainly restricted to star clusters, Galactic bulge and local group galaxies, where they are identified as a clear clump feature in CMDs of these regions due to their large number and located nearly at same distance. However, for numerous field stars of the Milky Way, it is not easy to single out  individual RC stars. Field RC stars are spread over a wide range of distances along the line of sight causing a vertical structure in the CMD rather than a clump. The differential extinction makes this feature more complicated to be identified easily from the CMD. Spectroscopic surveys are required for the reliable identification and study  of Milky Way field RC stars. 

In this section we discuss the recent studies in the application of RC method for distance estimation, which addresses some of the above mentioned limitations. In Sect.\ref{smc_rc} we describe a potential method to remove the contamination of the RGB stars from the RC LF. Then we discuss the identification of Milky Way field RC stars using the LAMOST spectroscopic sample in Sect.\ref{mw_rc}. 

\subsection{Modelling the near infrared RC Luminosity function of crowded regions in external galaxies}
\label{smc_rc}
As afore-mentioned, the main caveats of applying RC method for distance estimation to external galaxies are the population effects on their absolute magnitude calibration and how well we can model the observed RC LF by incorporating the effects of contaminants, especially the RGB stars. 

The population correction is estimated as follows: the solar neighbourhood RC sample and the external galaxy RC sample are modelled using stellar population models, assuming the star formation rate and age -- metallicity relation of the respective systems. The difference between the theoretically calculated absolute magnitudes of the solar neighbourhood sample and the external galaxy sample is taken as the population correction term. This correction is applied to the observed absolute magnitude of the Hipparcos RC sample and compared with the apparent magnitude of the RC in the external galaxy to estimate the distance modulus. The estimation of population correction term requires knowledge of detailed star formation history of the external galaxy. We note here that the theoretically calculated absolute RC magnitude of external galaxy sample is not directly used as reference for distance estimation due to the limitations in the models \citep{SG02} and for absolute calibration we still rely on highly precise observational values. These population effects strongly affect both $V$ and $I$ band magnitudes in a complex way (as demonstrated by \citealt{SG02,pgu03,Girardi2016}) and hence, limits the application of RC as distance indicator in optical bands. 

The population effects of RC stars in near infrared (NIR) bands, especially in $\it{K_\textnormal{s}}$-band, are less or even negligible  compared to optical bands (\citealt{SG02,pgu03,Girardi2016}). The effect of extinction is also less in $\it{K_\textnormal{s}}$-band. These properties make RC a better standard candle in K$_s$ band. But in the NIR CMDs of crowded regions, the RGB and RC stars are not well separated. \cite{subramanian13} studied the NIR data ($\it{JH}$ bands) of RC stars in the Large Magellanic Cloud (LMC) and showed that the profile fits to the observed RC LF in the crowded central regions are not satisfactory. 

The best way to remove the effect of contaminants in the observed LF of the RC and to obtain their apparent magnitude, is to model the LF using stellar population models, including the star formation rate and age -- metallicity relation obtained from the star formation history of the external galaxies. An attempt of such an elaborate modelling of the LF of the RC stars was preformed by \cite{WG13} in their study on X shaped bulge of our Galaxy.  Apart from the knowledge of the star formation history of the galaxy, this detailed modelling limits the simple application of the RC stars as standard candles. 

In this sub-section we discuss a simple method to remove the contamination of RGB stars from the RC stars, applicable in crowded fields,  using the NIR data from the VISTA survey of the Magellanic Cloud system (VMC). \cite{Nid13} had used a similar technique to remove the contamination of the RGB stars from RC stars from the optical CMD in sparse outer fields of the SMC.

The VMC survey is a continuous and homogeneous ongoing survey of the Magellanic system in the $\it{YJK_\textnormal{s}}$ (central wavelengths, $\lambda_c$ = 1.02 $\mu$m, 1.25 $\mu$m and 2.15 $\mu$m, respectively) near-infrared (NIR) bands using the 4.1 m VISTA telescope located at Paranal Observatory in Chile. The limiting magnitudes with signal-to-noise (S/N) ratio of 5, for single-epoch observations of each tile, in the $\it{Y, J} $ and $\it{K_\textnormal{s}}$ bands are $\sim$ 21.1 mag, 20.5 mag and 19.2 mag, respectively, in the Vega system. The stacked images can provide sources with limiting magnitudes of up to 21.5 mag in $\it{K_\textnormal{s}}$ with S/N = 5.  A detailed description of the VMC survey is given by \cite{cioni11}. The RC feature in the SMC ($\it{K_\textnormal{s}}$ $\sim$ 17.3 mag) is around 2 mag brighter than the 5 $\sigma$ detection limit of single-epoch observations. 

As the $\it{(Y-K_\textnormal{s})}$ colour gives the widest colour separation possible in the VMC data, which allows a better separation between RC and RGB stars, we use the $\it{Y}$ and $\it{K}_{s}$ band photometric data to do the RC selection and further analysis. 
Hess diagrams, with bin sizes of 0.01 mag in $\it{(Y-K_\textnormal{s})}$ colour and 0.04 mag in $\it{K_\textnormal{s}}$, representing the stellar density in the observed $\it{(Y-K_\textnormal{s})}$ vs $\it{K_\textnormal{s}}$ CMD of one of the central VMC tiles, SMC 4\_3 is shown in Figure~\ref{rc_cmd}.
The RC stars, at $(Y-K_\textnormal{s})$ $\sim$ 0.7 mag and $\it{K_\textnormal{s}}$ $\sim$ 17.3 mag, is easily identifiable in the CMD. 

First we defined a box with size, 0.5 $\le$ $(Y-K_\textnormal{s})$ $\le$ 1.1 mag in colour and 16.0 $\le$ $\it{K_\textnormal{s}}$ $\le$ 18.5 mag in magnitude, for the selection of the RC stars from the CMD. The selection box is also shown Figure~\ref{rc_cmd}. The range in magnitude was chosen to include the vertical extent (clearly seen in eastern regions of the SMC, refer \citealt{ss17}) of the RC regions and also to incorporate the shift towards fainter magnitudes due to interstellar  extinction.  

\begin{figure*}
\centering
\includegraphics[height=0.95\textwidth,width=0.95\textwidth]{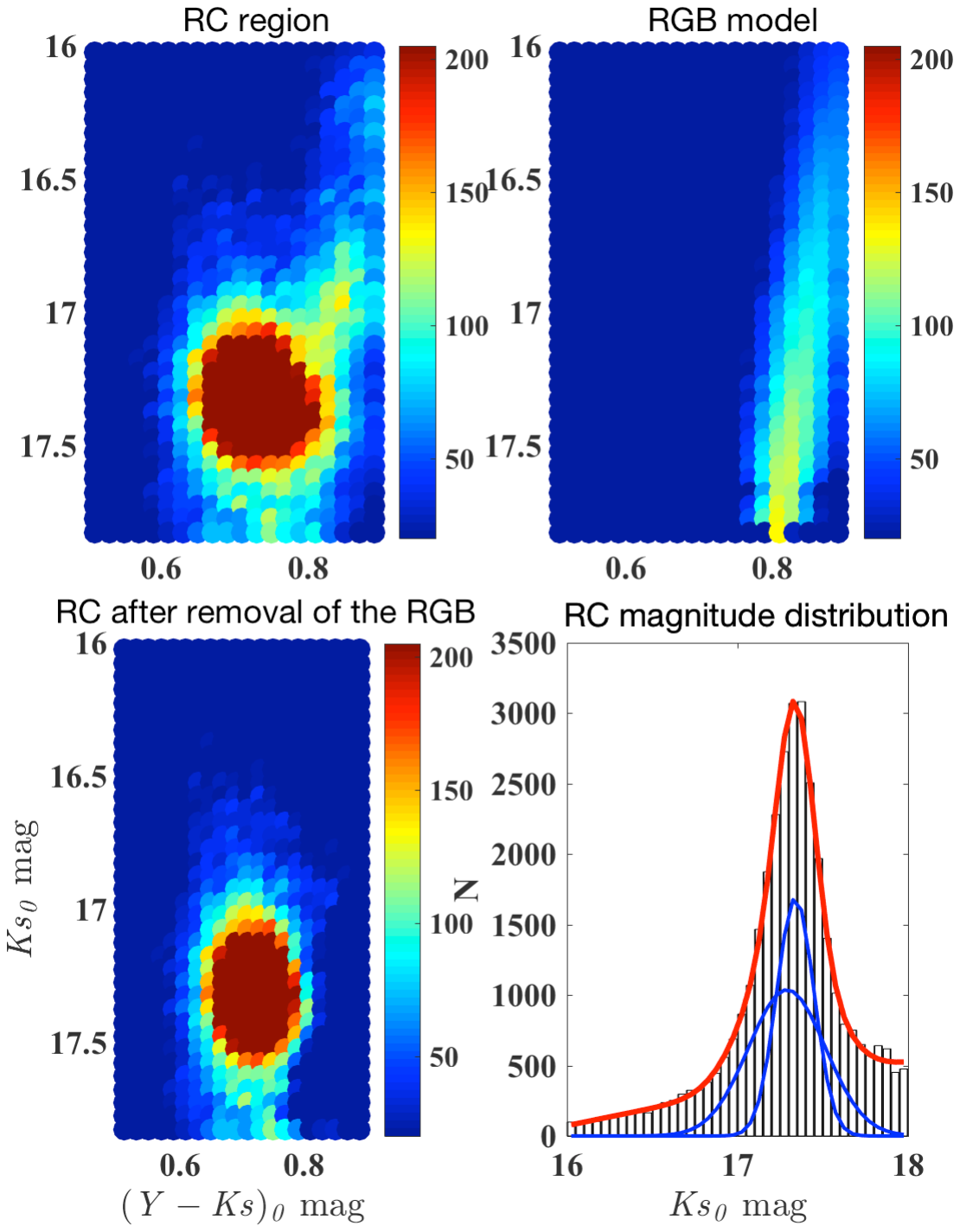} 
\caption{Steps involved in deriving the luminosity function of the RC stars and the profile fitting to the distribution for one sub-field of tile SMC 4\_3. The top-left and top-right plots show the reddening-corrected and MW-foreground-subtracted RC region and the RGB model obtained for the sub-field respectively. The bottom-left and bottom-right plots show the RGB subtracted RC region and the luminosity function of the RC stars with the profile fits respectively. The total fit (thick-red line) to the distribution and the separate components (thin-blue line) are also shown in the bottom-right plot.}
\label{rc_method}
\end{figure*}

The colour cut which separates the RC and RGB is not very well defined in the CMD of the crowded central region. Hence, RGB stars are also included in the initial selection box. Later we will model the RGB density and subtract it from the Hess diagrams to obtain the RC distribution. Before modelling the RGB stars, the RC region in the CMD is cleaned by removing the closest matching star corresponding to the colour and magnitude of each foreground Milky Way (MW) star from the TRILEGAL MW stellar population model \citep{Leo2005} which includes a model of extinction in the Galaxy. The cleaned CMD region is then corrected for foreground and internal extinction using the extinction map of \cite{.pdf2015}. 

We performed a careful analysis to remove RGB contamination in order to analyse the RC LF.  Tile SMC 4\_3 was divided into four sub-regions. The RC stars in each sub-region were identified and Hess diagrams corresponding to the RC region were constructed. The bin sizes in colour and magnitude were 0.01 mag and 0.04 mag, respectively. The separation of the RC from the RGB  based on their colour may not be reliable for the fainter magnitudes and we restrict the analysis of luminosity function up to $K_{s,0}$ = 18 mag. This fainter magnitude cut off of  $K_{s,0}$ = 18 mag is $\sim$ 0.6 -- 0.7 mag lower than the peak of the Classical RC. In a given magnitude bin (along a row in the Hess diagram), we performed a double Gaussian profile fit to the colour distribution and obtained the peak colours corresponding to the RC and the RGB distributions. This was repeated for all magnitude bins, over the magnitude range of 16.0 -- 18.0 mag in the $K_{s,0}$ band.  After this first step, we performed a linear least-squares fit to the RGB colours with the magnitude corresponding to each bin. Again, the first step was repeated with the constraint that the colour of the RGB corresponding to each magnitude is the same as that obtained from the linear fit. From the resultant Gaussian parameters corresponding to the RGB, we obtained a model for the RGB density distribution. This model was subtracted from the Hess diagram and a clean RC distribution is obtained. These steps are demonstrated in Figure~\ref{rc_method}. 

The figure clearly demonstrates that the contamination from the RGB stars is minimal in the final RC distribution.  The final RC density distribution is summed in the colour range 0.55 $\le$ $(Y-K_s)_{0}$ $\le$ 0.85 mag to obtain the LF of the RC stars. The resultant LF is initially modelled with a Gaussian profile, representing the RC, and a quadratic polynomial term to include the presence of any remaining RGB component. The fit improves by 30\% when an additional broad component is included. The LF and the profile fits are shown in the lower-right panel of Figure~\ref{rc_method}. The narrow component represents the Classical RC. The broad component could be due to the presence of young stars and/or the intrinsic line of sight depth of the SMC central regions. If the broad component is due to the presence of young stars, further division of the sub-regions into smaller regions would suppress this component owing to the smaller fraction of young stars compared to Classical RC. On the other hand, if this component is due to the intrinsic line of sight depth of the SMC then it would still show up. Earlier studies of the SMC central regions (using the RC as well as other tracers such as RR Lyrae stars, \citealt{kh12}) show large line of sight depth in the central region. So the observed broad component could be due to the line of sight depth effect. In any case, for distance estimation we are only interested in the peak of the Classical RC, which is dominant and represented by the narrow component. 

Thus, the method described above effectively removes the contamination of RGB stars, even in the crowded fields, and provides a clean RC distribution to estimate the  apparent magnitude of Classical RC. The final step in distance estimation is to apply the population correction term and compare with absolute RC magnitude of Hipparcos sample.

\cite{ss17} applied the above procedure to the NIR data of $\sim$ 20 deg$^2$ region of the SMC and estimated a mean K$_{s,0}$ band peak magnitude corresponding to Classical RC as 17.36 $\pm$ 0.01 mag. This magnitude is used to estimate the distance modulus to the central SMC by using the the absolute magnitude of the RC stars provided by \cite{Laney12}. Using high-precision observations of solar neighbourhood RC stars, they provide the absolute RC magnitudes in the $J$,$H$ and $K$ bands in 2MASS system. We converted them into the VISTA $K_{S}$ system using the transformation relations provided by \cite{Rubele2015}. The absolute magnitude of RC stars in the VISTA $K_{S}$ band, $M_{Ks}$ = $-$1.604$\pm$0.015 mag.  As indicated earlier, the absolute magnitudes of RC stars in the solar neighbourhood and in the SMC are expected to be different owing to the differences in metallicity, age and star formation rate between the two regions. \cite{SG02} estimated this correction term in the $K$ band to be $-0.07$ mag. We applied this correction and estimated the mean distance modulus to the SMC as  18.89 $\pm$ 0.04mag. 

Our estimate is higher than the previous estimates of SMC distance based on RC LF in optical bands, including those which incorporated population corrections (\citealt{cole98}). But it is lower than the recent accurate estimates of distance to the SMC (18.96 $\pm$ 0.02 mag, \citealt{g14,db15}). This difference could be mainly due to the inaccuracy of the population correction term. As suggested by \cite{pgu03} the population corrections calculated from models are not accurate enough for high-precision distance measurements. Better models and accurate star formation history estimates of the observed galaxy are essential to determine the population correction term and hence use of RC as accurate distance indicators.

Now we have stellar population models (such as PARSEC \citealt{Bressan2012}) with improved knowledge of opacities, solar abundance and microphysics. Better estimates of star formation history of nearby galaxies, using ground as well as space based deep observations, are now available. Gaia parallaxes and stellar parameters will improve the absolute calibration of the RC. Better estimates of population correction, combined with more accurate absolute RC magnitude will provide better results in future to use RC as a high-precision distance indicator. 

The best way to use RC as an accurate standard candle would be to perform complete CMD fitting procedure using improved and complete set of models. This fitting procedure will be naturally driven by the dominant intermediate-age population in the CMD and hence, the RC. This is not a simple procedure and for galaxies like the SMC, which has considerable amount of depth along the line of sight, requires a distance distribution instead of discrete distance grid for better results. 



\subsection{Selection of Milky Way field RC stars}
\label{mw_rc}
 For Galactic centre or the Local Group of galaxies, the advantage of RC stars as distance indicators is that there is a striking, easily identifiable feature in the CMD of stars in those distant objects due to the presence of a large number of RC stars at nearly the same distance. 
However, it is not easy to identify individual RC stars from the numerous field stars of the Milky Way. 
Field RC stars are spread over a wide range of distances and there is no apparent over-density in the CMD of field stars that can be used to identify the RC members.

Fortunately, high-resolution spectroscopic analysis of nearby RC stars show that they distribute in a relative ``small box'' in the $T_{\rm eff}$\,--\,log\,$g$ diagram (i.e. HR diagram) of $4800$\,$\leq$\,$T_{\rm eff}$\,$\leq$\,$5200$\,K and $2.0$\,$\leq$\,${\rm log}$\,$g$\,$\leq$\,$3.0$ (e.g. \citealt{2010MNRAS.408.1225P}).
This is simply because  $T_{\rm eff}$ is an excellent proxy of stellar colour  whereas  log\,$g$ is sensitive to the stellar absolute luminosity.
Recently, with values of $T_{\rm eff}$ and log\,$g$ available from large-scale spectroscopic surveys  for large numbers of field stars, it has become feasible to select a large number of  field RC candidates.
RC candidates thus selected have been  widely used to study the metallicity gradients and stellar kinematics of the Galactic disk(s) (e.g. \citealt{2012MNRAS.421.3362B,2011MNRAS.412.2026S,2013MNRAS.436..101W,2014A&A...571A..92B}).

However, stars falling inside  that ``small box'' are not purely RC stars and have significant ($\sim 60$\%; \citealt{2013MNRAS.436..101W}) contamination from RGB.
The differences in absolute magnitudes between RC and RGB stars can be larger than 1\,mag and this can lead to large systematic errors in distances of the selected  RC sample stars.
More recently, a new method has been proposed by \cite{2014ApJ...790..127B} (hereafter B14) to select a clean RC sample from large-scale spectroscopic data.
The method first separates the RC-like and RGB stars using cuts on a metallicity-dependent $T_{\rm eff}$\,--\,log\,$g$ diagram, assisted by theoretical stellar isochrones and calibrated using {\em Kepler} high quality asteroseismic log\,$g$ (e.g. \citealt{Creevey2013}), as well as high precision stellar atmospheric parameters from the APOGEE survey \citep{2014ApJS..215...19P}.
Secondary RC stars are then removed from the RC-like stars via cuts on a metallicity ($Z$)\,--\,colour  $(J-K_{\rm s})_{0}$ diagram.
The expected purity of the final RC sample is $\geq$\,$93$\,\% and the typical distance errors are within 5\,--\,10\,\%.
The key point of this new method in selecting a clean RC sample is the surface gravity log\,$g$.   

\begin{figure*}
\centering
\includegraphics[scale=0.35,angle=0]{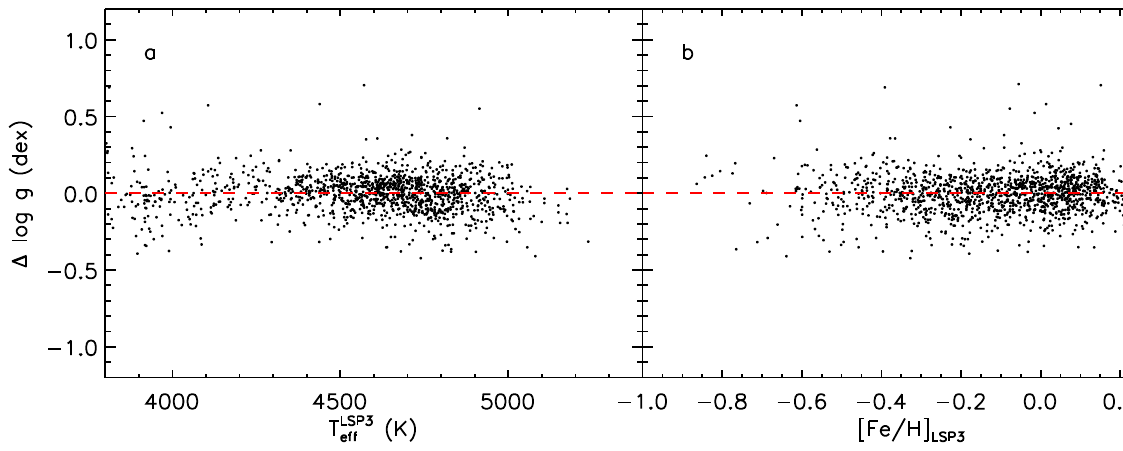}
\caption{Distributions of values of residuals, log\,$g_{\rm KPCA}$ $-$ log\,$g_{\rm AST}$ , of the training sample as a function of LSP3 $T_{\rm eff}$ and [Fe/H] (panels a and b, respectively) and of asteroseismic log\,$g$ (panel c). Panel (d) shows a histogram distribution of the residuals (black line). Also over plotted in red is a Gaussian fit to the distribution. The mean $\mu$ = 0.00 dex and dispersion $\sigma$ = 0.10 dex of the fit, as well as the number of stars used, N = 1355, are marked. For the current KPCA analysis, N$_{\rm PC}$ = 25 is assumed. See the text for details. The figure is taken from H15.}
\label{mwrc_1}
\end{figure*}

\begin{figure*}
\centering
\includegraphics[scale=0.35,angle=0]{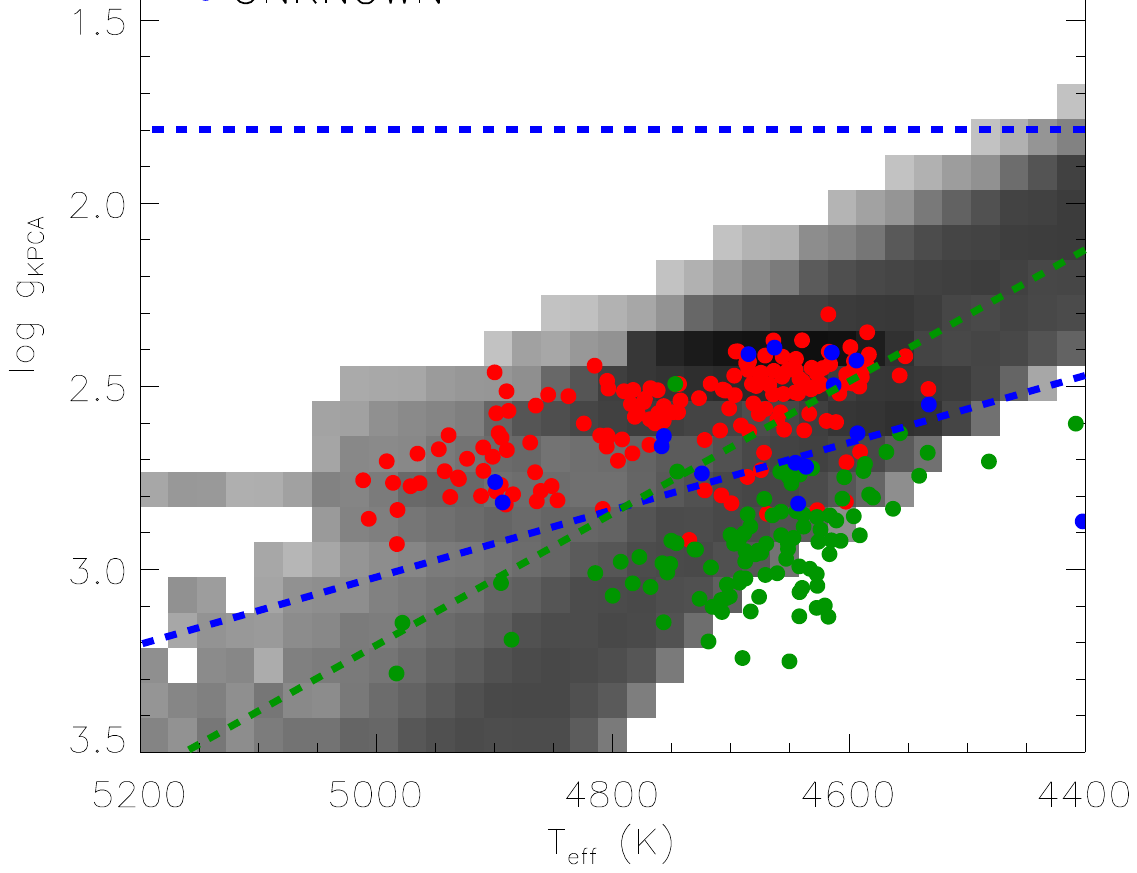}
\includegraphics[scale=0.35,angle=0]{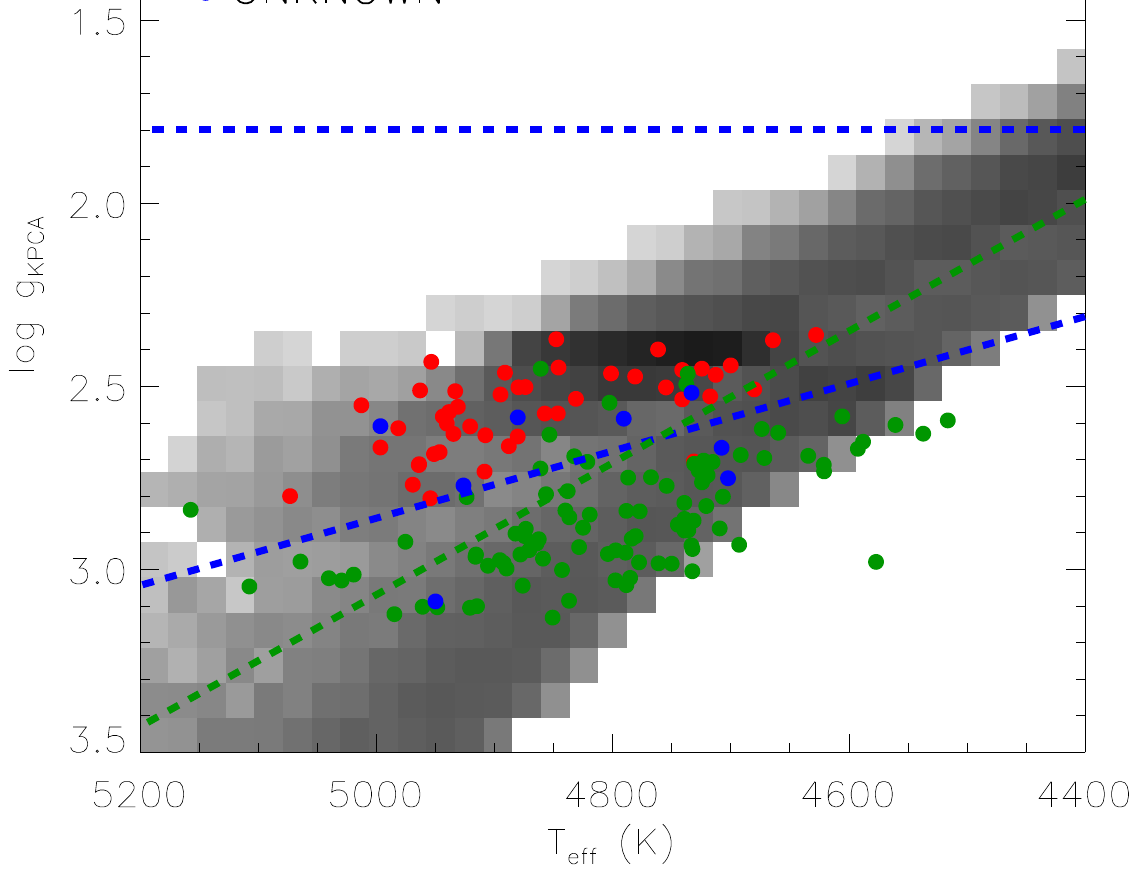}
\caption{
Distribution of stars on the $T_{\rm eff}$\,--\, $\log\,g$ , HR diagram  predicted by the PARSEC stellar evolution model \citep{Bressan2012}, compared to that of the  asteroseismic sample with evolution stages classified (coloured points, \citealt{Stello2013}, for two metallicity bins as marked.
Values of $T_{\rm eff}$ and [Fe/H] of the asteroseismic sample stars are from LSP3 and those of $\log\,g$ from the KPCA method.
Red, magenta and blue points represent stars of  RC, RGB and unknown evolutionary stages, as classified by \cite{Stello2013}, respectively.
The blue dashed lines represent the cuts that separate RC stars and  the less luminous RGB stars as given in  Eqs (1) and (2) in H15.
The green dashed lines give  the cuts adopted  by B14. The figure is taken from H15.}
\label{mwrc_2}
\end{figure*}

\begin{figure}
\centering
\includegraphics[scale=0.70,angle=0]{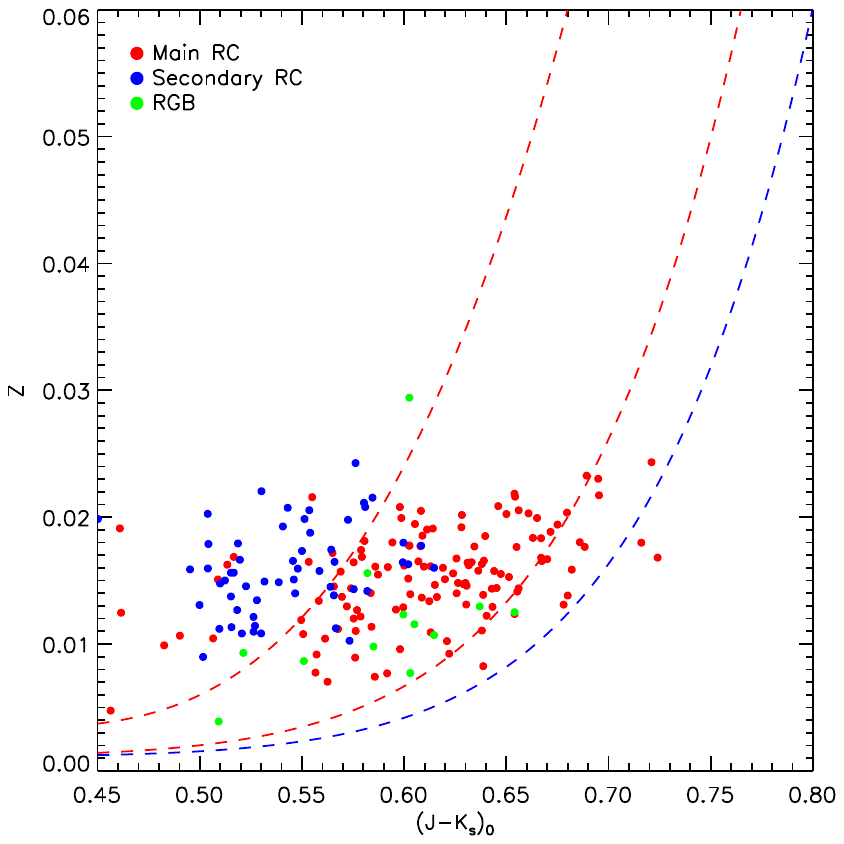}
\caption{Distribution of RC-like stars in the colour  $(J-K_{\rm s})_{0}$--Z plane, using 2MASS photometric measurements and LSP3 metallicities.
Dots in red, blue and green  represent main RC, secondary RC and RGB stars, respectively.
The red dashed lines are developed by B14 to eliminate the contamination of secondary RC stars.
The blue dashed line is the revised cut  developed in H15 in replacement of the lower red dashed line.
The figure is taken from H15.}
\label{mwrc_3}
\end{figure}

Unfortunately, it is difficult to apply this method to the LAMOST spectroscopic survey of the galactic anti-centre (LSS-GAC, \citealt{2014IAUS..310.321, 2015MNRAS.448..855Y}) data given the relatively large systematic plus random errors ($\sim$\,0.2\,--\,0.4\, dex) of  log\,$g$ estimates delivered by the LAMOST Stellar Parameter Pipeline at Peking University (LSP3; \citealt{Xiang2015}), as shown by a recent comparison of LSP3 and asteroseismic  log\,$g$ values for common objects in the LAMOST--$Kepler$ fields \citep{Ren2016}.    
To apply the similar method developed by B14 to select pure RC stars from LSS-GAC data, we have improved log\,$g$ measurements using  a Kernel Principal Component Analysis (KPCA, \citealt{kpca98}) method trained by accurate asteroseismic data from the LAMOST--Kepler fields (\citealt{2015RAA....15.1240H}, hereafter H15).
First, the training sample is constructed by cross-matching our sub-sample of about 50\,000 stars in the LAMOST--Kepler fields with the currently available largest asteroseismic
log\,$g$ sample ($\sim$16 000 stars) from \cite{Huber2014}. 
In total, 3562 common sources with SNR\,$\ge$\,10 are identified.
To further select red giant stars and ensure high quality of the stellar parameters yielded by LSP3, we take a sub-sample of 1355 stars from the 3562 common sources.
Secondly, we apply the KPCA method to the training sample to construct the relations between seismic log\,$g$ and the LAMOST spectra (see Figure~\ref{mwrc_1}).
Finally, We apply the constructed relations to the LAMOST spectra and re-determine log\,$g$ values for all LSS-GAC DR2 red giants. 
The accuracy of the newly estimated KPCA log\,$g$ is as high as 0.1-0.15\,dex and has no significant systematic error.
With the high accurate, newly estimated KPCA log\,$g$, we can apply the similar technique developed by B14 to select main RC from LAMOST data.
To show how well the technique works for LAMOST data, we present the two steps of the selections as mentioned above in Figs. \ref{mwrc_2} and \ref{mwrc_3}.

\subsubsection{The LSS-GAC RC sample and its applications}
\begin{figure*}
\centering
\includegraphics[scale=0.225,angle=0]{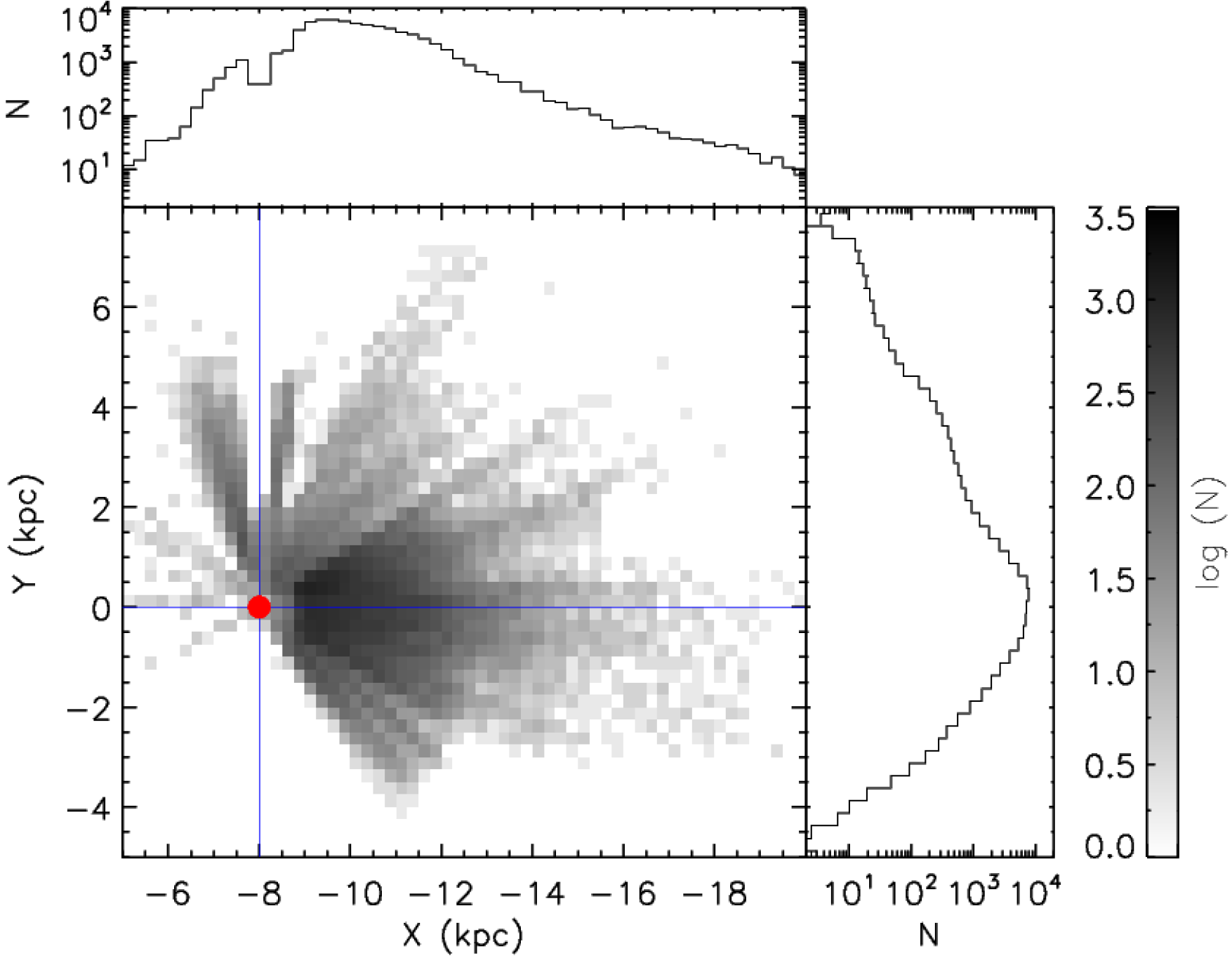}
\includegraphics[scale=0.225,angle=0]{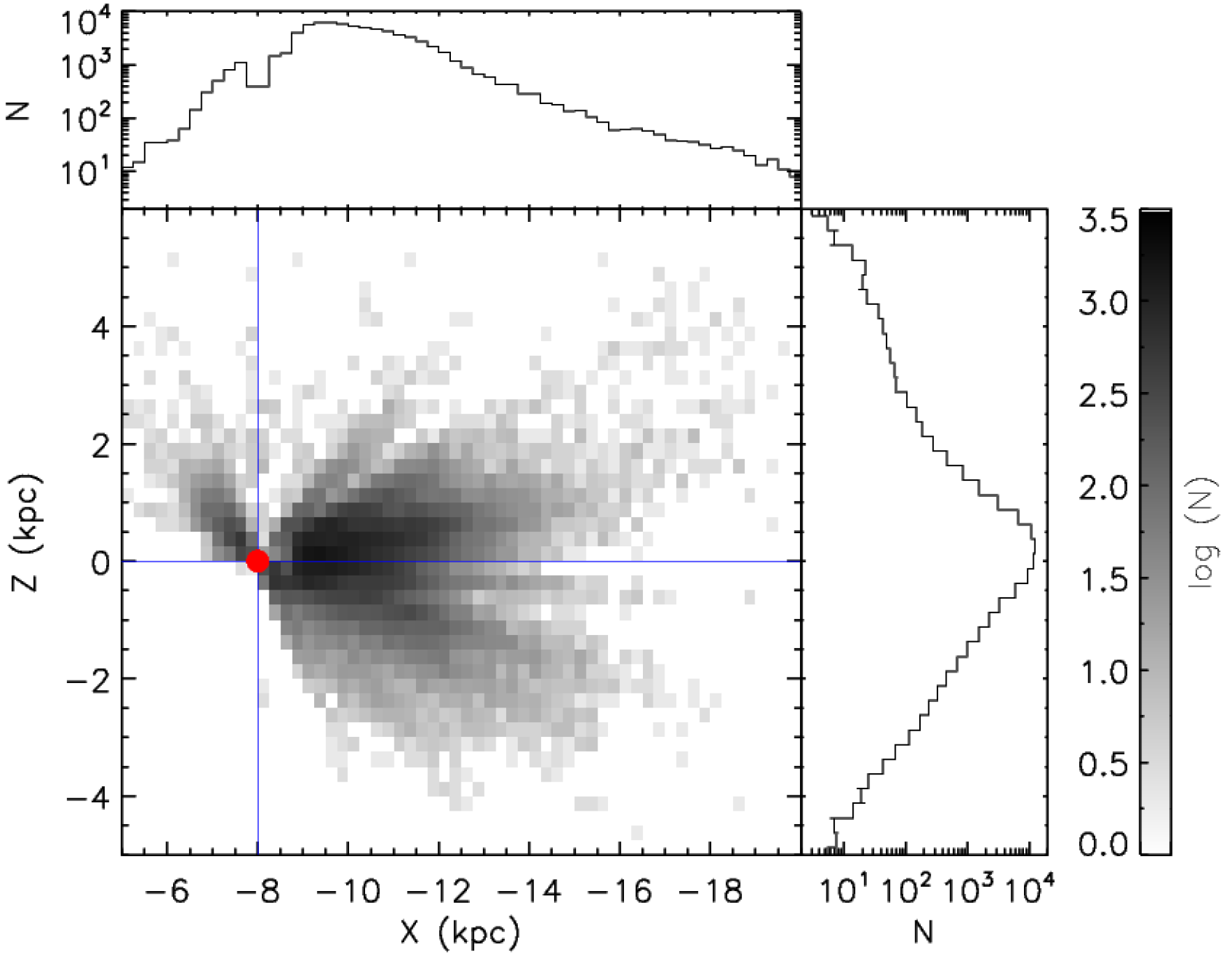}
\caption{Greys cale number density distribution of the 0.11 million LSS-GAC RC sample stars in the $X$\,--\,$Y$ (left panel) and $X$\,--\,$Z$ (right panel) planes. 
The Sun is located at ($X$, $Y$, $Z$) = ($-8.0$, 0.0, 0.0) kpc.
The stars are binned by 0.25$\times$0.25 kpc$^{2}$ in both diagrams. 
The densities are shown on a logarithmic scale.
Histogram distributions along the $X$, $Y$ and $Z$ axes are also plotted.}
\label{mwrc_4}
\end{figure*}

With $T_{\rm eff}$, [Fe/H] estimated with the LSP3, $\log\,g$, derived with the KPCA method and $(J-K_{\rm s})_{0}$ calculated from 2MASS photometry\footnote{Only stars with  {\it ph\_qual} flagged as `A' in both $J$ and $K_{\rm s}$ bands are included.} after corrected for extinction as estimated with the `star pair' method \citep{2015MNRAS.448..855Y}, we apply the similar technique developed by B14 to LSS-GAC DR2 red giant sample and obtain a clean RC sample of over 0.11 million stars of S/N\,(4650\,\AA)\,$\geq\,10$ (see Figure~\ref{mwrc_4}).
The distances of these RC stars are derived using the recent calibration, $M_{K_{\rm s}}\,=\,-1.61$\,mag, for nearby RC sample \citep{Laney12}.
Since the intrinsic scatter of absolute magnitudes of RC stars  is estimated to be within 0.1\,mag, the distances derived are expected to have uncertainties no more than 5\,--\,10\,\%, given  a typical photometric error of $\sim\,$0.05\,mag in $K_{\rm s}$-band and an extinction error of $\sim\,$0.04\,mag in $E(B-V)$ (see \citealt{2015MNRAS.448..855Y}).
Proper motions of  the sample stars are taken  from the U.S. Naval Observatory CCD Astrograph Catalog (UCAC4, \citealt{2013AJ....145...44Z}) and PPMXL \citep{2010AJ....139.2440R} catalogues
Given the high distance precision and large spatial coverage (Figure~\ref{mwrc_3}) of the current RC star sample, it is very useful to tackle a variety of problems with regard to the Galactic chemistry (e.g. stellar metallicity gradients of Galactic disk, listed below), structures  and  dynamics (e.g. Galactic rotation curve, listed below). 
The sample is publicly available at website:  ({http://162.105.156.249/site/RC\_Sample}). 

{\bf Galactic rotation curve}: 
Using the pure RC sample selected from LSS-GAC and the APOGEE surveys, combined with a sample of halo K giants selected from the  SDSS/SEGUE, the Galactic rotation curve (RC) between 8 and 100\,kpc in Galactocentric radius has been constructed \citep{2016MNRAS.463.2623H}.
The newly constructed RC has a generally flat value of $240$\,km\,s$^{-1}$ within a Galactocentric radius $r$ of 25\,kpc and then decreases steadily to $150$\,km\,s$^{-1}$ at $r\sim$\,100\,kpc.
On top of this overall trend, the RC exhibits two prominent localised dips, one at $r\sim$\,11\,kpc and another at $r\,\sim$\,19\,kpc.
The dips could be explained by assuming the existence of two massive (dark) matter rings in the Galactic plane.
From the newly constructed RC, combined with other data, we have built a mass model of the Galaxy,  yielding a virial mass of the Milky Way's dark matter halo of {$0.90^{+0.07}_{-0.08} \times 10^{12}$\,${\rm M}_{\odot}$} and a local dark matter density, { $\rho_{\rm \odot, dm} = 0.32^{+0.02}_{-0.02}$\,GeV\,cm$^{-3}$}.

{\bf Metallicity gradients of the Galactic disk}: 
We also have determined the radial and vertical metallicity gradients of the Galactic disk(s), again using the LSS-GAC pure RC sample (H15).
Our analysis shows that, near the solar circle ($7 \leq R \leq 11.5$\,kpc), the radial metallicity gradient exhibits a negative slope that flattens with increasing  $|Z|$.
In contrast,  in the outer disk ($11.5\,<\,R\,\leq\,14$\,kpc), the radial metallicity gradient has an essentially  constant, much less steep slope  at all heights from the mid-plane, suggesting that the outer disk may have experienced an evolutionary path different from that of the inner disk.
In addition, the vertical metallicity gradient is found to flatten roughly with increasing $R$. 
The flattening of the gradient in the lower disk ($0\,\leq\,|Z|\,\leq\,1$\,kpc) is much quicker than in the upper disk ($1\,<\,|Z|\,\leq\,3$\,kpc).

The parallaxes and stellar parameters of Galaxy RC sample from Gaia will improve the zero-point calibration. Combing the improved zero-point calibration with the results from on-going and future asterosesmic surveys such as PLAnetary Transits and Oscillations of stars \citep{Rauer2014},  the Transiting Exoplanet Survey Satellite \citep{Ricker2014}, and the second phase of Kepler (K2; \citealt{Stello2015}) will provide large sample of RC stars to perform Galactic archaeology.

\section{Final Remarks}
We provided an overview of the properties of young and intermediate-age stars, such as Cepheids, Mira stars and RC stars in the context of their application as distance indicators. Below given are the salient points. \\

\begin{itemize} 

\item {\bf Cepheids} 

\begin{itemize}

\item {\it NIR Cepheid distances} \\
\begin{itemize} 

\item NIR PW relations allow us to derive Cepheid distances with the accuracy to meet the precision of ongoing cosmological experiments. They are less affected by extinction and metallicity effects. By using NIR PW relations, \cite{inno13} obtained the most precise estimate of distances to the Magellanic Clouds based on Cepheids, with random errors at 1\% level and systematics at 5\% level. Independent calibrations from Gaia and HST will play a crucial role to reach a better accuracy. \\

\item New NIR light-curve templates can provide accurate mean magnitudes from a few or even one observation, allowing a significant saving of observing time and an optimal usage of data already available. \\

\end{itemize}
 
\item {\it MIR Cepheid distances} \\

\begin{itemize} MIR studies of Cepheids lead to significant advantages with respect to optical wavelengths. The $L$ band appears to be ideal wavelength range for these measurements, lacking the presence of variable CO absorption that is instead affecting the $M$ band.\\

\item The CHP project \cite{2011AJ....142..192F} has demonstrated that the 3.6~$\mu$m Cepheid PL relation can be effectively used to measure the Hubble constant and other cosmological parameters with a precision comparable to the accuracy (better than $\sim 3$\%) in Planck's CMB and BAO analysis \cite{2012ApJ...758...24F}. \\

\item Accurate parallaxes for Galactic Cepheids expected with the final Gaia data release, combined with future JWST L band observations of Cepheids in host galaxies of Type Ia Supernovae will enable 1\% measurements of H$_0$ as well as competitive measurements of other cosmological parameters. \\

\end{itemize} 

\item {\it Linearity of PL relation} \\

\begin{itemize}

\item The statistical analysis of non-linearities in the LMC PL relations provides evidence of a break in both fundamental mode and first-overtone mode relations. \\

\item The observed non-linearities suggest a correlation with sharp changes in the light curve structure of Cepheids and needs a theoretical investigation to look for additional resonance features. \\

\item The effect of these non-linearities in the slope of HST-based PL relations is minimal. Hence, does not have any statistically significant effect on the distance scale or the value of Hubble constant. But they will have a significant impact on the accuracies of distance scale with more precise PL relations, for example from James Webb Space Telescope. \\

\end{itemize}

\item {\it Metallicity effects in PL relation} \\

\begin{itemize}

\item NIR Surface Brightness method which is a geometrical method, enables us to independently estimate the distances to nearby Cepheids. \\

\item  The PL relations constructed from these independent distance estimates for Cepheids in Milky Way, LMC and SMC are ideal to investigate the effect of metallicity in Cepheid PL relations. \\

\item Preliminary results indicate that the effect of metallicity, especially in $K$ band PL relation, is very small. \\

\end{itemize}

\end{itemize}

\item {\bf Mira variables}\\

\begin{itemize}

\item For better calibration of Galactic PL relation of Mira variables and to understand their physical properties, very precise parallax measurements are essential. We present the strategy and initial results of an on-going VLBI astrometric survey of Galactic Miras. \\

\item The initial results show consistent PL relations for Mira variables between our Galaxy and the LMC. \\

\end{itemize}

\item{\bf Red Clump stars}\\

\begin{itemize}

\item {\it Galactic Bulge and External galaxies} \\
\begin{itemize} 

\item We present a simple method to remove the contamination of RGB stars from the RC LF function in crowded fields and hence provide better measurements of the apparent magnitude of Classical RC. This technique is used to estimate the mean distance modulus of the SMC as 18.89 $\pm$ 0.04 mag (including a population correction term). We note that the population correction terms obtained from models may not be accurate enough to use RC as high precision absolute distance indicator. \\

\item New stellar population models, improved star formation history measurements of the galaxies and improved calibration of absolute RC magnitude from Gaia will all together provide better results to use RC as accurate distance indicator. \\

\item Though not simple, another way to use RC as accurate standard candle is to perform complete CMD fitting procedure using improved and complete set of stellar population models. \\

\end{itemize}

\item{\it Field RC stars in Milky Way}\\

\begin{itemize}

\item Recent spectroscopic surveys allow us to select a large number of field RC stars in Milky Way. A technique to identify a clean sample of  RC stars from field stars in Milky Way using LAMOST spectroscopic survey is presented. This sample is used to derive the metallicity gradient and rotation curve of our Galaxy. \\

\item The parallaxes and stellar parameters from Gaia, combined with the results from ongoing and future asterosesmic surveys will provide a large sample of RC stars to perform Galactic archaeology. \\

\end{itemize} 

\end{itemize}

\end{itemize}

\begin{acknowledgements}
SS acknowledges research funding support from Chinese Postdoctoral Science Foundation (grant number 2016M590013). YH thanks the support by the by the National Key Basic Research Program of China 2014CB845700, the National Natural Science Foundation of China 11473001 and the LAMOST FELLOWSHIP. LI acknowledge support from the  Sonderforschungsbereich SFB 881 {\it The Milky Way System} (subproject A3) of the German Research Foundation (DFG).
\end{acknowledgements}

\end{document}